\documentclass[a4paper,11pt, preprint]{article}
\pdfoutput=1 

\usepackage{jcappub} 
\usepackage{braket}
\usepackage{dsfont}
\usepackage{overpic}
\usepackage{subcaption}
\usepackage{transparent}
\usepackage[many]{tcolorbox}

\usepackage[T1]{fontenc} 

\newcommand{\pd}{\text{d}}
\newcommand{\bx}{\boldsymbol{x}}
\newcommand{\bu}{\boldsymbol{u}}
\title{Gravitational waves from\\cosmic strings for pedestrians}

\author{Kai Schmitz,}
\author{Tobias Schröder}

\affiliation{Institute for Theoretical Physics, University of M\"unster\\
Wilhelm-Klemm-Straße 9, 48149 M\"unster, Germany}

\emailAdd{kai.schmitz@uni-muenster.de}
\emailAdd{schroeder.tobias@uni-muenster.de}

\abstract{Cosmic strings represent an attractive source of gravitational waves (GWs) from the early Universe. {However, numerical computation of the GW signal from cosmic strings requires the evaluation of complicated integral and sum expressions, which can become computationally costly in large parameter scans. This motivates us to} rederive the GW signal from a network of local stable cosmic strings in the Nambu--Goto approximation and based on the velocity-dependent one-scale model from a ``pedestrian'' perspective. That is, we derive purely analytical expressions for the total GW spectrum, which remain exact wherever possible and whose error can be tracked and reduced in a controlled way in crucial situations in which we are forced to introduce approximations. In this way, we obtain powerful formulas that, {unlike existing results in the literature}, are valid across the entire frequency spectrum and across the entire conceivable range of cosmic-string tensions. We provide an in-depth discussion of the GW spectra thus obtained, including their characteristic break frequencies and approximate power-law behaviors, comment on the effect of changes in the effective number of degrees of freedom during radiation domination, and conclude with a concise summary of our main formulas that can readily be used in future studies.}

\begin{document}

\hfill MS-TP-24-38

\maketitle
\flushbottom

\section{Introduction}
\label{sec:intro}
Cosmic strings are tube-like field configurations that can arise in cosmological phase transitions in the early Universe. The spontaneous symmetry breaking $G\to H$ associated with such a phase transition leads via the Kibble--Zurek mechanism to the formation of at least one topologically stabilized string per Hubble volume whenever the vacuum manifold $\mathcal{M}\cong G/H$ is not simply connected, i.e., $\pi_1\left(\mathcal{M}\right)\ncong \Set{\mathds{1}}$ \cite{Kibble_1976, Kibble_1980, Zurek_1985}. The simplest example for such a symmetry breaking is $U(1)\to \Set{\mathds{1}}$, while strings are also ubiquitous in symmetry-breaking chains of grand unified theories \cite{Jeannerot_2003}. Inside the string, the phase transition is prevented, and the symmetry remains unbroken. Correspondingly, the true vacuum after the phase transition is not realized inside the string, which, therefore, carries energy.

There is a large variety of observational consequences of cosmic strings. Firstly, strings can affect the cosmic microwave background (CMB) by introducing changes in the temperature anisotropy angular power spectrum~\cite{Dvorkin_2011, Ringeval:2012tk, Planck_2013, Charnock_2016} as well as step-like temperature discontinuities via the Gott--Kaiser--Stebbins effect~\cite{Gott_1984, Kaiser_1984, Bouchet_1988, Planck_2013}. Strings can also act as gravitational lenses~\cite{Vilenkin_1984, Mack:2007ae, Kuijken:2007ma, Bloomfield:2013jka} and enhance the 21\,cm signal due to overdensities inside the string wake~\cite{Brandenberger_2010, Maibach_2021}. Apart from these gravitational signatures, strings might be observed via cosmic rays~\cite{Vilenkin:1986zz, Bhattacharjee:1989vu, Garfinkle:1987yw, Auclair:2019jip} or, in the case of superconducting strings, via CMB spectral distortions~\cite{Tashiro:2012nb, Anthonisen:2015tra, Ramberg:2022irf}. Furthermore, they have been used to explain the baryon asymmetry of the Universe \cite{Brandenberger:1994bx, Brandenberger:1994mq} and recently attracted attention as possible seeds of structure formation explaining the overabundance of high-redshift galaxies observed by the James Webb Space Telescope~\cite{Jiao_2023}. One of the most promising ways to detect cosmic strings is, however, by means of their gravitational radiation.

Since strings cannot have endpoints, they are either closed loops or infinitely long. While the latter cannot decay due to winding-number conservation, there is no such obstruction for the former. The details of this decay, and in particular the dominant decay channel, depend on the particle physics model under consideration. If the strings arise from the breaking of a global symmetry, they will usually mainly decay into the associated Nambu--Goldstone bosons~\cite{Davis_1985}. In the case of a local symmetry breaking, the decay into gravitational radiation will be dominant \cite{Vachaspati:1984yi, Brandenberger:1986vj, Blanco-Pillado:1998tyu, Auclair:2019jip}. Henceforth, we will focus on the latter case. Local strings emanating from symmetry breakings at an energy scale $\eta$ typically carry an energy per unit length (string tension) of $\mu\propto \eta^2$ and have a width $\delta \propto \eta^{-1}$~\cite{Vilenkin_Shellard_2000}. Though very thin, strings typically form with infinite length or with super-horizon length. In string--string and string self-interactions, the strings intercommute, i.e., they exchange parts. This process leads to the production of string loops of sub-horizon length, which are chopped off from the long strings. These loops start to oscillate under their tension and thus emit gravitational radiation. Accordingly, they lose energy, implying that they shrink over time and finally evaporate. The long-string network also emits gravitational radiation \cite{Figueroa:2012kw, Figueroa:2020lvo, CamargoNevesdaCunha:2022mvg}. Its contribution to the SGWB is, however, strongly suppressed compared to the loop contribution \cite{Buchmuller:2013lra}. 

The energy loss of the long-string network ensures that it reaches a so-called scaling regime in which the energy density in strings becomes a constant fraction of the critical density. Since the long strings are topologically prohibited from decaying,\footnote{This is actually only true for infinitely long strings. Super-Hubble string loops are nonetheless prevented from decaying as they are stretched due to the cosmic expansion.} the long-string network will never vanish, continuously produce loops, and, hence, also gravitational waves (GWs). As a consequence, the stochastic gravitational wave background (SGWB) from cosmic strings will typically have contributions from early times at very large frequencies ($f>{\rm MHz}$) as well as contributions from late times, dominant at very low frequencies ($f<{\rm nHz}$), making it appealing in two regards: First, the GW spectrum contains information about the expansion history of the Universe from times as early as string formation until today. Second, it can be constrained by many different current and future experiments such as ground-based interferometers like LIGO \cite{LIGOScientific:2014pky}, Virgo \cite{VIRGO:2014yos} or KAGRA \cite{Somiya:2011np, Aso:2013eba}, space-based interferometers like LISA \cite{amaroseoane2017laserinterferometerspaceantenna}, DECIGO \cite{Seto:2001qf}, or BBO \cite{Corbin:2005ny}, as well as pulsar timing arrays (PTAs) like NANOGrav \cite{McLaughlin:2013ira}, EPTA \cite{Kramer:2013kea} or SKA \cite{Carilli:2004nx}. 

PTA experiments recently found strong evidence for the existence of an SGWB \cite{NANOGrav:2023gor, Xu:2023wog, EPTA:2023fyk, Reardon:2023gzh}. Since the spectral shape of the measurements disfavors stable strings as a sole explanation, these current PTA measurements give, most likely, only an upper bound on the string tension. Meanwhile, cosmic superstrings, which are fundamental strings or one-dimensional D-branes of cosmologically relevant size that arise at the end of brane inflation~\cite{Dvali:2003zj, Copeland:2003bj}, yield spectra that give an excellent fit to the NANOGrav 15-year data~\cite{NANOGrav:2023hvm, Ellis:2023oxs}. {The GW signal from superstrings as well as from color-flux tubes can, in the PTA band, to a first approximation, be obtained from that of ordinary field strings by rescaling the amplitude of the GW spectrum with a constant prefactor, although this is not true over a wider range of frequencies~\cite{Sousa_2016, Yamada_2022, Yamada:2022aax, Avgoustidis:2025svu}. The analytical results presented in this paper are, therefore, directly relevant for PTA searches for GWs from superstrings and color-flux tubes as well.} 

To draw any conclusions from the observation of a SGWB about the underlying physics, a proper analytical understanding of all features of the SGWB from cosmic strings for different parts of parameter space is vital. In this paper, we provide detailed derivations and analytical expressions of such features as well as analytical approximations of the total GW spectrum. {Previous work in this regard was mainly conducted in~\cite{Sousa_2020} with a focus on the fundamental-mode contribution to the GWB and in~\cite{Blanco-Pillado:2024aca} with a focus on the total GWB spectrum.}
The expressions we derive in the following extend over a wider range of parameters and cover qualitatively different spectra that can not be described by means of the expressions derived in earlier works~\cite{Sousa_2020, Blanco-Pillado:2024aca, Servant:2023tua}.
In addition to improving the understanding of all the parameter dependencies from the analytical expressions in detail, the typically long computation times to obtain the SGWB signal numerically are drastically reduced. In this sense,  the present paper can be regarded as the ``pedestrian'' route to the GW signal from cosmic strings. While the accuracy of our results will not always match the accuracy of exact numerical results, we achieve a purely analytical derivation that does not require the numerical evaluation of complicated integral and sum expressions. On top, in crucial situations in which we are forced to introduce approximations that decrease the accuracy of our results we are able to retain control over the induced error in the sense that it can be systematically reduced.

Our results are thus suited for fast numerical parameter scans at the cost of a minimally reduced accuracy. For other sources of GWs from the early Universe, GW templates of this form have already been available for a long time. In simple models, the GW signal from cosmic inflation can, e.g., be modeled in terms of a constant or running power law~\cite{Guzzetti:2016mkm}. Similarly, the GW signal from a cosmological phase transition can in many models be approximated by a broken or doubly broken power law~\cite{Caprini:2024hue}. For cosmic strings, simple analytical templates for the GW signal were, however, unavailable until now, at least no templates valid at all frequencies and capable of covering all possible types of GW spectra that can be realized in cosmic-string models. An important motivation behind the present paper is to close this gap and work out such analytical templates that no longer require further numerical steps. 

The remainder of this article is structured as follows: In Section~\ref{sec:SGWBfromLoops}, we will first review
the velocity-dependent one-scale model describing the evolution of a network of long strings. Assuming this evolution, we will present the steps involved in finding a general expression for the SGWB from string loops. We will furthermore show how the GW spectrum is affected by the finite width of cosmic strings. In Section~\ref{sec:FundamentalMode}, we will rigorously derive analytical expressions for the GW spectrum obtained from the fundamental excitation of string loops. We will discuss all qualitatively different spectra obtained in different regions of parameter space and analytically investigate their features in detail. Additionally, we will show how the effect of variations in the effective number of relativistic degrees of freedom (DOFs) in the early Universe can be included in the spectrum. Equipped with these results, we will continue in Section~\ref{sec:TotalSpectrum} by providing an accurate analytic estimate of the total spectrum. A handy summary of all expressions necessary to calculate the total spectrum is provided in Section~\ref{sec:Summary_of_Formulae}. \\\\
\textbf{Notation and conventions:}
We use $G$ to denote Newton's constant and units such that $c=\hbar = 1$. We consider a spatially flat Friedmann--Lema\^itre--Robertson--Walker (FLRW) background spacetime, $\pd s^2 = a^2\left(\eta\right)\left(\pd \eta^2-\pd \bx^2\right)$. The term ``horizon'' refers in this paper to the particle horizon that would occur in a Big Bang Cosmology model but which is not necessarily the actual causal horizon if one assumes an early inflationary phase.

\section{Stochastic gravitational-wave background from loops}
\label{sec:SGWBfromLoops}
In this Section, we are going to derive the expression for the GW spectrum emitted from decaying string loops, which form in interactions of long strings. For this, we require information about (i) the gravitational radiation from individual loops and (ii) the ensemble properties of a network of cosmic strings. 

Let us begin by considering the dynamics of strings in general. Cosmic strings are microscopically thin objects such that they can, for the purpose of describing their dynamics on cosmological scales, be treated as one-dimensional objects \cite{Vilenkin_Shellard_2000}. The associated effective action is the Nambu--Goto action 
\begin{align}
    S_{\rm NG} = -\mu \int_{\Sigma} \sqrt{-\gamma} \:\: \pd^2 \sigma \,. \label{eq:Nambu-Goto_action}
\end{align}
Here, $\mu$ is the string tension, and $\sqrt{-\gamma}\:\: \pd^2\sigma$ is the induced volume form on the string's two-dimensional worldsheet $\Sigma$. Let us denote the spatial comoving position of the string as a function of the time- and space-like worldsheet coordinates $\sigma^0$ and $\sigma^1$ as  $\bx\left(\sigma^0, \sigma^1\right)$. We, furthermore, fix the so-called transverse gauge associated with reparametrization invariance of the worldsheet by setting $\sigma^0 = \eta$ and $\dot{\bx}\cdot \bx ' =0$ where $\dot{f}$ and $f'$ denote for any function $f$ a derivative w.r.t.\ $\eta$ or $\sigma^1$, respectively. Considering the stress--energy tensor of the string, one can sensibly introduce the energy of the string as 
\begin{align}
    E = \mu a \int \epsilon \, \pd \sigma^1 && \text{with} && \epsilon =  \sqrt{\frac{\bx'^2}{1-\dot{\bx}^2}}\, . \label{eq:String_Energy}
\end{align}
Another quantity that will become central in the following considerations is the root-mean-square (RMS) velocity of the strings 
\begin{align}
    v = \sqrt{\Braket{\dot{\bx}^2}} \, , \label{eq:String_Velocity}
\end{align}
where we introduced the average for a function $f$ as
\begin{align}
    \Braket{f}= \frac{\int f \epsilon\, \pd \sigma^1}{\int \epsilon\, \pd\sigma^1} \, .
\end{align}

\subsection{Network evolution}
\label{subsec:String_Network_Evolution}
Equipped with these definitions, we want to turn to the evolution of a network of such Nambu--Goto strings. To describe the network, it is more suitable to work with the energy density of the strings $\rho\propto E/a^3$ instead of their energy~\eqref{eq:String_Energy}. Using this energy density, the velocity~\eqref{eq:String_Velocity}, and the equations of motion derived from~\eqref{eq:Nambu-Goto_action}, together with the transverse gauge condition, one finds
\begin{align}
    \frac{\pd \rho}{\pd t} + 2H\left(1+v^2\right)\rho = 0 \, . \label{eq:String_Network_Energy_Density}
\end{align}
The network evolution is, in principle, very complex, allowing for a variety of string configurations, including different microstructures of the strings. Perhaps astonishingly, the statistical properties of the string network and their evolution can, nevertheless, be well described by a semi-analytic model called the velocity-dependent one-scale (VOS) model. This model was mainly developed in Refs.~\cite{Kibble_85, Martins_Shellard_96, Martins_Shellard_2002}. 
The apparent caveat is the need to introduce phenomenological parameters for the VOS model, which cannot directly be related to the underlying microphysics. These parameters need to be fixed by matching the VOS model to numerical simulations.

\begin{figure}
\centering
\begin{overpic}[width = 0.6\textwidth]{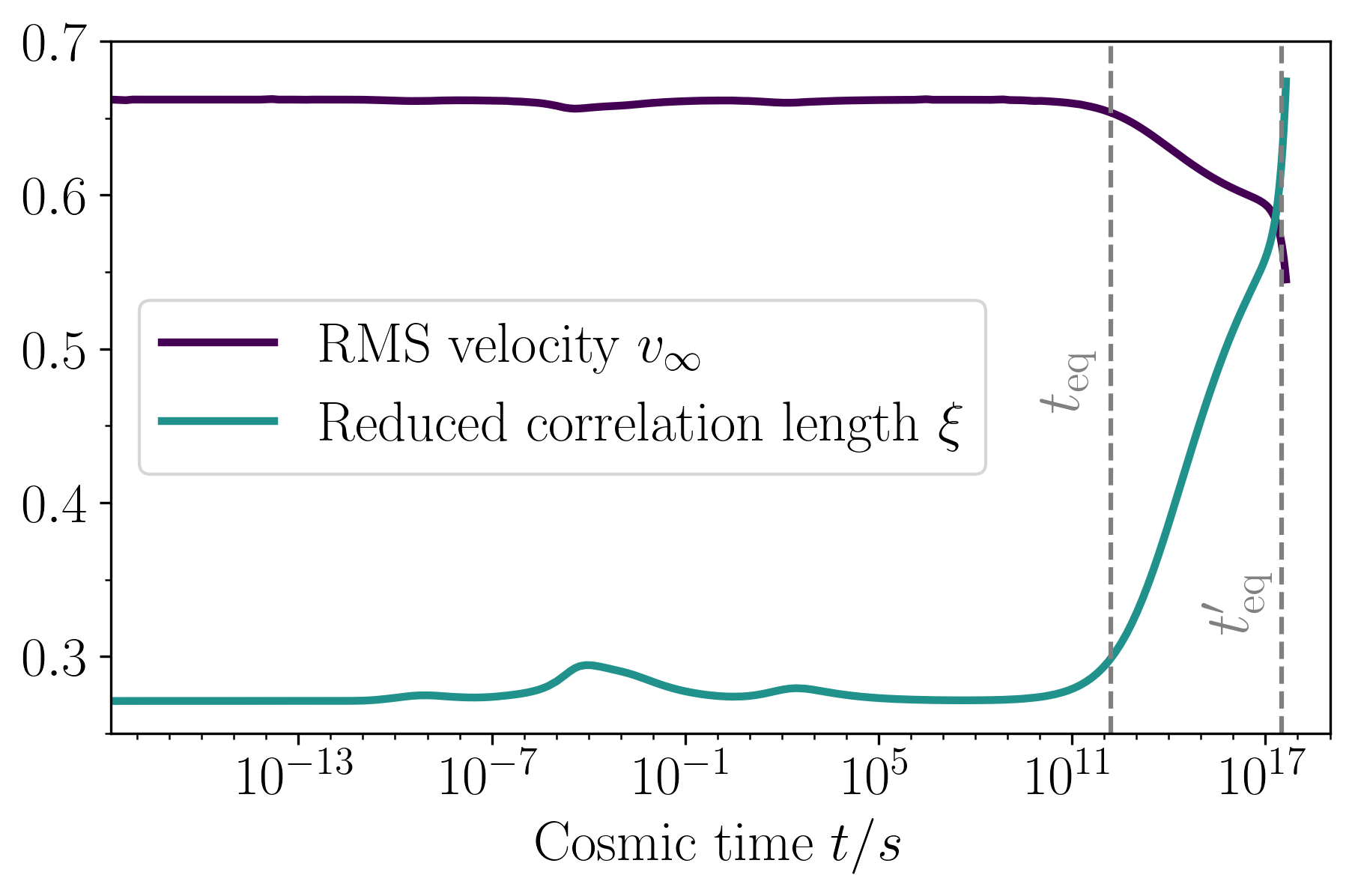}
\end{overpic}
    \caption{\footnotesize Time evolution of the reduced correlation length $\xi$ (teal) and the RMS velocity $v_\infty$ (deep purple) describing the long-string network. The evolution is obtained by numerically solving the coupled differential equations \eqref{eq:DifferentialequationsVOS}.  The vertical dashed lines indicate matter--radiation equality $t_{\rm eq}$ and cosmological constant--matter equality $t_{\rm eq}'$.}
    \label{fig:VOS}
\end{figure}

Generally, there are three different length scales that describe the statistical properties of the strings in the network. The first scale is the correlation length, which is the scale beyond which string directions become uncorrelated. The other two scales are the curvature radius of the string and the average inter-string distance. The central assumption of the one-scale model \cite{Kibble_85} is that these three length scales are the same, and we denote them by $L$. To describe the network evolution, it is then useful to distinguish between long strings with $l>L$, denoted by an index ``$\infty$'', and string loops with $l<L$. 
In this way, we can introduce the energy density of the network of long strings as 
\begin{align}
\rho_\infty = \frac{\mu}{L^2} \, . \label{eq:rho_and_L}
\end{align}
According to \eqref{eq:String_Network_Energy_Density}, we can then write
\begin{align}
    \frac{\pd\rho_\infty}{\pd t} + 2H(1+v_\infty^2)\rho_\infty + \frac{\pd \rho_{\infty}^{\rm loss}}{\pd t} = 0 \, ,\label{eq:Energy_Density_Evolution_Long_String_Network}
\end{align}
where $\pd \rho_{\infty}^{\rm loss}/\pd t $ describes the exchange rate of energy density between the long-string network and the string loops.  Since string loops will, in all probability, not gather to form new long strings, but long strings will interact with each other and themselves to form string loops, this ``exchange'' rate will, in practice, be a loss rate for the long-string network.  Following Ref.~\cite{Kibble_85} and using in addition to the one-scale assumption the numerically well-justified assumption that for typical string velocities (not too close to $1$), colliding strings inter-commute with probability $1$ \cite{Shellard_87, Matzner_89}, one finds 
\begin{align}
\frac{\pd \rho_{\infty}^{\rm loss}}{\pd t} =  \frac{\Tilde{c} \rho_\infty v_\infty}{L}  \label{eq:Energy_loss_Long_strings} \, .\end{align}
Here, we implicitly defined the constant $\Tilde{c}$, which is known as the ``loop-chopping efficiency'' and needs to be fixed by numerical simulations, which find the value \cite{Martins_Shellard_2002, Martins_2003}
\begin{align}
    \Tilde{c} = 0.23\pm 0.04 \, .
\end{align}
Using \eqref{eq:Energy_loss_Long_strings} together with \eqref{eq:rho_and_L} in \eqref{eq:Energy_Density_Evolution_Long_String_Network}, one finds that the evolution of the universal length scale $L$ is described by the equation
\label{eq:DifferentialequationsVOS}
\begin{align}
    \frac{\pd L}{\pd t} - \left(1+v_\infty^2\right) HL  - \frac{\Tilde{c}v_\infty}{2} = 0 \, . \label{eq:DifferentialequationLengthVOS}
\end{align}
Similarly, we can derive equations for the evolution of the network velocity. Differentiating \eqref{eq:String_Velocity} and using the Nambu--Goto equations in transverse gauge, one obtains
\begin{align}
    \dot{v}_\infty = \frac{1}{v_\infty}\left[-2\mathcal{H}v_\infty^2+\mathcal{H}\Braket{\dot{\bx}^4}_\infty+\mathcal{H}v_\infty^4+ \Braket{\frac{\bx''\cdot \dot{\bx}}{\epsilon^2}}_\infty\right], \label{eq:new}
\end{align}
where we used that $\bx'\cdot \dot{\bx}=0$. 
Furthermore, we make now the numerically verified \cite{Martins_Shellard_96} approximation  $\Braket{\dot{\bx}^4}_\infty=\Braket{\dot{\bx}^2}^2_\infty=v^4_\infty$. To get rid of the second derivative of $\bx$ in the last term of \eqref{eq:new}, let us first replace the derivative with respect to $\sigma^1$ by one with respect to the physical length $s$ along the string. They are related by $\pd s/\pd\sigma^1 = |\bx'|$. We can write $\pd^2 \bx /\pd s^2 = a\,\hat{\bu}/R$ for some unit vector $\hat{\bu}$ and an $R$ that is by definition the curvature radius.
Recall now that by the one-scale assumption $R=L$.
Defining furthermore $k\equiv \Braket{\left(1-\dot{\bx}^2\right)\dot{\bx}\cdot \hat{\bu}}_\infty/\left(v_\infty\left(1-v^2_\infty\right)\right)$, we can rewrite the differential equation for the network velocity as
\begin{align}
    \frac{\pd v_\infty}{\pd t} = \left(1-v^2_\infty\right) \left[\frac{k}{L}-2Hv_\infty\right] \, . \label{eq:DifferentialequationVelocityVOS}
\end{align}
The newly introduced $k$ can be considered quantifying the string's microstructure \cite{Martins_Shellard_96}. In order to solve the coupled system of differential equations \eqref{eq:DifferentialequationLengthVOS} and \eqref{eq:DifferentialequationVelocityVOS}, we need, in addition to the knowledge about the background evolution of the FLRW spacetime given by $H$, information about the phenomenological parameter $k$. This was extensively studied in Ref.~\cite{Martins_Shellard_2002}, which resulted in an expression that interpolates between the expressions obtained from simulations and known analytical string solutions for the non-relativistic and the ultra-relativistic regime. This expression reads
    \begin{align}
        k(v_\infty) = \frac{2\sqrt{2}}{\pi} \left(1-v^2_\infty\right) \left(1+2\sqrt{2}v^3_\infty\right) \frac{1-8v^6_\infty}{1+8v^6_\infty} \, .
    \end{align}
 The evolution of the string network is now solely described in terms of the variables $L$ and $v_\infty$. 
The VOS equations have attractor solutions which are scaling solutions, i.e., $L\propto t$ and $v_\infty = \text{const.}$, if the scale factor follows a power-law evolution $a(t) \propto t^\beta$ with $\beta\in(0,1)$. Denoting $L= \xi t$, 
\begin{align}
    \xi^2_\beta = \frac{k(v_\beta)\left(k(v_\beta)+\tilde{c}\right)}{4\beta\left(1-\beta\right)} && \text{and} && v^2_\beta= \frac{k(v_\beta)}{k(v_\beta)+\tilde{c}} \frac{1-\beta}{\beta} \label{eq:Xi_and_v_in_scaling}
\end{align}
give rise to stable fixed point solutions of the VOS equations \cite{Martins_Shellard_2002}. For our analytical calculations, we require the values of the reduced correlation length $\xi$ and average velocity $v_\infty$ during
matter domination ($\beta = 2/3$) and radiation domination ($\beta = 1/2$). Solving the above scaling equations, one finds
\begin{subequations}\label{eq:VOS_scaling_solution_values}
\begin{align}
    \xi_r = 0.27 \, , && v_r = 0.66 \, ,\\ \xi_m = 0.63 \, , && v_m = 0.58 \, . 
\end{align}
\end{subequations}
To assess the quality of our analytic approximations for the GW spectrum, we will always compare our expressions to results we obtain numerically.\footnote{The assumptions we make on the spacetime background evolution for our numerical computations are summarized in Section \ref{subsec:Numerical}.} The numerical solution to the VOS equations \eqref{eq:DifferentialequationLengthVOS} and \eqref{eq:DifferentialequationVelocityVOS} is shown in Fig.~\ref{fig:VOS}. Obviously, replacing the numerical solution with the constant scaling solutions \eqref{eq:VOS_scaling_solution_values} is a good approximation during radiation domination, but not during the matter-dominated phase. The reason for this is that, after matter--radiation equality, the network does not settle fast enough in the new attractor solution for matter domination before the onset of dark-energy domination. Luckily, this only leads to negligible deviations in the GW spectrum, as we will see in Section \ref{subsec:FullFundamentalSpectrum}.

\subsection{Loop evolution}
\label{subsec:Loop_Evolution}
In the next step, we want to describe the evolution of the string loops produced from the long-string network. Let us first consider individual loops. String loops oscillate due to their tension and decay predominantly into gravitational radiation. For now, we will neglect all other possible decay channels and come back to this point at the end of Section~\ref{subsec:Gravitational_Wave_Spectrum}. The total power emitted into gravitational radiation by a string loop is of the form~\cite{Vilenkin_Shellard_2000}
\begin{align}
    P = \Gamma G\mu^2 \,, \label{eq:TotalPower}
\end{align}
where for a large number of loops taken from network simulations, one finds, on average, the fairly peaked value of $\Gamma \simeq 50$ \cite{Blanco-Pillado_2017}. Since a string loop of length $l$ has energy $E=\mu l$, one can directly conclude that a loop, which had at time $t'$ a length $l'$, has at time $t$ the length
\begin{align}
    l = l' -\Gamma G\mu \left(t-t'\right) \, . \label{eq:LoopLengthEvolution}
\end{align}
The statistical information we need from the loops to describe the produced SGWB is contained in the function $n(l,t)\,\pd l$, which characterizes the number density of string loops within the length interval $\left[l, l+\pd l\right]$ at time $t$. To utilize the VOS model, it is useful to express this number density in terms of the loop production function $f(l,t)$ such that {\cite{Auclair:2019wcv}}
\begin{align}
    n(l,t)= \int_{t_{\rm ini}}^t  f\left(l'(l,t,t'),t'\right) \left(\frac{a\left(t'\right)}{a\left(t\right)}\right)^3 \, \pd t' \, . \label{eq:Loop_number_density_loop_production_function}
\end{align}
Correspondingly, $f\left(l,t\right)\, \pd l\, \pd t$ is the number density of loops produced during the time interval $\left[t, t+\pd t\right]$ within the length interval $\left[l, l+\pd l\right]$. The lower integration boundary $t_{\rm ini}$ can be viewed as the time when loop production from the long string network becomes significant. We will discuss it in more detail at the end of Section~\ref{subsec:Gravitational_Wave_Spectrum}. Since we previously assumed that the energy lost by the network of long strings is transferred into the production of string loops (cf.\ \eqref{eq:Energy_Density_Evolution_Long_String_Network}), we have to demand that {\cite{Auclair:2019jip}}
\begin{align}
    \frac{1}{\gamma}\frac{\pd \rho_{\infty}^{\rm loss}}{\pd t}= \mu \int_0^{L(t)} l f\left(l,t\right) \, \pd l \, .\label{eq:Loop_Production_Normalization}
\end{align}
Here, we included a Lorentz factor $\gamma$ accounting for the fact that loops are created with a non-vanishing center-of-mass velocity. Therefore, a fraction of the energy of the long-string network goes into the kinetic energy of the loops and is subsequently lost by redshifting~\cite{Martins_Shellard_96}. 
To fix the loop production function, we assume that all loops are created at all times with a length $l_*$, corresponding to a fixed fraction $\alpha$ of the one scale $L$: 
\begin{align}
    l_*(t)=\alpha\, L(t) \, .
\end{align}
This assumption is well justified by numerical network simulations \cite{Blanco-Pillado_2013}, which show that the distribution of $l_*(t)/t$ is sharply peaked around a value of $0.1$, such that we can fix $\alpha \xi(t) \simeq 0.1$ to a good approximation. We can then write the loop production function as $f(l,t)=\mathcal{N} \delta\left(l-\alpha L(t)\right)$. Using~\eqref{eq:Loop_Production_Normalization} together with \eqref{eq:Energy_loss_Long_strings}, we can, in principle, fix the normalization $\mathcal{N}$. In this way, we would, however, introduce a large error. This is because, while the majority of string loops are produced with lengths close to $l_*(t)/t\simeq 0.1$, most of the energy of the long strings still goes into loops with $l_*(t)/t<0.1$, namely in the form of kinetic energy. The error introduced by approximating the distribution of string lengths at production by a delta distribution was estimated in Ref.~\cite{Blanco-Pillado_2013} and can be accounted for by an additional efficiency factor $\mathcal{F}\sim 0.1$ into the loop production function. The average initial velocity of the remaining largest loops in the network has been determined in simulations to be $\simeq 1/\sqrt{2}$~\cite{Bennett_1989}, such that we can set $\gamma \simeq \sqrt{2}$. Combining these results, one arrives at a loop production function of the form {\cite{Auclair:2019wcv}}
\begin{align}
    f(l,t) = \mathcal{F}\, \frac{\Tilde{c}\,v_\infty(t)}{\sqrt{2}\,l\,L^3(t)}\, \delta\left(l-\alpha L(t)\right) \,. \label{eq:Loop_Production_Function}
\end{align}
We can now use this expression for the loop production function to calculate via \eqref{eq:Loop_number_density_loop_production_function} the loop number density and find\footnote{The loop number densities we use are derived from the above loop production functions matched to those extracted from simulations in Ref.~\cite{Blanco-Pillado_2013}. Another loop number distribution was developed in Ref.~\cite{Lorenz:2010sm} based on simulations in Ref.~\cite{Ringeval:2005kr}. Analytical results in this framework have, e.g., been developed in Refs.~\cite{Ringeval:2017eww, Auclair:2020oww}. A critical comparison of the two approaches can be found in Refs.~\cite{Auclair:2019zoz, Blanco-Pillado:2019tbi}. In our paper, we stick to the former approach, which is also supported by more current network simulations \cite{Wachter:2024zly} taking gravitational backreaction into account.}
\begin{align}
n(l,t)=\Theta(t-t_*)\Theta(t_*-t_{\rm ini})\underbrace{ \mathcal{F} \frac{\tilde{c}}{\sqrt{2}\alpha} \frac{v_\infty(t_*)}{\xi(t_*)^4 t_*^4} \left[\Gamma G \mu +\alpha \xi(t_*)+\alpha t_* \frac{\text{d} \xi}{\text{d} t}(t_*)\right]^{-1}\left(\frac{a(t_*)}{a(t)}\right)^3}_{=\tilde{n}(l,t)}. \label{eq:General_Loop_Number_Density}
\end{align}
Here, $t_*$ is implicitly defined as the solution to
\begin{align}
l-\Gamma G \mu (t_*-t)-\alpha \xi(t_*) t_*=0,
\end{align}
and since $\alpha \xi(t_*) t_*$ is the length of the loop at its birth, $t_*$ is the corresponding time of birth. Note that $t_*$ is a function of $l$ and $t$. {An expression similar to \eqref{eq:General_Loop_Number_Density}
was derived previously in \cite{Sousa:2013aaa}, neglected, however, the two Heaviside step functions. These arise automatically from the integration and ensure that the loops contributing to the loop number density at time $t$ were already produced ($t>t_*$) and were produced after $t_{\rm ini}$.}

To calculate the GW spectrum from decaying string loops in the following analytically, it is necessary to make simplifying assumptions about the expansion history of the Universe. In particular, we will assume that the Universe is at early times radiation-dominated, and after a time $t_{\rm eq}$, is purely matter-dominated until today. While rough, this approximation allows a precise calculation of the SGWB, as we will see later. In this spirit, we artificially introduce a factor \[1=\left[\Theta(t-t_\text{eq})+\Theta(t_\text{eq}-t)\right]\left[\Theta(t_*-t_\text{eq})+\Theta(t_\text{eq}-t_*)\right]\] such that we can write 
\begin{align}
    n(l,t)=n_{\rm rr}(l,t)+n_{\rm rm}(l,t)+n_{\rm mm}(l,t) \,, \label{eq:Splittingloopnumberdensity}
\end{align}
where
\begin{subequations}
\label{eq:Thetafunctions}
\begin{align}
 n_{\rm rr}(l,t)=\Theta(t_\text{eq}-t)\Theta(t-t_*)\Theta(t_*-t_{\rm ini})\tilde{n}(l,t), \label{eq:radiation}\\
    n_{\rm rm}(l,t)=\Theta(t-t_\text{eq})\Theta(t_\text{eq}-t_*)\Theta(t_*-t_{\rm ini})\tilde{n}(l,t),\label{eq:radiationmatter}\\
     n_{\rm mm}(l,t)=\Theta(t-t_*)\Theta(t_*-t_\text{eq})\tilde{n}(l,t).\label{eq:matter}
\end{align}
\end{subequations}
As we will see in a moment, the time $t$ at which we evaluate the loop number density plays the role of the emission time of gravitational radiation from the respective loops. Hence, the above splitting distinguishes between loops that were born during radiation domination and emitted GWs during the radiation-dominated era (RR), loops that were produced during radiation domination and emitted GWs during matter domination (RM), and loops produced and emitting during the matter-dominated era (MM). 

Consider the case that $t_*$ is a time during which $a(t_*)\propto t^\beta_*$. In this case, one finds that
\begin{align}
    t_*(l,t)=\frac{l+\Gamma G \mu t}{\alpha \xi_\beta+\Gamma G\mu} \,, \label{eq:t_birth}
\end{align}
and if moreover $t$ is also a time during which $a(t)\propto t^\beta$, one obtains
\begin{align}
    \tilde{n}_{\beta\beta}(l,t)=\mathcal{F}\frac{\tilde{c}}{\sqrt{2}\alpha} \frac{v_\beta}{\xi_\beta^4} \frac{(\alpha \xi_\beta + \Gamma G\mu)^{3(1-\beta)}}{t^{3\beta}(l+\Gamma G\mu t)^{4-3\beta}}\equiv \frac{C_\beta}{t^{3\beta}(l+\Gamma G\mu t)^{4-3\beta}}, \label{eq:Beta_Loop_Number_Density}
\end{align}
where we defined
\begin{align}
    C_\beta = \mathcal{F}\frac{\tilde{c}}{\sqrt{2}\alpha} \frac{v_\beta}{\xi_\beta^4} (\alpha \xi_\beta + \Gamma G\mu)^{3(1-\beta)},
\end{align}
with $v_\beta$ and $\xi_\beta$ from \eqref{eq:Xi_and_v_in_scaling}.
This covers the number density of loops that are both produced and decay during radiation \eqref{eq:radiation} or matter domination \eqref{eq:matter} with $\beta = 1/2$ and $\beta = 2/3$, respectively. Finally, we have to consider the number density of loops that were produced during radiation domination but only decay during matter domination. Fortunately, we can, in this case, just evaluate the expression for $n_{\rm rr}(l,t)$ at $t_\text{eq}$ and then redshift the expression to later times {\cite{Blanco-Pillado_2013}}, i.e.,
\begin{align}
    \tilde{n}_{\rm rm}(l,t)=\left(\frac{a_\text{eq}}{a(t)}\right)^{3}\tilde{n}_{\rm rr}(l_\text{eq}(l,t),t_\text{eq}).
\end{align}
Here, $l_\text{eq}(l,t)=l+\Gamma G\mu (t-t_\text{eq})$ is the length of a loop at $t_\text{eq}$ which has at time $t$ length $l$. Plugging this into \eqref{eq:Beta_Loop_Number_Density} yields
\begin{align}
    \tilde{n}_\text{rm}(l,t)=\frac{1}{t_\text{eq}^{3/2}}\left(\frac{\Omega_r}{\Omega_m}\right)^3 \left(\frac{a_0}{a(t)}\right)^3\frac{C_r}{\left(l+\Gamma G\mu t\right)^{5/2}} \label{eq:RM_Loop_Number_Density}
\end{align}
upon using that $a_\text{eq}/a_0=\Omega_r/\Omega_m$.

\subsection{Gravitational-wave spectrum}
\label{subsec:Gravitational_Wave_Spectrum}
We can use the previous results to calculate the GW spectrum (cf.\ e.g.\ {\cite{Vilenkin_Shellard_2000, Maggiore_2018, Auclair:2019wcv}}) emitted by decaying string loops. {Let us start from a general expression for the average power emitted in the form of gravitational radiation by a single string loop of length $l$, where the average is taken over different configurations in the considered loop population. This expression reads \cite{Vilenkin_Shellard_2000}}
\begin{align}
    \frac{\pd P_l^{\rm GW}}{\pd f'}(l,f') = G\mu^2 l \Delta\left(lf'\right) \,, \label{eq:Power_Distribution_intro}
\end{align}
where $f'$ is the frequency of the radiation at emission. Equation \eqref{eq:TotalPower} implies the normalization condition $\int_0^\infty \Delta\left(x\right)\pd x=\Gamma$. For the average energy density in gravitational radiation from a network of string loops, we can then write\footnote{The upper integration boundary could also be set to infinity, but the largest loops that are present in the network at time $t'$ are those formed at time $t'$. The loop production function \eqref{eq:Loop_Production_Function} therefore sets the maximum loop length at time $t'$ to $\alpha L(t')$. This fact is taken into account by the condition $t'>t_*(l,t')$, implemented via the first Heaviside function in the loop number density \eqref{eq:General_Loop_Number_Density}.}
\begin{align}
    \frac{\pd \rho_{\rm GW}}{\pd t'\pd f'}(t',f')= \int_0^{\alpha L(t')}  \frac{\pd P_l^{\rm GW}}{\pd f'}(l,f')\,n(l,t') \, \pd l\,,
\end{align}
where $t'$ is the time of emission. To find the GW spectrum today at $t_0$ as a function of the observed GW frequency $f$, we need to take into account that the energy density in radiation redshifts as $\propto a^{-4}$ and the frequency satisfies $f = \left(a(t')/a_0\right) f'$, such that we obtain in total {\cite{Vilenkin_Shellard_2000}}
\begin{align}
    \frac{\pd \rho_{\rm GW}}{\pd f}(f) = G\mu^2\int_{t_{\rm ini}}^{t_0} \,\left(\frac{a(t')}{a_0}\right)^3 \int_0^{\alpha L(t')}  n(l,t')\, l\, \Delta\left(\frac{a_0}{a(t')}fl\right) \, \pd l \, \pd t' \, .
\end{align}
As usual in cosmology, we re-express this spectrum in terms of the dimensionless density parameter
\begin{align}
    \Omega_{\rm GW}\left(f\right) = \frac{1}{\rho_c} \frac{\pd \rho_{\rm GW}\left(f\right)}{\pd \ln(f)} && \text{with} && \rho_c = \frac{3H_0^2}{8\pi G}\, 
\end{align}
and find 
\begin{align}
    \Omega_{\rm GW}\left(f\right) = \frac{8\pi G^2 \mu^2 f}{3H_0^2}\int_{t_{\rm ini}}^{t_0}  \left(\frac{a(t')}{a_0}\right)^3 \int_0^{\alpha L(t')} n(l,t')\, l\, \Delta\left(\frac{a_0}{a(t')}fl\right)\, \pd l \, \pd t' \, . \label{eq:Spectrum_Not_Yet_Mode_Decomposed}
\end{align}

Considering the equations of motion of string loops derived from the Nambu--Goto action, one observes that they have a periodicity of $T = l/2$
(cf.\ e.g.\ \cite{Vilenkin_Shellard_2000}). Accordingly, the oscillation modes and thus the spectrum can be decomposed into a Fourier series with frequencies $f_k = 2k/l, k\in \mathbb{N}$. We can then set {\cite{Sanidas:2012ee}} 
\begin{align}
    \Delta\left(x\right) = \sum_{k=1}^\infty \Gamma_k\, \delta\left(x-2k\right)\,, \label{eq:Power_Distribution_Scaling}
\end{align}
where $\Gamma_k/\Gamma$ is the fraction of the total power in radiation that comes from the $k$-th oscillation mode of string loops such that $\sum_{k} \Gamma_k = \Gamma$. 
Depending on the microstructure of the string loops, different scalings of $\Gamma_k$ in the limit of large mode numbers have been found. If the radiation comes from cusps, kinks, or kink--kink collisions, one observes the scaling 
$\Gamma_k\propto k^{-4/3},\, k^{-5/3},$ or $k^{-2}$, respectively \cite{Vachaspati_1984, Damour_2001, Binetruy_2009}. Henceforth, we will parameterize 
\begin{align}
\Gamma_k = \frac{\Gamma}{H_{n_{\rm max}}^{(q)}} k^{-q}.
\end{align}
Here, $H_{n_{\rm max}}^{(q)} = \sum_{k=1}^{n_{\rm max}} k^{-q}$ denotes the $n_{\rm max}$-th generalized harmonic number of order $q$ and accounts for the correct normalization. The reason for putting $n_{\rm max}$ is that, in general, the sum to infinity cannot be evaluated analytically and numerical evaluation requires a finite $n_{\rm max}$. This can lead to artificial modifications of the spectra at extremely large frequencies, on which we will comment where relevant. The choice $q=4/3$ approximates the result for $\Gamma_k$ that was found in numerical simulations \cite{Blanco-Pillado_2017} very well, in particular, in the case of $k\gg 1$ and, correspondingly, we pick it as our benchmark value. For all of our computations, we do not fix $q$ and our results remain general. More recent simulations \cite{Wachter:2024zly} take gravitational backreaction into account and find that a constant value of $q$ cannot completely reproduce their gravitational wave spectra. This effect can lead to deviations between a few percent and 30\%.

For our spectrum, we obtain then, after carrying out the $l$-integration, the form
\begin{align}
     \Omega_{\rm GW}(f) = \frac{16\pi \Gamma G^2\mu^2}{3 H_0^2 H_{n_{\rm max}}^{(q)}} \sum_{k=1}^{n_{\rm max}}\frac{1}{k^q} \frac{k}{f} \int_{t_{\rm ini}}^{t_0}  \left(\frac{a(t')}{a_0}\right)^5 n\left(k l(t',f), t'\right) \, \pd t' \, .
\end{align}
Here, $l(t',f)= 2\left(a(t')/a_0\right)/f $ is the length the loop had at the emission time $t'$ of GWs if $f$ is the frequency we observe today corresponding to those GWs from the fundamental string-loop excitation. It is now important to observe that, due to the functional dependence of the spectrum, we do not need to calculate the contributions from different oscillation modes separately. Instead, the total spectrum can be expressed in terms of the fundamental-mode contribution via
\begin{align}
    \Omega_{\rm GW}\left(f\right) =\frac{1}{H_{n_{\rm max}}^{(q)}}\sum_{k=1}^{n_{\rm max}} \frac{1}{k^q}\,\Omega^{(k)}(f) && \text{with} && \Omega^{(k)}(f) = \Omega_{\rm GW}^{(1)} \left(\frac{f}{k}\right) \,. \label{eq:Sum_Full_Spectrum}
\end{align}
The contribution from the fundamental oscillation mode is
\begin{align}
\Omega_\text{GW}^{(1)}(f)&=\frac{16\pi}{3} \left(\frac{G\mu}{H_0}\right)^2 \frac{\Gamma}{f} \int_{t_{\rm ini}}^{t_0} n(l(t',f), t') \left(\frac{a(t')}{a_0}\right)^5 \text{d} t'=\nonumber\\&= \frac{16\pi}{3} \left(\frac{G\mu}{H_0}\right)^2 \frac{\Gamma}{f} \int_{a_{\rm ini}/a_0}^{1} n(l(a,f), a) \left(\frac{a}{a_0}\right)^4 \frac{1}{H(a)}\,\pd\left(\frac{a}{a_0}\right). \label{eq:GW_Spectrum_Fundamental}
\end{align}

Up until now, we used the Nambu--Goto approximation and described strings as one-dimensional objects, albeit, in reality, strings have a finite width $\delta$. Apart from possible effects on the particle decay of string loops, which we will not consider in this work\footnote{The influence of particle radiation is discussed e.g. in Ref.~\cite{Auclair:2019jip}.}, the finite width will be irrelevant for most parts of the spectrum. Nevertheless, at extremely high frequencies that correspond to emission wavelengths that can resolve the string's microstructure, the non-vanishing string width will become relevant. We will now show how this can be taken into account. Consider a string loop of length $l$ and a corresponding oscillation mode $k$ with wavelength $\lambda_k = l/(2k)$. If this wavelength becomes smaller than the string width $\delta$, the Nambu--Goto approximation will break down, and we, furthermore, do not expect the emission of gravitational radiation. We account for this by modifying the function $\Delta$ introduced in \eqref{eq:Power_Distribution_intro} and expressed in \eqref{eq:Power_Distribution_Scaling}. In practice, we implement an additional Heaviside function $\Theta\left(l-2k\delta\right)$ and replace $\Delta$ in \eqref{eq:Spectrum_Not_Yet_Mode_Decomposed} by\footnote{In reality, the emission of radiation will not be unmodified above a length scale $\delta$ and completely die off below. We rather expect a smooth transition between the two regimes. Nonetheless, due to the absence of a better understanding of this transition, the modeling in terms of a Heaviside function should give a good first approximation.}
\begin{align}
    \tilde{\Delta}(x, f) = \sum_{k=1}^\infty \Gamma_k\, \delta\left(x-2k\right) \Theta\left(x-2k\delta\,\frac{a_0}{a(t')} f\right) \, .
\end{align}
Observe that this breaks the sole dependence of $\Delta$ on the variable $x$. The integration over $l$ or, equivalently, over $x$ in \eqref{eq:Spectrum_Not_Yet_Mode_Decomposed} leads to the same result as before, however, with an additional Heaviside function of the form $\Theta\left(a(t')/a_0 - \delta f\right)$. This signifies that any frequency $f$ observed nowadays in the spectrum is only allowed to correspond to an emission frequency at time $t'$ associated with a wavelength larger than $\delta$. This statement is, in particular, independent of the mode number. 
We can also view this as a constraint on the earliest possible time of emission depending on the frequency, which yields, upon introducing the Planck time $t_{\rm Pl}=G^{1/2}$,
\begin{align}
    \frac{a_{\rm min}(f)}{a_0} = \frac{t_{\rm Pl}f}{\left(G\mu\right)^{1/2}} \, , \label{eq:a_min}
\end{align}
where we used that the string width can be expressed in terms of the string tension as $\delta \sim \mu^{-1/2}$, up to prefactors of order one \cite{Vilenkin_Shellard_2000}. This effect should be included by replacing $a_{\rm ini}$ (and correspondingly in $t_{\rm ini}$) with \begin{align}
    a_{\rm ini} \to \max\left\{a_{\rm min}(f), a_{\rm ini}\right\} \, .\label{eq:a_min2}
\end{align}
It is important to pay attention to the fact that, while frequency-dependent, this minimum scale factor is independent of the mode number. Thus, the simple relation between the spectra from higher harmonics and the fundamental mode giving rise to \eqref{eq:Sum_Full_Spectrum} does no longer hold. Fortunately, we can simply treat $a_{\rm ini}$ as an additional variable, carry out the sum over harmonics in \eqref{eq:Sum_Full_Spectrum}, and only replace $a_{\rm ini}$ with the new frequency-dependent expression in the end. This is the route we will take in the following. As we will see later, this change in the initial time can, indeed, have an impact on the GW spectrum at extremely high frequencies. 

Eventually, we want to briefly discuss $t_{\rm ini}$ itself -- the time when GW emission from decaying string loops becomes significant. $t_{\rm ini}$ is a parameter that can, in principle, be determined from other, microscopic model parameters. Nonetheless, there are substantial theoretical uncertainties in choosing the correct expression for $t_{\rm ini}$. More detailed discussions of this can be found in \cite{Gouttenoire:2019kij, Servant:2023tua, Schmitz_2024}. The weakest constraint is that $t_{\rm ini}$ must be large enough such that the phase transition leading to the network formation has already happened. {Furthermore, thermal friction will suppress the motion
of the long strings and GW emission at early times. Loop production, however, can become significant earlier due to the large number density of strings (see Ref.~\cite{Mukovnikov:2024zed}). The so-called friction-dominated era ends roughly at a time 
\begin{align}
    t_{\rm fric} \propto \frac{t_{\rm Pl}}{\left(G\mu\right)^2} \, ,
\end{align}
which provides a typical and well-motivated choice for the initial time.} It is also the earliest time when the network can reach the scaling regime, which we rely on in our calculations. Apart from that, particle decay might also affect the initial time since particle decay due to cusp formation or kink--kink collisions becomes the dominant decay channel of string loops below a critical loop length.\footnote{In fact, this statement has been the subject of a long-lasting debate and no consensus has been reached yet. See, e.g., \cite{Baeza-Ballesteros:2024otj} for a recent article in this regard.} The initial time must, hence, be chosen late enough such that the loop production length has already become large enough to make the gravitational decay dominant. While we will keep $t_{\rm ini}$ a free parameter for all of our analytical expressions, we will adopt $t_{\rm ini} =t_{\rm fric}$ for most of our plots. {As was argued in detail in \cite{Gouttenoire:2019kij, Servant:2023tua, Schmitz_2024}, $t_{\rm ini}$ may also be much larger than $t_{\rm fric}$, if one accounts for the possibility of loop evaporation into particle radiation, which dominates over loop decay into gravitational radiation below a critical loop length. Especially for very low string tensions, this critical length can be large such that loops decay predominantly into particle radiation, even at times much later than $t_{\rm fric}$. This motivates choosing considerably larger values for $t_{\rm ini}$.}

\subsection{Numerical spectra \label{subsec:Numerical}}
In order to assess the quality of our analytical results later on, we need a comparison. In all of our plots, we will, therefore, show numerical spectra in addition to the analytical spectra.
To obtain the fundamental spectrum numerically, we evaluate equation \eqref{eq:GW_Spectrum_Fundamental} with the loop number density \eqref{eq:General_Loop_Number_Density} 
and Heaviside functions adapted to the time interval (RR, RM, or MM) under consideration. 
The time dependence of the scale factor is obtained by solving the Friedmann equation numerically, using a model in which
\begin{align}
    H(a) = H_0 \sqrt{\Omega_\Lambda + \Omega_m \left(a/a_0\right)^{-3} + \rho_r(a)/\rho_c^0}
\end{align}
with Hubble constant $H_0 = 67.4 \, {\rm km}\, {\rm s^{-1}}\, {\rm Mpc}^{-1}$, and density parameters $\Omega_\Lambda = 0.685$ and $\Omega_m = 0.315$ \cite{Planck:2018vyg}. For the comparison to analytical results in which we take changes in the effective number of relativistic DOFs into account, we use in our numerical computation the time evolution of $\rho_r(a)$ based on tabulated data for temperature $T$, energetic DOFs $g_\rho$ and entropic DOFs $g_s$ from \cite{Saikawa:2020swg}. If we assume instead a constant number of effective DOFs, we only set $\rho_r^0 = \pi^2/30 g_\rho(T_0) T_0^4$ for a temperature of $T_0 = 2.73 \, {\rm K}$ today and use the scale factor dependence $\rho_r(a)\propto a^{-4}$. 
The necessary input from the VOS model is obtained by numerically solving the VOS equations \eqref{eq:DifferentialequationLengthVOS} and \eqref{eq:DifferentialequationVelocityVOS} based on the evolution of the scale factor obtained from the previously described calculation. 
Having thus found the fundamental-mode contribution to the spectrum, we obtain the total spectrum by explicitly carrying out the sum from mode $1$ to $n_{\rm max}$ using the relation in \eqref{eq:Sum_Full_Spectrum}.

\section{Fundamental-mode contributions}
\label{sec:FundamentalMode}

In this Section, we shall derive analytical expressions for the fundamental-mode contribution to the GW spectra from decaying string loops (cf.\ \eqref{eq:GW_Spectrum_Fundamental}). 
We will discuss all qualitatively different parts of parameter space and, moreover, provide analytical expressions for all relevant features of these spectra. 
Following the splitting \eqref{eq:Splittingloopnumberdensity}, we decompose the GW spectrum according to the production time of loops and the emission time of GWs,
\begin{align}
    h^2\Omega_{\rm GW} = h^2\Omega_{\rm rr} + h^2\Omega_{\rm rm} + h^2\Omega_{\rm mm}
\end{align}

\subsection{RR loops}
\label{subsec:RR_Loops_Fundamental}
Let us begin with the GWs produced during radiation domination stemming from the fundamental oscillation mode of string loops. Using the loop number density \eqref{eq:radiation} with \eqref{eq:Beta_Loop_Number_Density} in the general expression for the spectrum \eqref{eq:GW_Spectrum_Fundamental},
we obtain
\begin{align}
    h^2\Omega_{\rm rr}^{(1)} = h^2 \frac{16\pi}{3}& \left(\frac{G\mu}{H_0}\right)^2 \frac{\Gamma}{f} C_r \Theta\left(a_{\rm end} - a_{\rm start}\right)\times  \label{eq:RRLoopIntegral1}\\ &\times\int_{a_{\rm start}/a_0}^{a_{\rm end}/a_0} \frac{\left(\frac{a}{a_0}\right)^4 \pd \left(\frac{a}{a_0}\right)}{t^{3/2}_{r}(a)\left(\frac{2}{f}\frac{a}{a_0} + \Gamma G\mu t(a)\right)^{5/2} H_{r}(a)}.\nonumber
\end{align}
The integration boundaries $a_{\rm start/end}$ are set by the Heaviside functions occurring in the loop number density, and we will discuss their explicit values in a moment. Since we only integrate over times, or equivalently scale factors, during radiation domination, we were able to simplify the expressions for the Hubble parameter and the cosmic time as a function of $a$ occurring in the integrand,
\begin{align} 
H_{r}\left(a\right) = H_r^0 \left(\frac{a_0}{a}\right)^2 && \text{and} &&
t_{r}\left(a\right) = \frac{1}{2H_r^0}\left(\frac{a}{a_0}\right)^2 \, ,
\end{align}
where we defined $H_r^0 = H_0 \Omega_r^{1/2}$. For the moment, we keep the effective number of relativistic DOFs constant and consider the impact of changes in it later.

For the further discussion, it will be beneficial not to work directly with the scale factor but with the variable 
\begin{align}
x_r = \frac{f}{f_r} && \text{with} && f_r = \frac{a_0}{a}\,h_r^0
\end{align}
instead. Note that the reference frequency $f_r$ carries the dependence on the scale factor. Furthermore, we introduced 
\begin{align}
\qquad h_r^0 = \frac{4H_r^0}{\Gamma G\mu} \simeq 1.68\times 10^{-11}\,\textrm{Hz} \left(\frac{50}{\Gamma}\right)\left(\frac{10^{-10}}{G\mu}\right) \,.
\end{align}
 Using the new variable $x_r$, the above integral \eqref{eq:RRLoopIntegral1} takes the form
\begin{align}
    h^2\Omega_{\rm rr}^{(1)}=\frac{64\pi}{3} C_r h^2\Omega_r \left(\frac{G\mu}{\Gamma}\right)^{1/2} \Theta\left(x_{\rm rr}^{\rm end} - x_{\rm rr}^{\rm start}\right)\int_{x^{\rm start}_{\rm rr}}^{x^{\rm end}_{\rm rr}} \frac{x_r^{1/2}}{\left(1+x_r\right)^{5/2}}\,\pd x_r \label{eq:RR_xr_integral}
\end{align}
and evaluates to the simple expression
\begin{equation}
\label{eq:Orr1}
h^2\Omega_{\rm rr}^{(1)} = \mathcal{A}_{\rm rr}\, \Theta(x_{\rm rr}^{\rm end} - x_{\rm rr}^{\rm start})\left[S_{\rm rr}^{(1)}(x_{\rm rr}^{\rm end}) - S_{\rm rr}^{(1)}(x_{\rm rr}^{\rm start})\right] \,,
\end{equation}
with amplitude
\begin{align}
\mathcal{A}_{\rm rr} &= \frac{128\pi}{9}\,C_r\,h^2\Omega_r \left(\frac{G\mu}{\Gamma}\right)^{1/2}\\ &\simeq 4.52 \times 10^{-10} \left(\frac{C_r}{0.171}\right)\left(\frac{50}{\Gamma}\right)^{1/2}\left(\frac{G\mu}{10^{-10}}\right)^{1/2} \,, \nonumber
\end{align}
and spectral shape function
\begin{equation}
S_{\rm rr}^{(1)}\left(x\right) = \left(\frac{x}{1+x}\right)^{3/2} \,.
\end{equation}
{A similar expression for the fundamental RR spectrum has been found in \cite{Sousa_2020}. However, it differs from our result due to the presence of a Heaviside function in Eq.~\eqref{eq:Orr1} and the values of the integration boundaries $x_{\rm rr}^{\rm start, end}$, which we will discuss now.} The integration boundaries are set by the Heaviside functions in the loop number density \eqref{eq:radiation}. First, the condition $t_{\rm eq}>t$ gives rise to an upper integration boundary
\begin{align}
x_{\rm rr}^{\rm end} = 
x_r^{\rm eq} =\frac{f}{f_r^{\rm eq}} && \text{with} && f_r^{\rm eq} = \frac{a_0}{a_{\rm eq}}\,h_r^0 \simeq 5.74 \times 10^{-8}\,\textrm{Hz} \left(\frac{50}{\Gamma}\right)\left(\frac{10^{-10}}{G\mu}\right)\,,
\end{align}
where $a_{\rm eq}/a_0 \simeq 2.92 \times 10^{-4}$ is the scale factor at matter--radiation equality.
Meanwhile, the two conditions $t>t_*>t_{\rm ini}$ set competing lower integration boundaries
\begin{equation}
x_{\rm rr}^{\rm start}  = \textrm{max}\left\{\chi_r^{-1}, \varphi_2\left(x_r^{\rm ini},\chi_r\right)\right\} \,. \label{eq:x_rr_start}
\end{equation}
Here, we introduced
\begin{equation}
\chi_r = \frac{\alpha\xi_r}{\Gamma G\mu} \,, \qquad x_r^{\rm ini} = \frac{f}{f_r^{\rm ini}} \,, \qquad f_r^{\rm ini} = \frac{a_0}{a_{\rm ini}}\,h_r^0 \,,
\end{equation}
and $\varphi_2$ is the positive solution of the quadratic equation
\begin{equation}
\varphi^2 + \varphi - (1+\chi_r) \left(x_r^{\rm ini}\right)^2 = 0 \,. \label{eq:def_phi2}
\end{equation}
Explicitly, we have
\begin{equation}
\varphi_2\left(x,\chi\right) = \left[\frac{1}{4} + \left(1+\chi\right)x^2\right]^{1/2} - \frac{1}{2} \,. \label{eq:phi2_explicit}
\end{equation}
While keeping $t_{\rm ini}$ as a free parameter, we can safely assume it to be a time deep in radiation domination such that we can write to a good approximation $a_{\rm ini}/a_0= \left(2H_r^0 t_{\rm ini}\right)^{1/2}$ and find
\begin{align}
    f_r^{\rm ini}  = \frac{2\sqrt{2}}{\Gamma G\mu}\left(\frac{H_r^0}{t_{\rm ini}}\right)^{1/2} \simeq 8.19 \times 10^{10}\, {\rm Hz} \left(\frac{50}{\Gamma}\right)\left(\frac{10^{-10}}{G\mu}\right) \left(\frac{10^{-24}\, {\rm s}}{t_{\rm ini}}\right)^{1/2}. 
\end{align}
As mentioned before, in most of our plots, we will assume that $t_{\rm ini} = t_{\rm fric} \propto t_{\rm Pl}/(G\mu)^2$ and we will set proportionality factors to be $1$ for the purpose of plotting. With this, we find
\begin{equation}
f_r^{\rm fric} = \frac{a_0}{a_{\rm fric}}\,h_r^0 
\simeq 3.53 \times 10^{10}\,\textrm{Hz}\left(\frac{50}{\Gamma}\right)\,.
\end{equation}

\begin{figure}
\centering
\begin{overpic}[width = \textwidth]{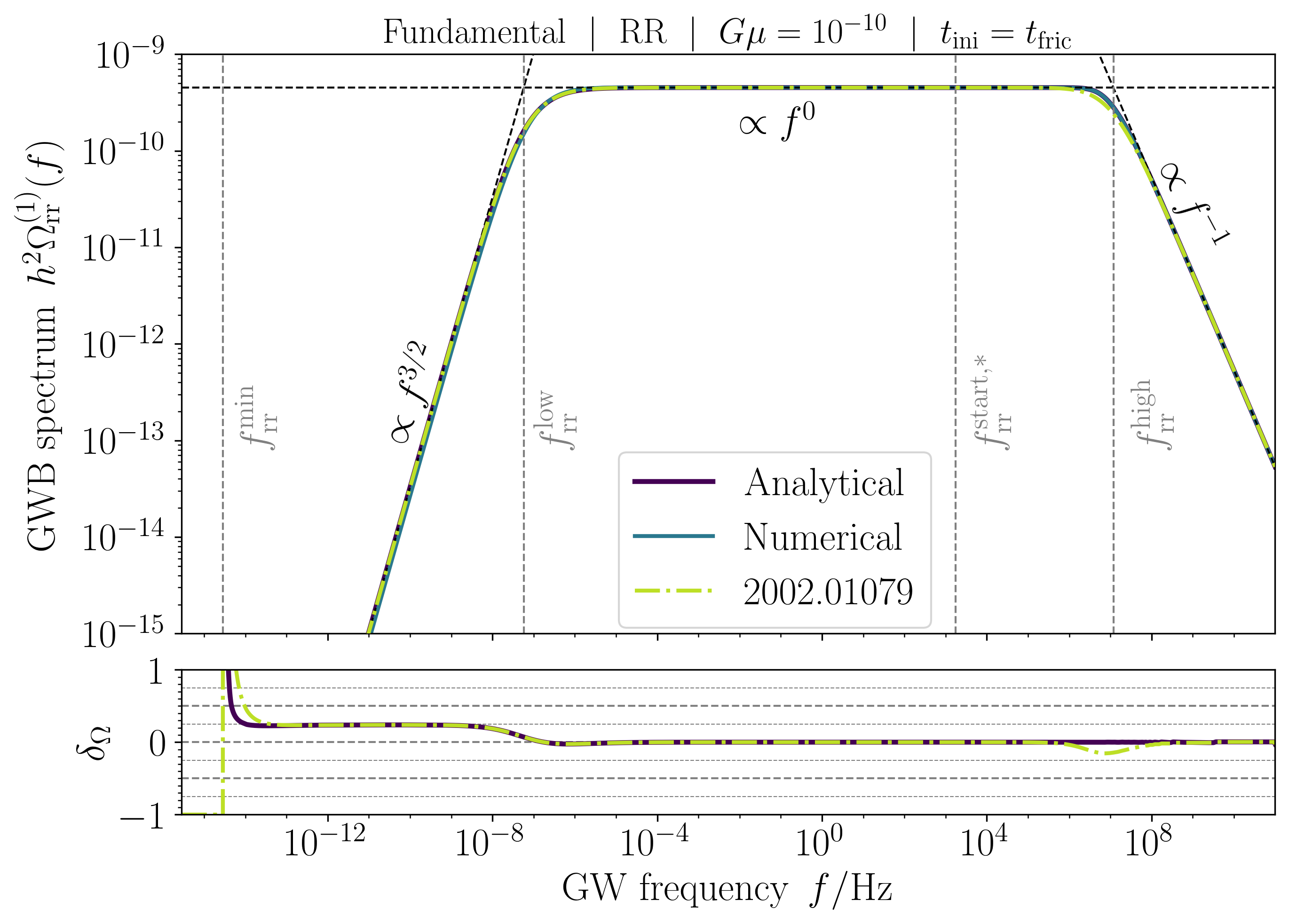}
\put(18.75,43){\rotatebox{90}{\eqref{eq_f_rr_min}}}
\put(42,43){\rotatebox{90}{\eqref{eq:f_rr_low}}}
\put(75.75,45){\rotatebox{90}{\eqref{eq:f_rr_star_start}}}
\put(88.25,44){\rotatebox{90}{\eqref{eq:f_rr_high}}}
\put(32.5,46.75){\rotatebox{72}{\eqref{eq:Orr1_low_simplified}}}
\put(64.5,61.25){\rotatebox{0}{\eqref{eq:Orr1_plateau}}}
\put(93.25,51){\rotatebox{-67}{\eqref{eq:Orr1_high}}}
\end{overpic}
    \caption{\footnotesize
    Fundamental GW spectrum for the RR case, a string tension of $G\mu = 10^{-10}$ and an initial time $t_{\rm ini}=t_{\rm fric}$. Upper panel: Our exact numerical result based on the VOS model accounting for the full time dependence of all relevant quantities and for a fixed effective number of DOFs as explained in Section \ref{subsec:Numerical} (teal), our analytical result derived in this paper (deep purple), and the analytical result derived in Ref.~\cite{Sousa_2020} (light green, dash-dotted). Grey dashed lines show characteristic frequencies of the spectrum, and black dashed lines indicate different power-law behaviors of the spectrum.
     Lower panel: Relative deviation $\delta_\Omega$ (cf. \eqref{eq:DeltaOmega}) of the two analytical results from our numerical result.}
    \label{fig:RR1}
\end{figure}

\begin{figure}[htbp]
\centering
\begin{subfigure}{.5\textwidth}
  \centering
\begin{overpic}[width = \textwidth]{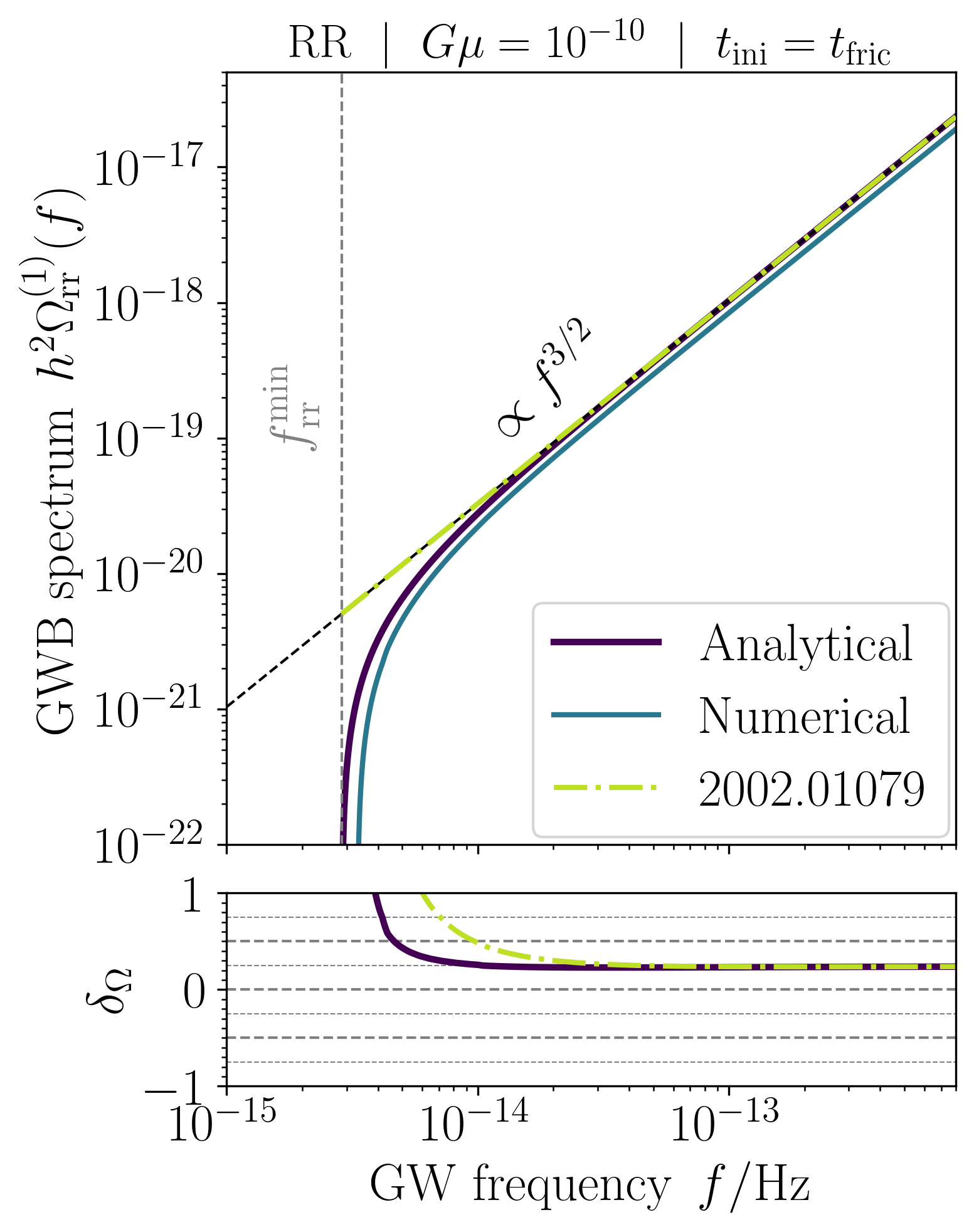}
\put(31,73){\rotatebox{90}{\eqref{eq_f_rr_min}}}
\put(54.5,74){\rotatebox{45}{\eqref{eq:Orr1_low_simplified}}}
\end{overpic}
\end{subfigure}%
\begin{subfigure}{.5\textwidth}
  \centering
\begin{overpic}[width = \textwidth]{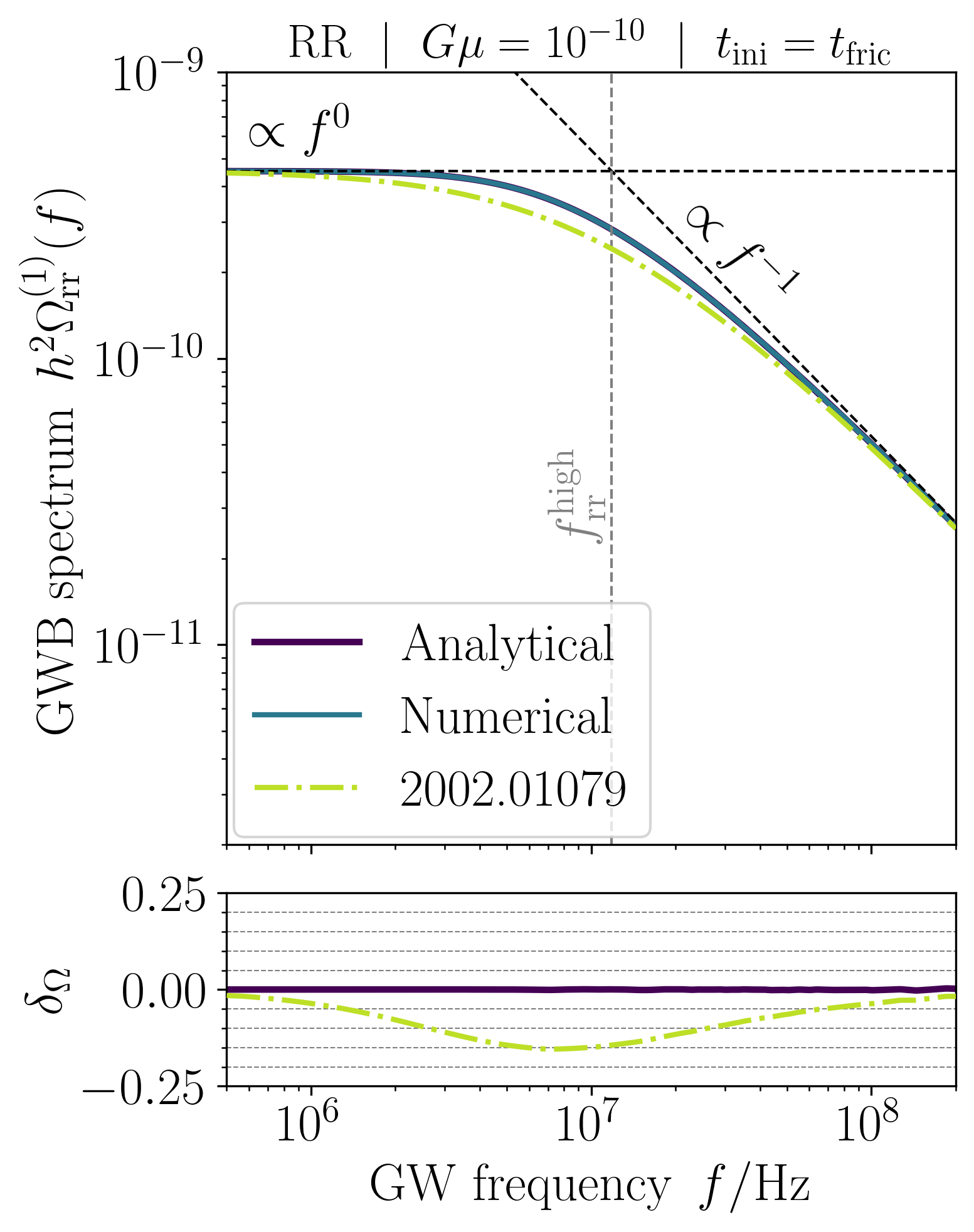}
\put(45,66){\rotatebox{90}{\eqref{eq:f_rr_high}}}
\put(31,88.25){\rotatebox{0}{\eqref{eq:Orr1_plateau}}}
\put(65,73){\rotatebox{-47}{\eqref{eq:Orr1_high}}}
\end{overpic}
\end{subfigure}
\caption{\footnotesize Left panel: Very-low-frequency region of the spectra shown in Fig.~\ref{fig:RR1}. {Our numerical spectrum and the analytical spectrum of this work drop to zero as they approach $f_{\rm rr}^{\rm min}$, while the result of Ref.~\cite{Sousa_2020} are cut off by hand.} Right panel: Region of the spectra shown in Fig.~\ref{fig:RR1} in which the transition from the plateau to the high-frequency regime occurs. The spectrum determined in Ref.~\cite{Sousa_2020} deviates in this regime by $\sim 20\%$ from our numerically calculated one, while our analytical result shows excellent agreement.}
\label{fig:RR_Zoom_1}
\end{figure}

\begin{figure}
\centering
\begin{overpic}[width = \textwidth]{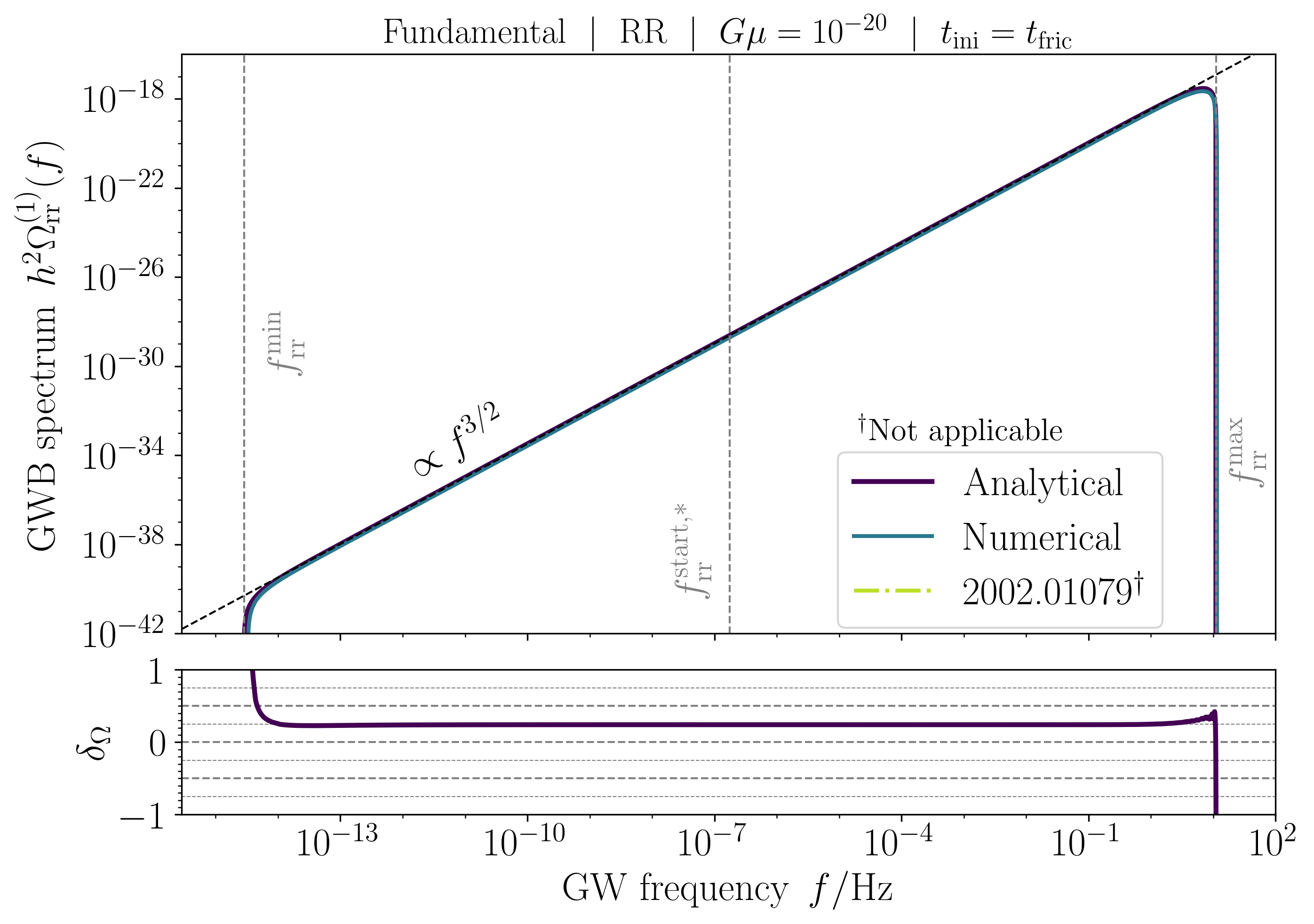}
\put(20.5,48){\rotatebox{90}{\eqref{eq_f_rr_min}}}
\put(51.7,33.5){\rotatebox{90}{\eqref{eq:f_rr_star_start}}}
\put(93.75,40){\rotatebox{90}{\eqref{eq:f_rr_max}}}
\put(39,38.5){\rotatebox{28}{\eqref{eq:Orr1_low_simplified}}}
\end{overpic}
    \caption{\footnotesize
  \footnotesize
    Fundamental GW spectrum for the RR case, a string tension of $G\mu = 10^{-20}$ and an initial time $t_{\rm ini}=t_{\rm fric}$. 
    Panels and color codes are the same as in Fig.~\ref{fig:RR1}. The spectrum derived in Ref.~\cite{Sousa_2020} does not appear in the plot since it turns negative.}
    \label{fig:RR2}
\end{figure}

The GW spectra obtained with the above formulae are shown in in Figs.~\ref{fig:RR1} and \ref{fig:RR2} for different parts of parameter space. The different characteristic frequencies and features of the spectra are discussed below. Before turning to this discussion, we would like to point out that our analytical expressions lead to an improvement over those found in Ref.~\cite{Sousa_2020} as can be seen in the plots when comparing the analytical results to our fully numerical calculation described in Section \ref{subsec:Numerical}. 
We quantify the deviation of the analytical from the numerical spectrum in our plots in terms of $\delta_\Omega$, defined as
\begin{align}
    \delta_{\Omega}(f) = \frac{\Omega_{\rm ana}(f) - \Omega_{\rm num}(f)}{\Omega_{\rm num}(f)} \, . \label{eq:DeltaOmega}
\end{align}
In particular, our spectra capture the transition from the plateau to the high-frequency part of the spectrum better, as can be seen in the right panel of Fig.~\ref{fig:RR_Zoom_1}. In the left panel, one can also observe that our analytical expression for the spectrum goes to zero at a frequency $f_{\rm rr}^{\rm min}$ as does the numerically calculated spectrum. {The result found in Ref.~\cite{Sousa_2020} extends, in principle, to arbitrarily low frequencies and needs to be cut off by hand. In Ref.~\cite{Sousa_2020}, this cutoff was implicitly given and can be shown to evaluate to the same value of $f_{\rm rr}^{\rm min}$ as derived here in the following. Nonetheless, this still leads to large deviations between the numerical spectra and those computed in Ref.~\cite{Sousa_2020} close to the minimum frequency. 
Finally, our expressions are applicable in the case of very low string tensions $G\mu\lesssim 10^{-20}$ and reproduce the numerical spectra as can be seen in Fig.~\ref{fig:RR2}. The analytical expressions of Ref.~\cite{Sousa_2020}, on the other hand, turn negative and, therefore, unphysical in this part of parameter space.} 
{The reason their spectrum turns negative lies in the boundaries of the integral over GW emission times. Our integral runs from $x_{\rm rr}^{\rm start}$ to $x_{\rm rr}^{\rm end}=x_r^{\rm eq}$ or, correspondingly, from the scale scale factor $a_{\rm rr}^{\rm start}(f)$ to $a_{\rm eq}$. By construction, our spectrum is only non-vanishing if $a_{\rm rr}^{\rm start}(f)< a_{\rm eq}$, which is explicitly ensured by the occurring Heaviside functions. In \cite{Sousa_2020}, the integration runs from $a_{\rm min}=a_0\left(2H_0 \Omega_r^{1/2}(1+\chi_r)t_{\rm ini}\right)^{1/2}$ to $a_{\rm eq}$.}\footnote{{The expression $a_{\rm min}/a_0$ in \cite{Sousa_2020} is given for $t_{\rm ini} = t_{\rm fric}= t_{\rm Pl}/G\mu^2$, which we have adapted to make it applicable to the case of general $t_{\rm ini}$ considered here.}} {For sufficiently small values of $G\mu$ and large values of $t_{\rm ini}$, it is clear that $a_{\rm ini}>a_{\rm eq}$, independently of the considered frequency. Indeed, this causes the spectrum of \cite{Sousa_2020} to turn negative for the parameter point chosen in Fig.~\ref{fig:RR2}.}
\\\\

\noindent
\textbf{Discussion of the spectrum:} Next, let us discuss the characteristics of the spectrum, first the minimum and maximum frequencies $f_{\rm rr}^{\rm min/max}$, between which the spectrum is non-vanishing. From \eqref{eq:Orr1}, it is obvious that these frequencies are reached when 
\begin{align}
x_{\rm rr}^{\rm start} = x_{\rm rr}^{\rm end} \,.
\label{eq:drop_off_condition_rr}
\end{align} 

While $x_{\rm rr}^{\rm end} = x_r^{\rm eq}$ is always true, for $x_{\rm rr}^{\rm start}$, we have to distinguish between two cases as can be seen from \eqref{eq:x_rr_start}. We denote the frequency at which we have to switch from one case to the other by $f_{\rm rr}^{{\rm start},*}$. Using \eqref{eq:x_rr_start} together with \eqref{eq:def_phi2}, one finds 
\begin{align}
\label{eq:f_rr_star_start}
f^{{\rm start},*}_{\rm rr} = \frac{f_{r}^{\rm ini}}{\chi_r}.
\end{align}

Let us begin with the minimum frequency $f_{\rm rr}^{\rm min}$ and, thus, low frequencies. Considering \eqref{eq:x_rr_start}, we note that $\chi_r^{-1}>0$ and that $\varphi_2(x_r^{\rm ini}, \chi_r)$ is monotonically increasing as a function of $f$, starting from $\varphi_2 = 0$ at $f = 0$ as can be recognized from the explicit form \eqref{eq:phi2_explicit}. Therefore, we find at sufficiently low frequencies that $x_{\rm rr}^{\rm start} = \chi_r^{-1}$. Here, sufficiently low means $f\leq  f^{{\rm start},*}_{\rm rr}$. Assuming that the lowest possible frequency satisfies $f_{\rm rr}^{\rm min}\leq f^{{\rm start},*}_{\rm rr}$  allows us to replace $x^{\rm start}_{\rm rr}$ by $\chi_r^{-1}$ and we find from \eqref{eq:drop_off_condition_rr} that 
\begin{align}
f_{\rm rr}^{\rm min} = \frac{f_r^{\rm eq}}{\chi_r}. \label{eq_f_rr_min}
\end{align}
Since $a_{\rm ini} < a_{\rm eq}$ and accordingly $f_r^{\rm eq}<f_r^{\rm ini}$, it follows that $f_{\rm rr}^{\rm min} < f^{{\rm start},*}_{\rm rr}$ and hence, our assumption is always satisfied.

The physics behind this result can easily be understood. The lowest frequencies correspond to the largest loop sizes. The largest loops are those with both the longest initial length and the least time to decay, and hence, those produced at the latest times. For the RR spectrum, these are the string loops produced at $t_{\rm eq}$ with length $l_{\rm rr}^{\rm max} = \alpha \xi_r t_{\rm eq} = \Gamma G\mu \chi_r t_{\rm eq}$. The corresponding observed frequency today is 
\begin{align}
f_{\rm rr}^{\rm min} = \frac{a_{\rm eq}}{a_0}\frac{2}{l_{\rm rr}^{\rm max}} = \frac{a_0}{a_{\rm eq}}\frac{4H_r^0}{\Gamma G\mu} \frac{1}{\chi_r}= \frac{f_{r}^{\rm eq}}{\chi_r}
\end{align}
where we assumed perfect radiation domination to express $a_{\rm eq} = a(t_{\rm eq})$. This sets the lowest possible frequency for which the spectrum does not yet vanish and, thus, reproduces \eqref{eq_f_rr_min}.

Let us now turn to $f_{\rm rr}^{\rm max}$. In the frequency range in which $f\geq  f^{{\rm start},*}_{\rm rr}$, equation \eqref{eq:drop_off_condition_rr} reads $x_r^{\rm eq} = \varphi_2(x_r^{\rm ini}, \chi_r)$. Using this equality in \eqref{eq:def_phi2}, it can be brought into the form
\begin{align}
f_{\rm rr}^{\rm max} = \frac{f_r^{\rm eq}}{(1+\chi_r)\left(\frac{a_{\rm ini}}{a_{\rm eq}}\right)^2-1} \,. \label{eq:f_rr_max}
\end{align}
This maximum has to be a non-negative number and, therefore, only exists if the condition $(1+\chi_r)\left(a_{\rm ini}/a_{\rm eq}\right)^2>1$ holds true. If it exists, our initial assumption $f_{\rm rr}^{\rm max} \geq f^{{\rm start},*}_{\rm rr}$ is, upon using again that $a_{\rm eq}> a_{\rm ini}$, equivalent to $\chi_r > - \left(a_{\rm ini} + a_{\rm eq}\right)/a_{\rm ini}$ and thus always satisfied. 

As before, the condition for the existence of the maximum can be understood intuitively. The ratio $(1+\chi_r) \left(a_{\rm ini}/a_{\rm eq}\right)^2$ is, in perfect radiation domination, the ratio between the time of complete decay of the loops born at $t_{\rm ini}$ and matter--radiation equality $t_{\rm eq}$. The condition that this ratio be larger than $1$ amounts to the statement that the earliest loops have not completely decayed before $t_{\rm eq}$. These loops are, however, the ones with the smallest initial loop size and have, in addition, the longest time to decay. Correspondingly, they are the smallest loops that contribute to the RR spectrum and, therefore, lead to the highest frequencies. In case they decay completely, the frequencies in the RR spectrum can (at least in the framework considered here) become infinitely large. If there is not enough time for the decay to complete, the highest frequency will correspond to the smallest loop size reached before matter--radiation equality, thus explaining why there is a maximum frequency.  We can calculate this highest allowed frequency. The loops born at $t_{\rm ini}$ with length $l_{\rm ini} = \alpha \xi_r t_{\rm ini}$ have, according to \eqref{eq:LoopLengthEvolution}, at $t_{\rm eq}$ a length 
\begin{align}
l_{\rm rr}^{\rm min} = \alpha \xi_r t_{\rm ini} - \Gamma G\mu \left( t_{\rm eq} -t_{\rm ini}\right) = \Gamma G\mu \left[(1+\chi_r) t_{\rm ini} - t_{\rm eq}\right] \, .
\end{align}
Under the assumption of perfect radiation domination, i.e., $t_r(a)=\left(a/a_0\right)^2/\left(2H_r^0\right)$,
this corresponds to a maximum observed frequency of
\begin{align}
    f_{\rm rr}^{\rm max} =\frac{a_{\rm eq}}{a_0} \frac{2}{l_{\rm rr}^{\rm min}} =   \underbrace{\frac{a_0}{a_{\rm eq}} \frac{4 H_r^0}{\Gamma G\mu}}_{=f_r^{\rm eq}} \frac{1}{(1+\chi_r)\left(\frac{a_{\rm ini}}{a_{\rm eq}}\right)^2-1},
\end{align}
and, hence, reproduces our previous result \eqref{eq:f_rr_max}.

Having found the minimum and maximum frequency for the spectrum, the Heaviside function in \eqref{eq:Orr1} can be replaced by the function
\begin{align}
    \left[1-\Theta(f-f_{\rm rr}^{\rm max})\Theta(f_{\rm rr}^{ \rm max})\right]\Theta(f-f_{\rm rr}^{\rm min}). \label{eq:Theta-RR}
\end{align}

Let us continue by discussing the shape of the spectrum and start with its low-frequency behavior. For $f \ll f_r^{\rm eq}$ and $f\leq f_{\rm rr}^{{\rm start},*}$, the spectrum reads
\begin{align}
    h^2 \Omega_{\rm rr, low}^{(1)} = \mathcal{A}_{\rm rr} \left(\left(x_{\rm rr}^{\rm end}\right)^{3/2} - \left(\frac{1}{1+\chi_r}\right)^{3/2}\right).
\end{align}
If furthermore $f\gg f_r^{\rm eq}/\left(1+\chi_r\right)$, which can with equation \eqref{eq_f_rr_min} be understood as the condition that we are at low frequencies but still sufficiently far above the minimum frequency, 
the spectrum is dominated by the first term. Correspondingly, the spectrum goes like 
\begin{align}
h^2\Omega_{\rm rr, low}^{(1)} &=\mathcal{A}_{\rm rr}\left(\frac{f}{f_r^{\rm eq}}\right)^{3/2} \label{eq:Orr1_low_simplified}  \\ &\simeq 3.29\times 10^{-14}\left(\frac{C_r}{0.171}\right)\left(\frac{\Gamma}{50}\right)\left(\frac{G\mu}{10^{-10}}\right)^2 \left(\frac{f}{ 10^{-10}\,\textrm{Hz}}\right)^{3/2}\,. \nonumber
\end{align}
For intermediate frequencies  $f_{\rm rr}^{\rm end} \ll f \ll f_{\rm rr}^{\rm start}$, the term $S_{\rm rr}^{(1)}(x_{\rm rr}^{\rm end})$ in \eqref{eq:Orr1} is roughly one, while the term $S_{\rm rr}^{(1)}(x_{\rm rr}^{\rm start})$ is still negligible.
This gives rise to a plateau of height
\begin{equation}
h^2\Omega_{\rm rr, plateau}^{(1)} = \mathcal{A}_{\rm rr} \,. \label{eq:Orr1_plateau}
\end{equation}
Finally, at large frequencies $f_{\rm rr}^{\rm start} \ll f$, both $S_{\rm rr}^{(1)}(x_{\rm rr}^{\rm start})$ and $S_{\rm rr}^{(1)}(x_{\rm rr}^{\rm end})$  are relevant and the value of the spectrum follows from their impartial cancellation. Asymptotically, it approaches the form
\begin{align}
    h^2 \Omega_{\rm rr, high}^{(1)} = \frac{3\mathcal{A}_{\rm rr}}{2(1+\chi_r)^{1/2}x_r^{\rm ini}} \left(1-\left(1+\chi_r\right)^{1/2}\frac{a_{\rm ini}}{a_{\rm eq}}\right) \, . \label{eq:Orr1_high}
 \end{align}
This can, of course, only hold if the high-frequency region is not cut off by the appearance of a maximum frequency. If, moreover, $(1+\chi_r)^{1/2} a_{\rm ini}/a_{\rm eq}\ll 1$, which is always the case in the absence of a maximum frequency, this asymptotic expression can further be simplified to 
\begin{align}
    h^2 \Omega_{\rm rr, high}^{(1)}\simeq \frac{3 \mathcal{A}_{\rm rr}}{2\left(1+\chi_r\right)^{1/2}x_r^{\rm ini}} \, . \label{eq:Orr1_high_simple2}
\end{align}
%
For all physically relevant cases, $\chi_r \gg 1$ holds. In this regime, we find
\begin{align} \label{eq:Orr1_high_simpl}
h^2\Omega_{\rm rr, high}^{(1)}  &\rightarrow \frac{3\,\mathcal{A}_{\rm rr}}{2\,\chi_r^{1/2}}\left(\frac{f_r^{\rm ini}}{f}\right)\\ \nonumber &\simeq 5.34 \times 10^{-18} \left(\frac{C_r}{0.171}\right)\left(\frac{50}{\Gamma}\right)\left(\frac{G\mu}{10^{-10}}\right)\left(\frac{10^{15}\,\textrm{Hz}}{f}\right) \,.
\end{align}

Having derived expressions for the three regions of the spectrum, we can also find the frequencies at which we transition from one region to another. The low-frequency regime ends, and the plateau begins at $f_{\rm rr}^{\rm low}$ where $\Omega_{\rm rr, low}^{(1)}(f_{\rm rr}^{\rm low}) = \Omega_{\rm rr, plateau}^{(1)}(f_{\rm rr}^{\rm low})$, which means due to expressions \eqref{eq:Orr1_low_simplified} and \eqref{eq:Orr1_plateau} that 
\begin{align}
f_{\rm rr}^{\rm low} = f_{r}^{\rm eq}. \label{eq:f_rr_low}
\end{align}
Similarly, the plateau region ends and the high-frequency regime begins at the frequency $f_{\rm rr}^{\rm high}$ which satisfies $\Omega_{\rm rr, high}^{(1)}(f_{\rm rr}^{\rm high}) = \Omega_{\rm rr, plateau}^{(1)}(f_{\rm rr}^{\rm high})$. This translates via \eqref{eq:Orr1_plateau} and \eqref{eq:Orr1_high} to 
\begin{align}
f_{\rm rr}^{\rm high}= \frac{3 f_r^{\rm ini}}{2(1+\chi_r)^{1/2}}\left(1-\left(1+\chi_r\right)^{1/2}\frac{a_{\rm ini}}{a_{\rm eq}}\right) \simeq \frac{3}{2\left(1+\chi_r\right)^{1/2}}\,f_r^{\rm ini} \,. \label{eq:f_rr_high}
\end{align}
In summary, we see that, typically, $h^2\Omega_{\rm rr}^{(1)}$ first rises like $f^{3/2}$ until it reaches a flat plateau and then decays like $f^{-1}$. This shape of the spectrum can be seen in Fig.~\ref{fig:RR1}. If a maximum frequency occurs in the spectrum, the high-frequency part and the plateau might disappear completely, and the spectrum is described by the low-frequency $f^{3/2}$-behavior only, dropping rapidly to zero as it approaches the minimum and maximum frequencies; see Fig.~\ref{fig:RR2}.

\begin{figure}
\centering
\begin{subfigure}{.5\textwidth}
  \centering
\begin{overpic}[width = \textwidth]{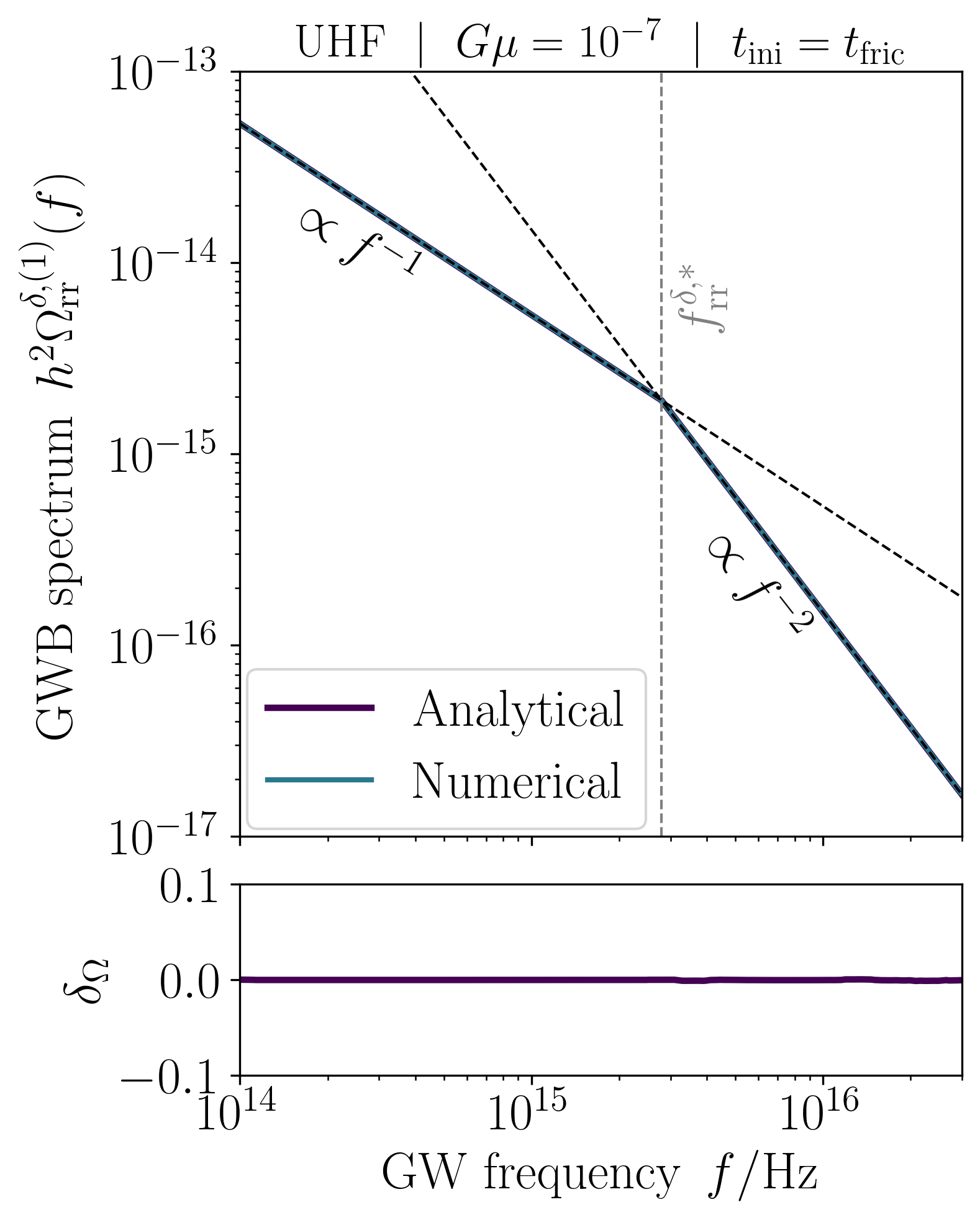}
\put(35,74){\rotatebox{-32}{\eqref{eq:Orr1_high}}}
\put(56,80){\rotatebox{90}{\eqref{eq:f_finite_width}}}
\put(64.75,45){\rotatebox{-53}{\eqref{eq:RR_UHF}}}
\end{overpic}
\end{subfigure}%
\begin{subfigure}{.5\textwidth}
  \centering
\begin{overpic}[width = \textwidth]{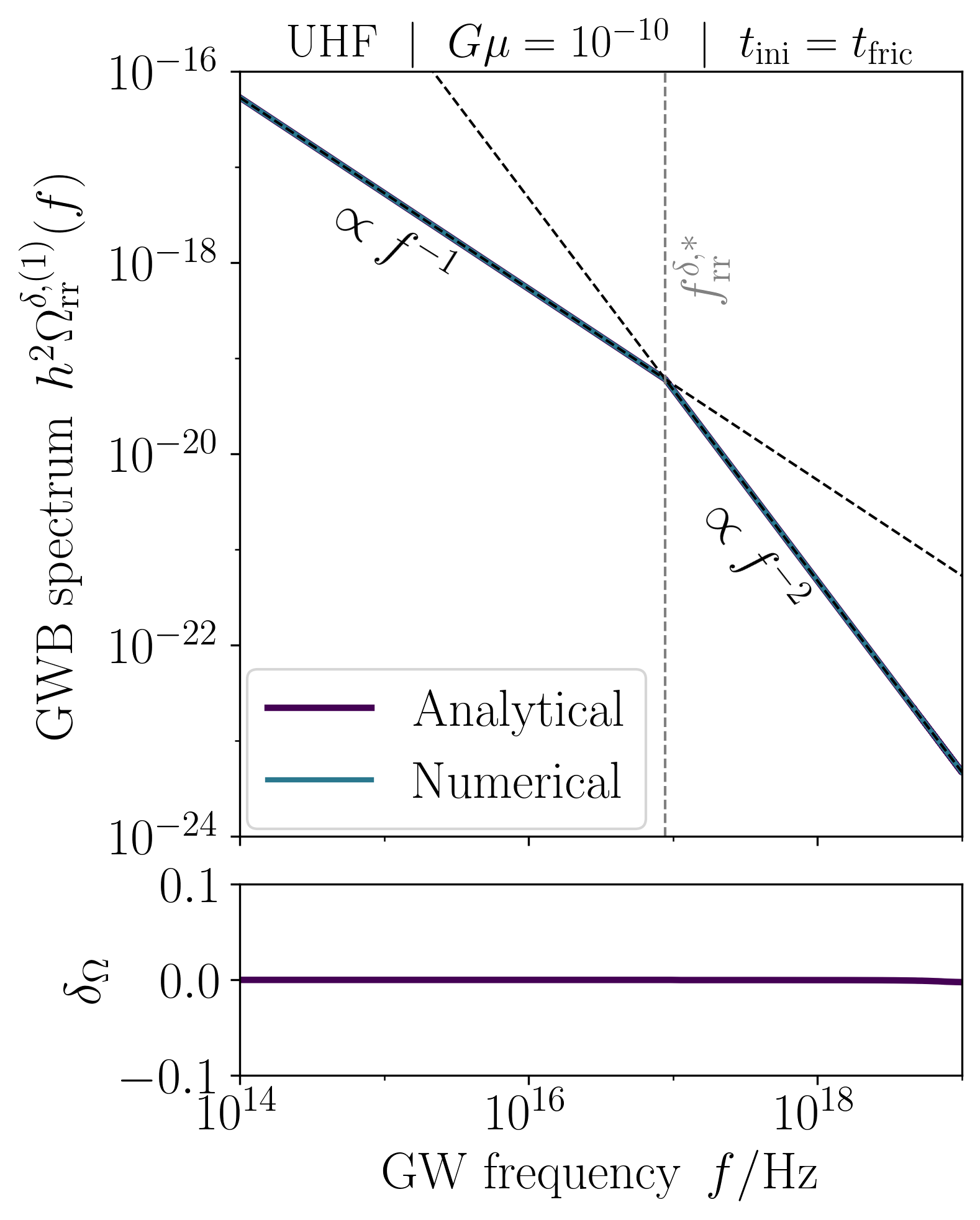}
\put(37,75){\rotatebox{-32}{\eqref{eq:Orr1_high}}}
\put(56,82.5){\rotatebox{90}{\eqref{eq:f_finite_width}}}
\put(64,47.5){\rotatebox{-53}{\eqref{eq:RR_UHF}}}
\end{overpic}
\end{subfigure}
\caption{\footnotesize  UHF regime of the fundamental GW spectrum for the RR case, string tensions of $G\mu = 10^{-7}$ (left) and $G\mu = 10^{-10}$ (right) and initial times given by $t_{\rm ini} = t_{\rm fric}$. Panels and color codes are the same as in Fig.~\ref{fig:RR1}. The grey dashed line shows the frequency at which the UHF effects become relevant.}
\label{fig:RR_UHF}
\end{figure}

\medskip\noindent
\textbf{Ultrahigh-frequency (UHF) regime:} Before moving on to the next type of loops, we want to return to the discussion at the end of Section \ref{subsec:Gravitational_Wave_Spectrum}. There, we included the fact that we do not expect the emission of GWs with wavelengths smaller than the string width. We found that this leads to a frequency-dependent earliest emission time, which is, in terms of the scale factor, expressed in equation \eqref{eq:a_min}. While the general way to include this in the total spectrum is clear from the discussion in the mentioned Section, we want to see how this affects the spectrum explicitly. Since the effect concerns the initial time, it will influence the spectrum at extremely large frequencies. As discussed before, a typical choice for the initial time is the moment when the friction regime ends and the scaling regime begins, which is roughly at $t_{\rm fric}=t_{\rm Pl}/\left(G\mu\right)^2$. The initial time would then be
\begin{align}
    t_{\rm ini} = \max\left\{t_{\rm min}(f), t_{\rm fric}\right\} \,,
\end{align}
such that our consideration of the finite string width becomes only relevant at frequencies larger than
\begin{align} \label{eq:f_finite_width}
    f_{\rm rr}^{\delta, *} = \left(\frac{2H_r^0}{G\mu\,t_{\rm Pl}}\right)^{1/2} = 8.82 \times 10^{16} {\rm Hz}\left(\frac{10^{-10}}{G\mu}\right)^{1/2} \, .  
\end{align}
Above this frequency, the power law will also change since $a_{\rm ini}/a_0$ in the high-frequency power-law \eqref{eq:Orr1_high_simpl} is now frequency-dependent as well. Just replacing $a_{\rm min}$ with the expression in \eqref{eq:a_min} yields
\begin{align}\label{eq:RR_UHF}
    h^2 \Omega_{\rm rr, high}^{\delta,(1)} &\simeq \frac{3\mathcal{A}_{\rm rr}}{2}\left(\frac{G\mu}{\chi_r}\right)^{1/2} \frac{h_r^0}{t_{\rm Pl} f^2} = \\ \nonumber & =4.71 \times 10^{-26} \left(\frac{C_r}{0.171}\right) \left(\frac{50}{\Gamma}\right) \left(\frac{G\mu}{10^{-10}}\right)^{1/2} \left(\frac{10^{20}{\rm Hz}}{f}\right)^2 \, .
\end{align}
Both the $f^{-1}$ power law derived before, as well as the $f^{-2}$ decay of the spectrum at ultrahigh frequencies, can be seen in the plots in Fig.~\ref{fig:RR_UHF} for two different string tensions. Looking at the relative deviation between the analytical expression and the numerically computed spectrum, we find that the agreement is excellent (up to numerical noise).

Similar considerations can be found in Ref.~\cite{Servant:2023tua}, which reaches the conclusion that only higher harmonic contributions are affected. This is in contrast to the results that we arrive at. We will compare our results to those of Ref.~\cite{Servant:2023tua} when turning to the total spectrum.  

Finally, it is clear that the spectrum will drop to zero once $a_{\rm min}(f) = a_{\rm eq}$. That is, there is always a maximum frequency. Solving this condition for $f$, one finds this to be at
\begin{align} \label{eq:UHF_start_frequency}
    f_{\rm rr}^{\delta, {\rm max}} = \frac{a_{\rm eq}}{a_0} \frac{\left(G\mu\right)^{1/2}}{t_{\rm Pl}} = 5.42 \times 10^{34} {\rm Hz} \left(\frac{G\mu}{10^{-10}}\right)^{1/2} \,,
\end{align}
which is, for any reasonable string tension, far out of reach of any current experiment.

\subsection{Varying degrees of freedom}
\label{subsec:VaryingDOFs}
From the previous discussion, it is clear that there is no bijection between the GW frequency occurring in the spectrum and the time the radiation was emitted. At each time, there is a large number of loops emitting at different frequencies and, furthermore, the frequency of emission is redshifted from the time of emission until today. Nevertheless, it is also clear that high frequencies in the spectrum are predominantly associated with gravitational radiation produced at early times and low frequencies with gravitational radiation produced at late times. In particular, changes in the effective number of DOFs during radiation domination will lead to modifications of the plateau region of the RR spectrum: A larger number of DOFs at early times suppresses the plateau at larger frequencies. As pointed out in many articles (cf.\ e.g.\ Refs.~\cite{Battye:1997ji, Blanco-Pillado_2017, Cui:2018rwi, Caldwell:2018giq, Antusch:2024ypp}), modeling these modifications properly is of great importance since a measurement of the spectrum allows then to draw conclusions on the thermal history of the Universe, including phase transitions after the formation of the strings.

Let us now try to approximate the changes due to variations in the effective number of relativistic DOFs. These enter the GW spectrum through the scale factor and, hence, it is advisable to first reflect on the Friedmann equation, which during radiation domination reads
\begin{align}
    H =\frac{1}{a}\frac{\pd a}{\pd t}= H_r^0\,\mathcal{G}^{1/2}(t) \left(\frac{a_0}{a}\right)^2 && \text{with} && \mathcal{G}(t) = \frac{g_{\rho}(t)}{g_\rho^0} \left(\frac{g_s^0}{g_s(t)}\right)^{4/3} \,
\end{align}
where $g_{\rho, s}$ are the effective number of energetic and entropic DOFs at a given time. We define
\begin{align}
    \Bar{H}_r(t) \equiv \frac{H_r^0}{t}\int_0^t \sqrt{\mathcal{G}(t')} \,\pd t'
\end{align}
such that we can rewrite the Friedmann equation as
\begin{align}
    \frac{a(t)}{a_0} =\left(2 \Bar{H}_r(t)\,t\right)^{1/2}.
\end{align}
If we were to consider no changes in the effective DOFs, we would simply have $\Bar{H}_r(t) \to H_r^0$.

This modification of the scale factor evolution enters the GW spectrum now in three places. First, it alters the loop number density \eqref{eq:General_Loop_Number_Density}, such that we obtain
\begin{align}
    \tilde{n}_{\rm rr, DOF}(l,t) = \left(\frac{\Bar{H}_r(t_*(l,t))}{\Bar{H}_r(t)}\right)^{3/2} \tilde{n}_{\rm rr}(l,t)
\end{align}
where the time of loop production $t_*$ is given in \eqref{eq:t_birth}. Second, the integrand of the fundamental spectrum \eqref{eq:GW_Spectrum_Fundamental} contains, in addition to the loop number density, scale factor ratios accounting for redshifting between the time of GW emission and observation. Finally, we need to express the length $l$ in the loop number density in the integrand in terms of frequency and scale factor, $l(f,t)=2\sqrt{2\bar{H}_r(t) t}/f$. Putting all of this together in \eqref{eq:GW_Spectrum_Fundamental}, we find
\begin{align}
    \Omega_{\rm GW}^{(1)}(f) = \frac{16 \pi}{3}\left(\frac{G\mu}{H_0}\right)^2 \frac{\Gamma}{f} \int_{t_{\rm min}}^{t_{\rm max}} \left(\bar{H}_r\left(t_*\right)\right)^{3/2} \bar{H}_r(t)\, \frac{ 2^{5/2} C_r t\, \pd t}{\left(\frac{2\sqrt{2\bar{H}_r(t)t}}{f} + \Gamma G\mu t\right)^{5/2}} \,. \label{eq:DOFs_Step1}
\end{align}
In order to carry out this integral as well as for the later purpose of mode summation, we simplify the expansion history in the already familiar manner: We assume that we can divide the radiation-dominated era into $N$ subintervals during which $\mathcal{G}$ is constant, i.e., we further divide the product of Heaviside functions occurring in \eqref{eq:radiation}
\begin{align} \label{eq:DOFs_Heaviside}
  \Theta\left(t_{\rm eq} - t\right)\Theta\left(t-t_*\right)\Theta\left(t_* - t_{\rm ini}\right) =  \sum_{i=0}^{N-1} \Biggl(\Theta\left(t_{(i+1)} -t\right)\Theta\left(t_*-t_{(i)}\right)\Theta\left(t-t_*\right) +\\ + \sum_{j=0}^{i-1} \Theta\left(t_{(i+1)}-t\right)\Theta\left(t-t_{(i)}\right)\Theta\left(t_{(j+1)}-t_*\right)\Theta\left(t_*-t_{(j)}\right)\Biggr) \, \nonumber 
\end{align}
with $t_0 = t_{\rm ini}$ and $t_{N} = t_{\rm eq}$. The above sum contains $N(N+1)/2$ terms, namely one term for each combination of birth and emission time interval, with the restriction that the birth time cannot be in an interval later than the one the emission time is in. We denoted the emission time interval with an index $i$ and the birth time interval with an index $j$. Restricting to this simplification, we are able to carry out the integral in \eqref{eq:DOFs_Step1} and find\footnote{We set $\bar{H}_r(t)$ constant during an interval with constant number of DOFs, which is not exact but a very good approximation since these intervals have, on order of magnitude, the same length in logarithmic time. For the linear time integral determining $\bar{H}_r(t)$, earlier values of the DOFs are, therefore, negligible.}
\begin{align}
      h^2\Omega_{ij}^{(1)}(f) =\Theta\left(x^{\rm end}_{ij} - x^{\rm start}_{ij}\right)\mathcal{A}_{\rm rr}\,\bigg(\frac{\bar{H}_r^j}{H_r^0}\bigg)^{3/2} \left(\frac{\bar{H}_r^i}{H_r^0}\right)^{1/2} S_{\rm rr}^{(1)}(x)\bigg\vert_{x^{\rm start}_{ij}}^{x^{\rm end}_{ij}}.
\end{align}
 Let us now specify the form of the integration boundaries. These derive, just as before, directly from the Heaviside functions in \eqref{eq:DOFs_Heaviside}. There are clearly two distinct cases, namely, either $i=j$ or $i>j$, which we label $A$ and $B$. {Case $A$ corresponds to GW emission occurring within the same constant-DOF interval in which the emitting loop was produced --- analogous to the RR and MM cases. Case $B$, on the other hand, corresponds to the emission taking place during a later constant-DOF interval than the one in which the loop was formed --- analogous to the RM case, although the intervals corresponding to $i$ and $j$ need not be adjacent.} In case $A$, the structure is exactly the same as before for the RR spectrum upon replacing $t_{\rm eq}\to t_{(i+1)}$ and $t_{\rm ini} \to t_{(i)}$, yielding
\begin{align}
    x_{A, i}^{\rm start} = \max\left\{\chi_r^{-1}, \varphi_2(x_i^{(i)},   \chi_r)\right\} && \text{and} && x_{A, i}^{\rm end} = x_i^{(i+1)} \,,
\end{align}
with $x_k^{(i)} = f/f_k^{(i)}$ and $f_k^{(i)} = \frac{4 \mathcal{G}_k^{1/2} H_r^0}{\Gamma G\mu} \frac{a_0}{a_{(i)}}$. In case $B$, we find
\begin{align}
    x_{B,ij}^{\rm start} = \max\left\{x_i^{(i)}, \varphi_2(x_i^{(j)}, \chi_r)\right\} && \text{and} && x_{B, ij}^{\rm end} = \min\left\{x_i^{(i+1)}, \varphi_2(x_i^{(j+1)}, \chi_r)\right\}.
\end{align}
We can then replace the previous expression for the fundamental RR spectrum by 
\begin{align} \label{eq:RR_Varying_DOFs_Fundamental_Spectrum}
   h^2\Omega_{\rm rr}^{(1)} = \mathcal{A}_{\rm rr}\,\sum_{i=0}^{N-1}\Bigg(& \mathcal{G}_i\, S_{\rm rr}^{(1)}(x)\Big\vert_{x_{A,i}^{\rm start}}^{x_{A,i}^{\rm end}} \:\Theta(x_{A,i}^{\rm end} - x_{A,i}^{\rm start}) +\\ &+ \sum_{j=0}^{i-1}\mathcal{G}_j^{3/4}\mathcal{G}_i^{1/4} \, S_{\rm rr}^{(1)}(x)\Big\vert_{x_{B,ij}^{\rm start}}^{x_{B,ij}^{\rm end}}\: \Theta(x_{B,ij}^{\rm end} -x_{B, ij}^{\rm start}) \Bigg) \, .\nonumber
\end{align}
{The first line of \eqref{eq:RR_Varying_DOFs_Fundamental_Spectrum} covers the GWB contributions from case $A$, the second line those from case $B$.}
Approximating the numerical evolution of the standard model effective DOFs taken from Ref.~\cite{Saikawa:2020swg}, we will use $N=4$ with $a_{(0)} = a_{\rm ini}$ and $a_{(4)}=a_{\rm eq}$ as well as
\begin{align}
  a_{(1)}/a_0 =1.6\times 10^{-15} , && a_{(2)}/a_0=3.4\times 10^{-12}, && a_{(3)}/a_0= 2.0\times 10^{-9},
\end{align}
and 
\begin{align*}
    \mathcal{G}_0 = 0.39 && \mathcal{G}_1 = 0.43&& \mathcal{G}_2 = 0.83&& \mathcal{G}_3 = 1.
\end{align*}
One needs to take care of the fact that $a_{\rm ini}$ can become large for large values of $t_{\rm ini}$. Keeping the labeling $a_{(0)}=a_{\rm ini}$ fixed, this means that $N$ can decrease due to this. The earliest value of $\mathcal{G}$, still labeled $\mathcal{G}_0$, will then be the one that is present in the time interval starting with $a_{\rm ini}$. 

A final ingredient that will be useful when turning to the total spectrum are the switch, minimum and maximum frequencies. For the former, we have 
\begin{align}
    f_{A,i}^{{\rm start}, *} = \frac{f_i^{(i)}}{\chi_r}, &&    f_{B,ij}^{{\rm start}, *}=\frac{f_i^{(i)}}{(1+\chi_r)\left(\frac{a_{(j)}}{a_{(i)}}\right)^2-1}, &&  f_{B,ij}^{{\rm end}, *} = \frac{f_i^{(i+1)}}{(1+\chi_r)\left(\frac{a_{(j+1)}}{a_{(i+1)}}\right)^2-1}.
\end{align}
With these (and knowledge about the low frequency-behavior of $\varphi_2$), we can write
\begin{align}
    x_{A,i}^{\rm start} &= \begin{cases} \chi_r^{-1} & \text{if } f< f_{A,i}^{{\rm start}, *}, \\ \varphi_2(x_i^{(i)}, \chi_r) & \text{else}
    \end{cases}, \\
    x_{B,ij}^{\rm end} &= \begin{cases} \varphi_2(x_i^{(j+1)}, \chi_r) & \text{if } f<f_{B,ij}^{{\rm end}, *}, \\
    x_r^{i+1} & \text{else}
    \end{cases}, \\ x_{B,ij}^{\rm start} &= \begin{cases} x_i^{(i)} & \text{if } f<f_{B, ij}^{{\rm start}, *}, \\ \varphi_2(x_i^{(j)}, \chi_r) & \text{else} \end{cases}. 
\end{align}
For the latter, we find
\begin{align}
    f_{A,i}^{\rm min} = \frac{f_i^{(i+1)}}{\chi_r}, && f_{A,i}^{\rm max} =  \frac{f_i^{(i+1)}}{\left(1+\chi_r\right)\left(\frac{a_{(i)}}{a_{(i+1)}}\right)^2 - 1},\\
    f_{B,ij}^{\rm min} = \frac{f_i^{(i)}}{\left(1+\chi_r\right) \left(\frac{a_{(j+1)}}{a_{(i)}}\right)^2 -1}, && f_{B,ij}^{\rm max} = \frac{f_i^{(i+1)}}{\left(1+\chi_r\right)\left(\frac{a_{(j)}}{a_{(i+1)}}\right)^2 -1}.
\end{align}
Finally, let us also mention, that since a larger effective number of DOFs at high energies will suppress the amplitude of the spectrum at high frequencies, the high-frequency power law will be modified. Instead of \eqref{eq:Orr1_high_simple2}, we find
\begin{align}
    h^2 \Omega_{\rm rr, high}^{(1)} \simeq \frac{3\mathcal{A}_{\rm rr} \mathcal{G}_0}{2\left(1+\chi\right)^{1/2}x_0^{(0)}} \, .
\end{align}

\subsection{RM loops}
\label{subsec:RM_Loops_Fundamental}
We continue with the consideration of string loops produced during radiation domination, but investigate now the gravitational radiation that they emit during the matter-dominated era. Correspondingly, this time, we use in the expression for the GW spectrum \eqref{eq:GW_Spectrum_Fundamental} the loop number density \eqref{eq:RM_Loop_Number_Density} with \eqref{eq:radiationmatter}, which yields
\begin{align} \label{eq:RM_First_Formula}
    h^2 \Omega_{\rm rm}^{(1)} = h^2 \frac{16\pi}{3} &\left(\frac{G\mu}{H_0}\right)^2 \frac{\Gamma}{f} \frac{1}{t_{\rm eq}^{3/2}} \left(\frac{\Omega_r}{\Omega_m}\right)^3 C_r \Theta\left(a_{\rm end} - a_{\rm start}\right)\times \\ &\times \int_{a_{\rm start}/a_0}^{a_{\rm end}/a_0} \frac{\frac{a}{a_0}\pd \left(\frac{a}{a_0}\right)}{\left(\frac{2}{f}\frac{a}{a_0} + \Gamma G\mu t_{m}(a)\right)^{5/2} H_{m}(a)} \,. \nonumber
\end{align}
The above integral runs only over scale factors in the matter-dominated era, and due to our assumption of pure matter domination during this time, we were able to simplify the expressions for the Hubble parameter and the cosmic time as a function of the scale factor
\begin{align} 
H_{m}\left(a\right) = H_m^0 \left(\frac{a_0}{a}\right)^{3/2} && \text{and} &&
t_{m}\left(a\right) = \frac{2}{3H_m^0}\left(\frac{a}{a_0}\right)^{3/2} \, , \label{eq:H_mat+t_mat}
\end{align}
where we defined $H_m^0 = H_0 \left(\Omega_m\right)^{1/2}$. 

As before, it is advantageous to change integration variables to
\begin{align}
    x_m = \frac{f}{f_m} && \text{with} && f_m = \left(\frac{a_0}{a}\right)^{1/2}\,h_m^0,
\end{align}
where $f_m$ carries the dependence on the scale factor and we introduced the frequency
\begin{align}
     h_m^0 = \frac{3H_m^0}{\Gamma G\mu} \simeq 7.36 \times 10^{-10}\,\textrm{Hz} \left(\frac{50}{\Gamma}\right)\left(\frac{10^{-10}}{G\mu}\right) \,.
\end{align}
With this substitution, the above integral takes the form 
\begin{align} h^2 \Omega_{\rm rm}^{(1)} &= \sqrt{96}\pi  C_r \frac{h^2 \Omega_m\left(\frac{a_{\rm eq}}{a_0}\right)^3}{\left( H_m^0 t_{\rm eq}\right)^{3/2}} \left(\frac{G\mu}{\Gamma} \frac{h_m^0}{f}\right)^{1/2}\Theta\left(x_{\rm rm}^{\rm end} - x_{\rm rm}^{\rm start}\right) \int_{x_{\rm rm}^{\rm start}}^{x_{\rm rm}^{\rm end}} \frac{x_m\, \pd x_m }{(1+x_m)^{5/2}}
\end{align}
and evaluates to the following expression for the spectrum
\begin{equation}
\label{eq:Orm1}
h^2\Omega_{\rm rm}^{(1)} = \mathcal{A}_{\rm rm} \Theta\left(x_{\rm rm}^{\rm end} - x_{\rm rm}^{\rm start}\right) \left[S_{\rm rm}^{(1)}(x_{\rm rm}^{\rm start}) -S_{\rm rm}^{(1)}(x_{\rm rm}^{\rm end})\right] \, .
\end{equation}
The amplitude introduced above reads
\begin{align}
\label{eq:A_rm}
\mathcal{A}_{\rm rm} &= \sqrt{\frac{128}{3}}\,\pi\,C_r\,\frac{h^2\Omega_m}{\left(H_m^0 t_{\rm eq}\right)^{3/2}}\left(\frac{a_{\rm eq}}{a_0}\right)^3 \left(\frac{G\mu}{\Gamma}\right)^{1/2} \\ &\simeq 6.51 \times 10^{-9} \left(\frac{C_r}{0.171}\right)\left(\frac{50}{\Gamma}\right)^{1/2}\left(\frac{G\mu}{10^{-10}}\right)^{1/2} \, , \nonumber
\end{align}
where $t_{\rm eq} \simeq 5.05 \times 10^4\,\text{yr}$ denotes the time of matter--radiation equality.
The spectral shape function for RM loops takes the form
\begin{equation} \label{eq:RM_Shape_Function}
S_{\rm rm}^{(1)}\left(x\right) = \frac{2+3\,x}{(x_m^0)^{1/2}\left(1+x\right)^{3/2}} \, .
\end{equation}
%
{A similar expression for the fundamental RM spectrum was computed in \cite{Sousa_2020}. As for the RR spectrum, the difference between our result and the previous result lies in the presence of the Heaviside function in Eq.~\eqref{eq:Orm1} and the values of the integration boundaries $x_{\rm rm}^{\rm start, end}$, at which the spectral shape function needs to be evaluated.}
These boundaries occur, as before, due to the Heaviside function in the loop number density \eqref{eq:radiationmatter}. The conditions $t_*>t_{\rm ini}$ and $t>t_{\rm eq}$ contribute to the lower integration boundary $x_{\rm rm}^{\rm start}$, while the conditions $t_*<t_{\rm eq}$ and $t<t_0$ are taken into account by $x_{\rm rm}^{\rm end}$. To write down explicit expressions for these boundaries, it is helpful to introduce another variable $\tilde{x}_m$ that does not involve the scale factor $a$, but instead the (would-be) scale factor 
\begin{align}
\frac{\tilde{a}_m\left(t\right)}{a_0} = \left(\frac{3}{2}H_m^0 t\right)^{2/3} \, ,
\end{align}
which describes the evolution of a purely matter-dominated Universe. We define in this way
\begin{align}
\tilde{x}_m = \frac{f}{\tilde{f}_m}  && \text{with} && \tilde{f}_m = \left(\frac{a_0}{\tilde{a}_m}\right)^{1/2}\,h_m^0 \, . \label{eq:def:x_m}
\end{align}
During the matter dominated era, the function $\tilde{a}_m\left(t\right)$ and the actual scale factor $a\left(t\right)$ in the standard $\Lambda$-cold-dark-matter ($\Lambda$CDM) model are (nearly) identical. However, in the following, we will also evaluate $\tilde{a}_m\left(t\right)$ at times outside the matter dominated era, where $\tilde{a}_m\left(t\right)$ and $a\left(t\right)$ no longer coincide with each other. 
%
Utilizing this, the integration boundaries for $x_m$ can be expressed as
\begin{equation}
\label{eq:x_rm_start_end}
x_{\rm rm}^{\rm start} = \max\left\{x_m^{\rm eq},\varphi_3\left(\tilde{x}_m^{\rm ini},\chi_r\right)\right\} \,, \qquad 
x_{\rm rm}^{\rm end} = \min\left\{x_m^0,\varphi_3\left(\tilde{x}_m^{\rm eq},\chi_r\right)\right\} \,,
\end{equation}
where $\varphi_3(x, \chi)$ is the real and positive solution of the cubic equation 
\begin{equation}
\varphi^3 + \varphi^2 = \left(1+\chi\right)x^3 \,. \label{eq:def_phi3}
\end{equation}
Explicitly, one finds\footnote{That this solution is always real can be seen as follows. First, consider \eqref{eq:phi}. As long as $y \geq \frac{4}{27}$, $\phi(y)$ is real and positive and $\varphi_3$ is (upon picking the real solution of $\phi^{1/3}$) real and positive as well.  In the case $0<y<\frac{4}{27}$, $\phi(y)$ acquires a non-vanishing imaginary part. At the same time, we find $|\phi(y)|^2 = 1$. This implies that $\phi^*(y) = 1/\phi(y)$ and hence $\varphi_3(x, \chi) = \frac{1}{3} [2\,{\rm Re}(\phi^{1/3}((1+\chi)\,x^3))-1]$, which is manifestly real.}
\begin{equation}
\varphi_3\left(x,\chi\right) = \frac{1}{3}\left[\phi^{1/3}\left((1+\chi)x^3\right) + \frac{1}{\phi^{1/3}\left((1+\chi)x^3\right)} -1\right] \,, \label{eq:phi3}
\end{equation}
with
\begin{equation}
\phi\left(y\right) = \frac{27}{2}\,y -1 + 3^{3/2} \left(\frac{27}{4}\,y^2 - y\right)^{1/2} \,. \label{eq:phi}
\end{equation}
In equation \eqref{eq:x_rm_start_end}, the function $\varphi_3$ is evaluated for $\chi = \chi_r$ as well as
\begin{align}
\tilde{x}_m^{\rm ini}  = \frac{f}{\tilde{f}_m^{\rm ini}} && \text{and} && \tilde{x}_m^{\rm eq}  = \frac{f}{\tilde{f}_m^{\rm eq}} \, . \label{eq:def:tilde_x_m_ini_eq}
\end{align}
Here, we defined the frequencies 
\begin{align}
\tilde{f}_m^{\rm ini}  &= \left(\frac{a_0}{\tilde{a}_m^{\rm ini}}\right)^{1/2} h_m^0 \simeq 6.00\times 10^4 {\,\rm Hz}\left(\frac{50}{\Gamma}\right) \left(\frac{10^{-10}}{G\mu}\right)\left(\frac{10^{-24}{\,\rm s}}{t_{\rm ini}}\right)^{1/3}\, , \\
\tilde{f}_m^{\rm eq} &= \left(\frac{a_0}{\tilde{a}_m^{\rm eq}}\right)^{1/2} h_m^0 \simeq 5.14 \times 10^{-8}{\,\rm Hz} \left(\frac{50}{\Gamma}\right)\left(\frac{10^{-10}}{G\mu}\right)\, ,
\end{align}
where we denoted $\tilde{a}_m^{\rm ini} = \tilde{a}_m\left(t_{\rm ini}\right)$ and $\tilde{a}_m^{\rm eq}  = \tilde{a}_m\left(t_{\rm eq}\right)$. 
In Eq.~\eqref{eq:x_rm_start_end}, we also need
\begin{align}
x_m^{\rm eq} = \frac{f}{f_m^{\rm eq}} && \text{and} && 
x_m^0 = \frac{f}{f_m^0} \label{eq:def:x_m_0_eq}
\end{align}
with the additional frequencies 
\begin{align}
f_m^{\rm eq} = \left(\frac{a_0}{a_{\rm eq}}\right)^{1/2} h_m^0\simeq 4.30\times 10^{-8} {\, \rm Hz}\left(\frac{50}{\Gamma}\right)\left(\frac{10^{-10}}{G\mu}\right),
\end{align}
and $f_m^0 = h_m^0$. Returning to $t_{\rm fric}=t_{\rm Pl}/\left(G\mu\right)^2$ as the standard choice for $t_{\rm ini}$, $\tilde{f}_m^{\rm ini}$ becomes 
\begin{align}
    \tilde{f}_m^{\rm fric} =3.42 \times 10^{4} {\,\rm Hz} \left(\frac{50}{\Gamma}\right)\left(\frac{10^{-10}}{G\mu}\right)^{1/3} \, .
\end{align}

\begin{figure}
\tiny
    \centering
    \begin{subfigure}{0.49\textwidth}
     \begin{overpic}[width = \textwidth]{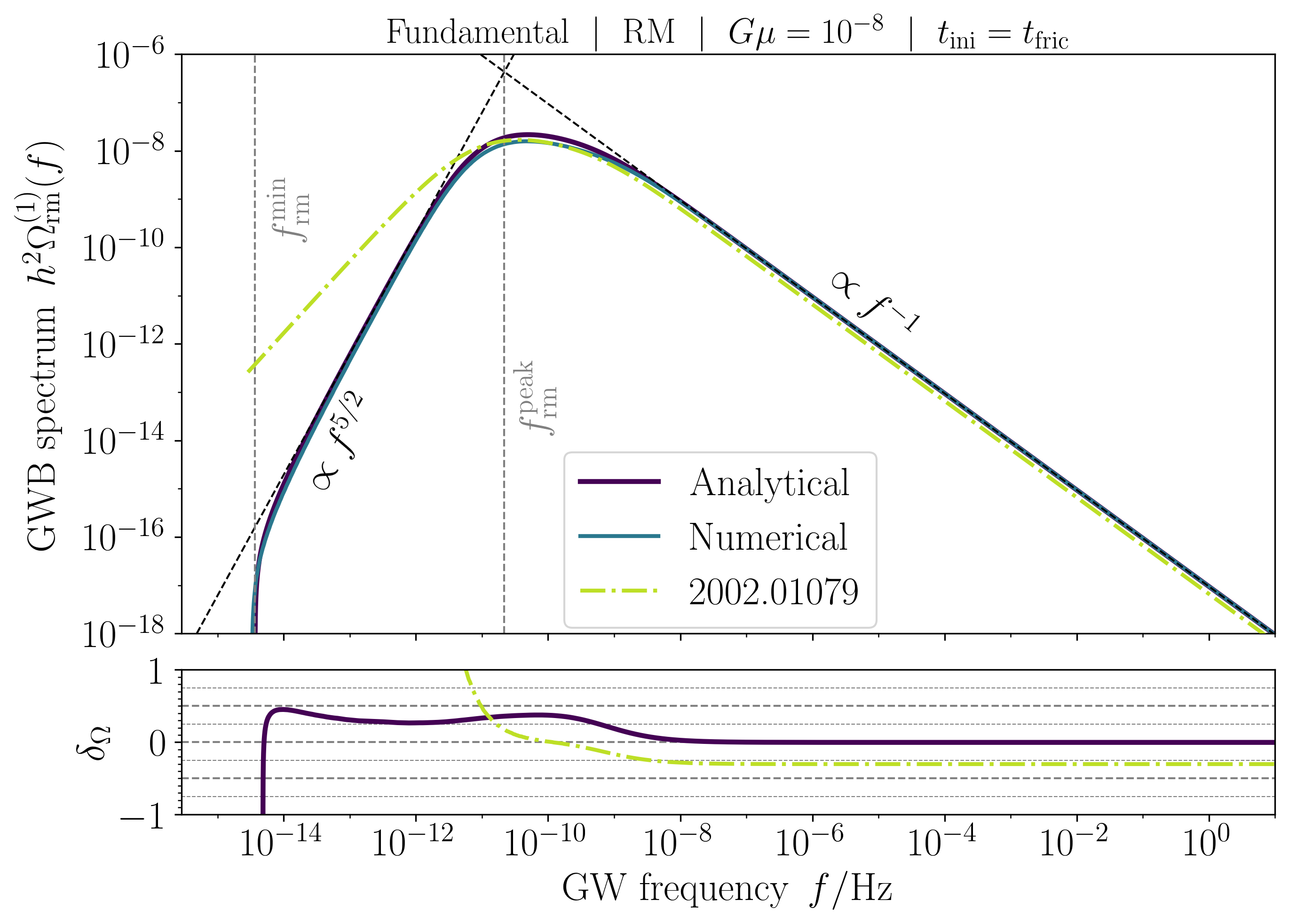}
\put(21.35,58.5){\rotatebox{90}{\eqref{eq:f_rm_min}}}
\put(40.5,45){\rotatebox{90}{\eqref{eq:f_rm_peak_1}}}
\put(71,44){\rotatebox{-36}{\eqref{eq:Orm_high1}}}
\put(28,41.5){\rotatebox{61}{\eqref{eq:Orm_low1}}}
\end{overpic}
\caption{ }
\label{fig:RM1}
    \end{subfigure}
    \begin{subfigure}{0.49\textwidth}
    \begin{overpic}[width = \textwidth]{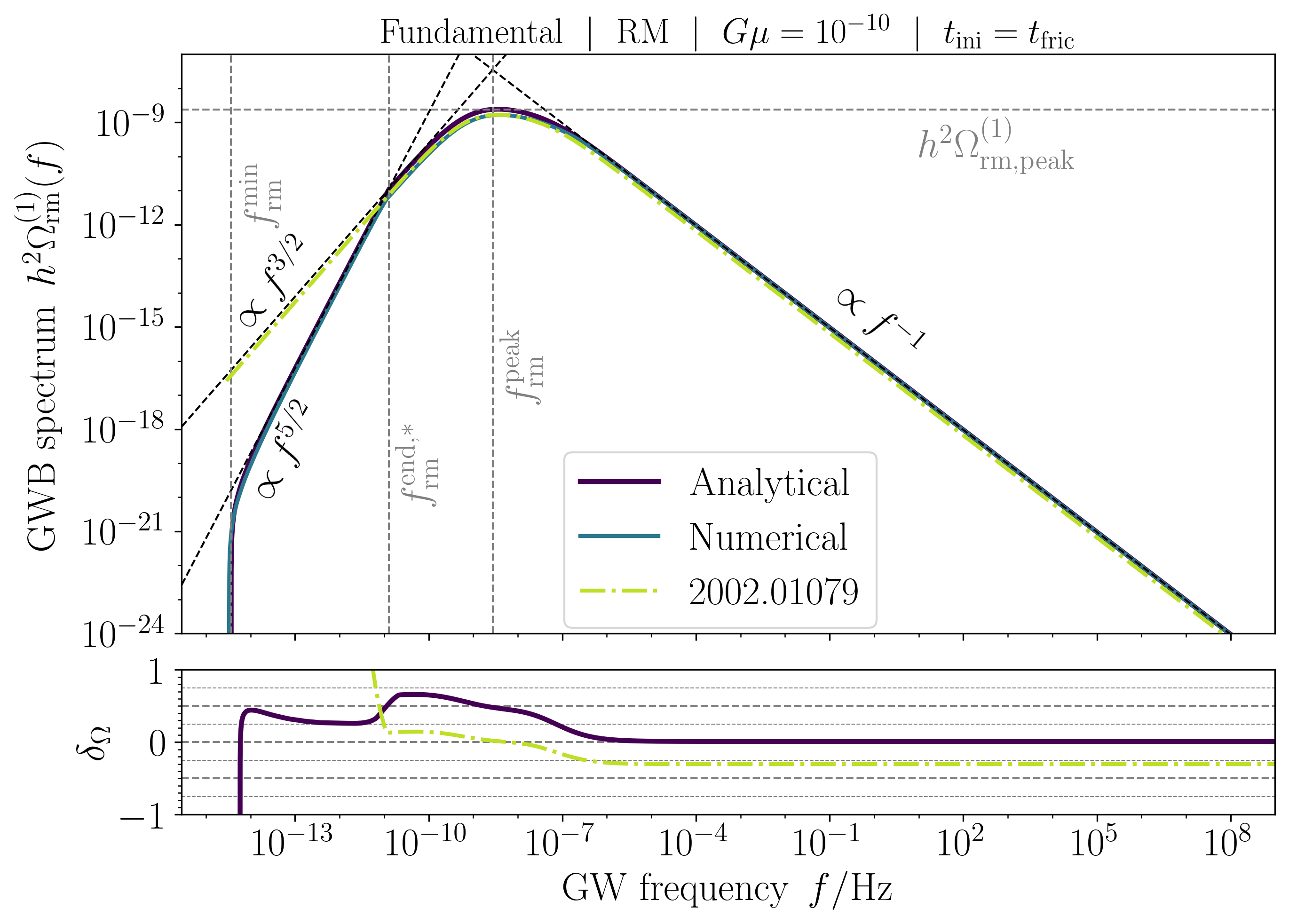}
    \put(14.75,52.5){\rotatebox{90}{\eqref{eq:f_rm_min}}}
    \put(31.3,40){\rotatebox{90}{\eqref{eq:f_rm_switch}}}
\put(39.25,47.75){\rotatebox{90}{\eqref{eq:f_rm_peak_2}}}
\put(71.15,43.15){\rotatebox{-38}{\eqref{eq:Orm1_HF2}}}
\put(86,60){\rotatebox{0}{\eqref{eq:Orm_peak2}}}
\put(23,41){\rotatebox{64}{\eqref{eq:Orm_low1}}}
\put(24.25,53.75){\rotatebox{47}{\eqref{eq:Orm_low2}}}
\end{overpic}
\caption{  }
\label{fig:RM2}
    \end{subfigure}

\bigskip
    
    \begin{subfigure}{0.49\textwidth}
    \begin{overpic}[width = \textwidth]{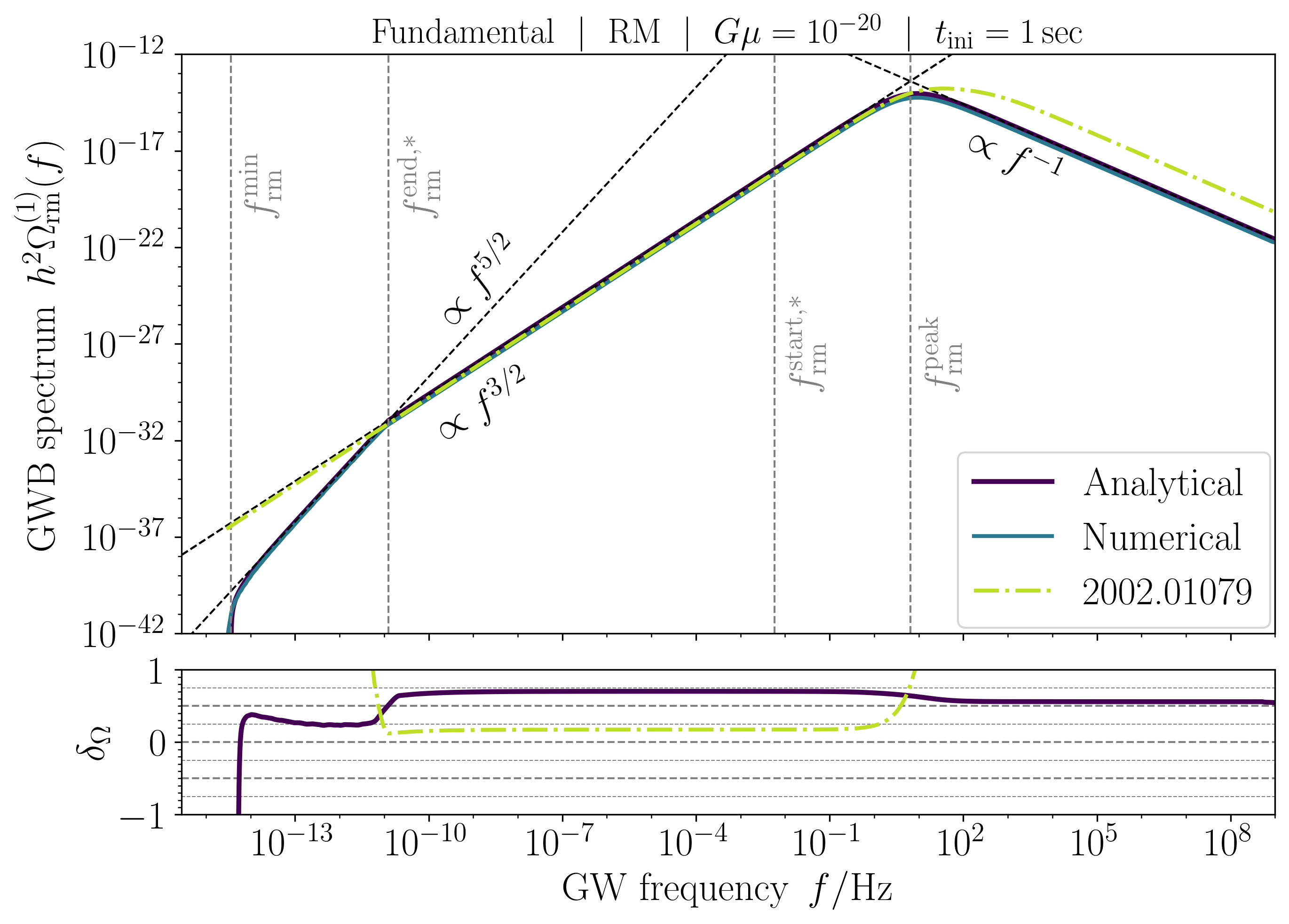}
    \put(14.75,53.25){\rotatebox{90}{\eqref{eq:f_rm_min}}}
    \put(61.25,49){\rotatebox{90}{\eqref{eq:f_rm_switch}}}
    \put(27,53.25){\rotatebox{90}{\eqref{eq:f_rm_switch}}}
\put(71.75,48){\rotatebox{90}{\eqref{eq:f_rm_peak_3}}}
\put(82,56.5){\rotatebox{-23}{\eqref{eq:Orm1_HF3}}}
\put(39.5,53){\rotatebox{48}{\eqref{eq:Orm_low1}}}
\put(41,42){\rotatebox{33.5}{\eqref{eq:Orm_low2}}}
\end{overpic}
\caption{  }
\label{fig:RM3}
    \end{subfigure}
    \begin{subfigure}{0.49\textwidth}
   \begin{overpic}[width = \textwidth]{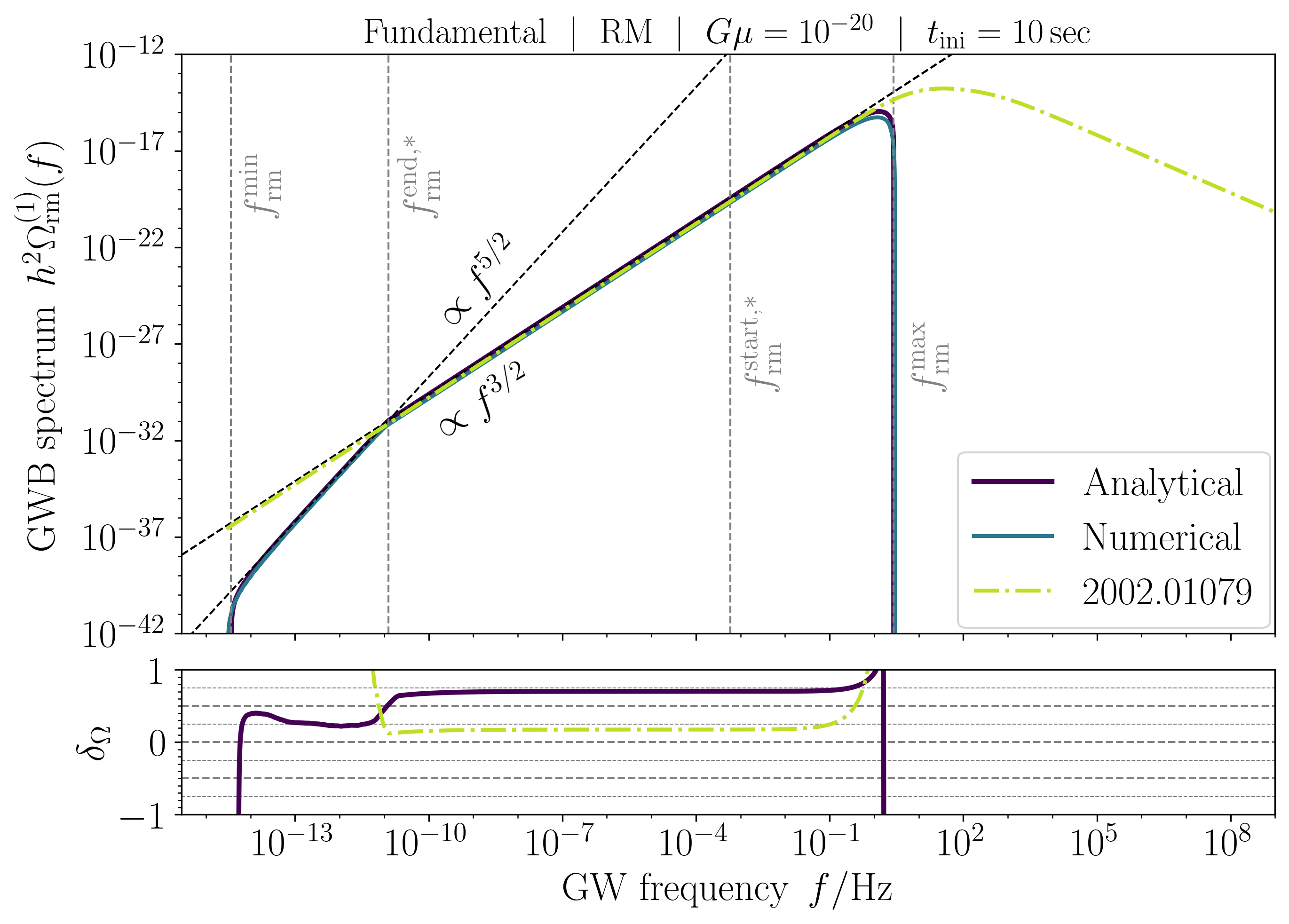}
       \put(14.75,53.25){\rotatebox{90}{\eqref{eq:f_rm_min}}}
    \put(53.5,40.5){\rotatebox{90}{\eqref{eq:f_rm_switch}}}
    \put(27,53.25){\rotatebox{90}{\eqref{eq:f_rm_switch}}}
\put(71,48){\rotatebox{90}{\eqref{eq:f_rm_max}}}
\put(39.5,53){\rotatebox{48}{\eqref{eq:Orm_low1}}}
\put(41,42){\rotatebox{33.5}{\eqref{eq:Orm_low2}}}
\end{overpic}
\caption{  }
\label{fig:RM4}
    \end{subfigure}
    \caption{\footnotesize
    Four qualitatively different fundamental GW spectra in the RM case. While Figs.~\ref{fig:RM2} and \ref{fig:RM3} exhibit the same power laws, they are qualitatively different due to the occurrence of $f_{\rm rm}^{{\rm start},*}$ as discussed at the very end of Sec.~\ref{subsec:RM_Loops_Fundamental}.  
    Panels and color codes are the same as in Fig.~\ref{fig:RR1}.
    {Recall from our discussion at the end of Sec.~\ref{subsec:Gravitational_Wave_Spectrum} that large initial times such as $t_{\rm ini}=1 \, {\rm sec}$ or $10\, {\rm sec}$ chosen in the lower panels are well motivated, especially for very low string tensions (see also Refs.~\cite{Gouttenoire:2019kij, Servant:2023tua, Schmitz_2024}). Note that the RM spectra provided in Ref.~\cite{Sousa_2020} are independent of $t_{\rm ini}$. Following the discussion in Refs.~\cite{Sousa_2020, Blanco-Pillado:2024aca}, we cut their RM spectra off at the frequency $f_{\rm rr}^{\rm min}$.}
    }
\end{figure}

The RM spectra obtained as a result of the above computation are depicted in Figs.~\ref{fig:RM1} to \ref{fig:RM4}, and we will turn to a discussion of their features and qualitative differences shortly. Before that, let us briefly comment on improvements over the RM spectrum derived in Ref.~\cite{Sousa_2020} that are visible in those plots. A difference common to all four figures is that our spectrum exhibits a minimum frequency $f_{\rm rm}^{\rm min}$ below which it strictly drops to zero. This is the same behavior one finds in the numerical calculation. {In contrast, the analytical spectrum predicted in earlier works extends, in principle, to arbitrarily low frequencies, which would physically correspond to GWs from arbitrarily large string loops. As explained in Ref.~\cite{Sousa_2020}, their spectrum must be cut off by hand. In later work by some of the same authors in Ref.~\cite{Blanco-Pillado:2024aca}, the cutoff was given explicitly and coincides with our expression for $f_{\rm rr}^{\rm min}$, which is slightly different from the actual cutoff $f_{\rm rm}^{\rm min}$, as can be seen in Figs.~\ref{fig:RM1} to \ref{fig:RM4}. However, more importantly, since the cutoff is not contained in the spectral shape of the expressions given in  Ref.~\cite{Sousa_2020}, these largely deviate at low frequencies from the numerically computed spectra. This deviation is clearly visible in all four plots.} For certain regions of parameter space, namely low string tensions and late initial times, our spectrum also contains a maximum frequency $f_{\rm rm}^{\rm max}$ and drops to zero above, as illustrated in Fig.~\ref{fig:RM4}. Such a high-frequency cutoff was not present in earlier work either. This maximum occurs since none of the string loops has fully decayed until today, so we cannot obtain GWs from arbitrarily small loops. Therefore, the expressions derived in this paper are, in contrast to those obtained in Ref.~\cite{Sousa_2020}, able to reproduce the numerical GW spectra found in Ref.~\cite{Schmitz_2024}. The remaining differences come from a change in the power law describing the spectrum at low frequencies, as we will discuss below.

\medskip\noindent
\textbf{Discussion of the spectrum:} We are now going to examine the details of the RM spectrum. Two of the most important characteristics are the minimum and maximum frequency, which are reached when the spectrum \eqref{eq:Orm1} vanishes. This clearly is the case if
\begin{align}
    x_{\rm rm}^{\rm start} = x_{\rm rm}^{\rm end}. \label{eq:RM_Max_Min}
\end{align}
From the explicit expressions for the integration boundaries \eqref{eq:x_rm_start_end}, we see that it is necessary to distinguish between two cases for each of them. Let us begin by determining the frequencies $f^{{\rm start}, *}_{\rm rm}$ and $f^{{\rm end}, *}_{\rm rm}$ at which we have to switch from one behavior in $x_{\rm rm}^{\rm start}$ or $x_{\rm rm}^{\rm end}$ respectively to another one. In the former case, the switch occurs when $x_m^{\rm eq} = \varphi_3 (\tilde{x}_m^{\rm ini}, \chi_r)$, in the latter case when $x_m^0 = \varphi_3(\tilde{x}_m^{\rm eq}, \chi_r)$. Using these conditions to substitute $\varphi_3$ in equation \eqref{eq:def_phi3} yields
\begin{align}
    f^{{\rm start}, *}_{\rm rm}= \frac{f_m^{\rm eq}}{\left(\frac{\tilde{a}^{\rm ini}_{m}}{a_{\rm eq}}\right)^{3/2}(1+\chi_r) - 1}  && \text{and} && f^{{\rm end}, *}_{\rm rm} = \frac{f_m^0}{\left(\frac{\tilde{a}_{m}^{\rm eq}}{a_0}\right)^{3/2}\left(1+\chi_r\right)-1} \,. \label{eq:f_rm_switch}
\end{align}
Now, consider first the switch at $f^{{\rm start}, *}_{\rm rm}$, which only occurs if this frequency is positive or, equivalently, if $(1+\chi_r)^{2/3}\tilde{a}_{m}^{\rm ini} >a_{\rm eq}$. 
In typical scenarios, this is not fulfilled since a combination of very large values of $t_{\rm ini}$ and very low string tensions $G\mu$ is necessary. To see which solution is realized below and above the switch frequency, consider the behavior at high frequencies first. At those, we have $\varphi_3(x, \chi) \longrightarrow (1+\chi)^{1/3} x$ and, accordingly, find the general asymptotic behavior
\begin{align}
    x_{\rm rm}^{\rm start} \longrightarrow \begin{cases} \left(1+\chi_r\right)^{1/3} \tilde{x}_{\rm m}^{\rm ini} & \text{if } \left(1+\chi_r\right)^{2/3}\tilde{a}_{\rm m}^{\rm ini}> a_{\rm eq} \\ x_{\rm m}^{\rm eq} & \text{else} 
    \end{cases} \, .
\end{align}
In the upper case, a switch from one solution to the other must occur since $f^{{\rm start}, *}_{\rm rm}$ exists. Since we just found that the high-frequency behavior corresponds to $\varphi_3 (\tilde{x}_m^{\rm ini}, \chi_r)$, this is the solution relevant above $f^{{\rm start}, *}_{\rm rm}$. Accordingly, $x_m^{\rm eq}$ is the relevant solution below the switch frequency. In the lower case, there is no switch frequency and since $x_m^{\rm eq}$ is relevant at high frequencies, it is relevant everywhere. Note that, even though we have used a large-frequency expansion, the result of this argument holds exactly. In total, we find
\begin{align}
    \text{If } \left(1+\chi_r\right)^{2/3}\tilde{a}_{m}^{\rm ini}> a_{\rm eq}, \text{ then }  && x_{\rm rm}^{\rm start} &= \begin{cases} x_{m}^{\rm eq} & \text{if } f<f^{{\rm start}, *}_{\rm rm} \\ \varphi_3(\tilde{x}_m^{\rm ini}, \chi_r) & \text{else}
    \end{cases}.\nonumber\\ \text{Otherwise} && x_{\rm rm}^{\rm start} &=x_m^{\rm eq}. \label{eq:x_rm_start_summary}
\end{align}
Next, consider the switch at $f^{{\rm end}, *}_{\rm rm}$, which requires $(1+\chi_r)^{2/3} \tilde{a}_m^{\rm eq} > a_0$. Unlike the condition for the first case, this is satisfied for most of the physically relevant parameter choices.\footnote{In particular, for our standard parameter values, this is satisfied as long as $G\mu <5.85 \times 10^{-9}$.} To identify which solution is realized at which frequency, we progress in a similar manner as before, resulting in the statement:
\begin{align}
    \text{If } (1+\chi_r)^{2/3} \tilde{a}_m^{\rm eq} > a_0, \text{ then } && x_{\rm rm}^{\rm end} &= \begin{cases} \varphi_3(\tilde{x}_m^{\rm eq}, \chi_r) & \text{if } f< f^{{\rm end}, *}_{\rm rm} \\
    x_m^0 & \text{else} \end{cases}.\nonumber\\
    \text{Otherwise} && x_{\rm rm}^{\rm end} &= \varphi_3(\tilde{x}_m^{\rm eq}, \chi_r).  \label{eq:x_rm_end_summary}
\end{align}
If both switch frequencies exist, which is automatically the case if $f_{\rm rm}^{{\rm start}, *}$ does, then we can also show that usually
\begin{align}
f^{{\rm end}, *}_{\rm rm}<f^{{\rm start}, *}_{\rm rm}. \label{eq:f_rm_start_end_star_order}
\end{align}
To this end, first note that $f_m^0/f_m^{\rm eq}=\left(a_{\rm eq}/a_0\right)^{1/2} < 1$. Besides, since we typically\footnote{This is true for initial times as late as $t_{\rm ini}\sim 5\times 10^6 {\rm s}$. For much larger $t_{\rm ini}$ the order can be inverted.} have $\tilde{a}_m^{\rm ini}/a_{\rm eq} <\tilde{a}_m^{\rm eq}/a_0$, we know that 
\begin{align}
1< \frac{\left(\frac{\tilde{a}_m^{\rm eq}}{a_0}\right)^{3/2}(1+\chi_r)-1}{\left(\frac{\tilde{a}_m^{\rm ini}}{a_{\rm eq}}\right)^{3/2}(1+\chi_r)-1} \, .
\end{align}
Putting these two results together, we immediately obtain \eqref{eq:f_rm_start_end_star_order}.

Given a frequency $f$ at which we want to investigate the spectrum, we have to distinguish in total three possible cases, which are summarized in table \ref{table:1}.
\begin{table}
\begin{center}
\begin{tabular}{ |c||c|c|c|  }
 \hline
 \textbf{Case} & \textbf{Condition} & $\boldsymbol{x_{\rm rm}^{\rm start}}$ & $\boldsymbol{x_{\rm rm}^{\rm end}}$ \\
 \hline
1   & Either $f^{{\rm end}, *}_{\rm rm}$ does not exist or $f<f_{\rm rm}^{{\rm end}, *}$    & $x_m^{\rm eq}$ &  $\varphi_3(\tilde{x}_m^{\rm eq}, \chi_r)$\\
2&   $f^{{\rm end}, *}_{\rm rm}<f$ and either $f^{{\rm start}, *}_{\rm rm}$ does not exist or $f< f^{{\rm start}, *}_{\rm rm}$ & $x_m^{\rm eq}$   &  $x_m^0$\\
3 & $f^{{\rm start}, *}_{\rm rm}$ exists and $f^{{\rm start}, *}_{\rm rm}<f$ & $\varphi_3(\tilde{x}_m^{\rm ini}, \chi_r)$ &  $x_m^0$\\
 \hline
\end{tabular}
\end{center}
\caption{Case distinction for $x_{\rm rm}^{\rm start}$ and $x_{\rm rm}^{\rm end}$ relevant for the RM spectrum.}
\label{table:1}
\end{table}
With the distinction between these three cases at hand, we can return to the initial problem of determining the maximum and minimum frequencies of the spectrum. The criterion for them to occur is given in \eqref{eq:RM_Max_Min}. Let us begin with case 1 from table \ref{table:1} in which we assume that either $f^{{\rm end}, *}_{\rm rm}$ exists and $f_{\rm rm}^{\rm min}<f^{{\rm end}, *}_{\rm rm}$ or $f^{{\rm end}, *}_{\rm rm}$ does not exist. Then, \eqref{eq:RM_Max_Min} becomes
\begin{align}
    x_m^{\rm eq} = \varphi_3(\tilde{x}_m^{\rm eq}, \chi_r). 
\end{align}
With the same argument as in the derivation of $f^{{\rm start}, *}_{\rm rm}$, upon replacing $\tilde{x}_m^{\rm eq} \leftrightarrow \tilde{x}_m^{\rm ini}$, one finds
\begin{align}
    f_{\rm rm}^{\rm min} = \frac{f_m^{\rm eq}}{\left(\frac{\tilde{a}_m^{\rm eq}}{a_{\rm eq}}\right)^{3/2}(1+\chi_r)-1} \,. \label{eq:f_rm_min}
\end{align}
Hence, a minimum can only exist if $\tilde{a}_m^{\rm eq} (1+\chi_r)^{2/3} > a_{\rm eq}$ which is always realized for physically reasonable string tensions.\footnote{The non-existence of a minimum, however, does not mean that the spectrum extends to $f=0$. This would not be reasonable either since the radiation would need to stem from arbitrarily large string loops. It is rather the case that non-satisfaction of the previous criterion implies $x_{m}^{\rm eq} \geq \varphi_3\left(\tilde{x}_m^{\rm eq}, \chi_r\right)$ and, thus, the Heaviside function in \eqref{eq:Orm1} vanishes and there is no RM contribution. From a physical viewpoint, this is only possible if loops produced during radiation domination have decayed fast enough so that none reach matter domination. This, in turn, is only possible if there is a gap between the epoch of radiation and matter domination, which in reality does not occur. Nonetheless, it appears as an artifact of our treatment because of the splitting of the history of the Universe into a pure matter and a pure radiation part due to which $\tilde{a}_m^{\rm eq}/a_{\rm eq}<1$, while at the same time we require loop production before $\tilde{a}_m^{\rm eq}$ and loop decay after $a_{\rm eq}$. {Since $\tilde{a}_{m}^{\rm eq}/a_{\rm eq}= (2-\sqrt{2})^{2/3}\simeq 0.7$, this happens only if $\chi_r< (2-\sqrt{2})^{-1}-1\simeq 0.7$, which corresponds, under the assumption that $\alpha \xi_r=0.1$, to huge string tensions, $\log_{10}\left(G\mu\right) \gtrsim -2.6$. However, if one allows $\alpha\xi_r$ to be several orders of magnitude smaller than $1$, this can actually become relevant for more realistic string tensions.}}  Before we continue with the next case, we have to make sure that, given the existence of $f^{{\rm end}, *}_{\rm rm}$, our assumption $f_{\rm rm}^{\rm min}<f^{{\rm end}, *}_{\rm rm}$ is actually satisfied. Let us start from this assumption and bring it into the form
\begin{align}
    \frac{f_m^{\rm eq}}{f_m^0} = \left(\frac{a_0}{a_{\rm eq}}\right)^{1/2} < \frac{\left(\frac{\tilde{a}_m^{\rm eq}}{a_{\rm eq}}\right)^{3/2}(1+\chi_r)-1}{\left(\frac{\tilde{a}_m^{\rm eq}}{a_0}\right)^{3/2}(1+\chi_r)-1} \,.\label{eq:rm_min_assumption}
\end{align}
First, note that the derivative of the right side w.r.t.\ $\chi_r$ is always negative 
and hence monotonically decreasing in $\chi_r$. This function reaches a minimum of $\left(a_0/a_{\rm eq}\right)^{3/2}$ for $\chi_r\to \infty$. In this case, the inequality \eqref{eq:rm_min_assumption} is indeed satisfied, and due to the monotonicity of the right-hand side, it also holds true for any other allowed value of $\chi_r$ implying that our initial assumption was justified. Correspondingly, we do not need to consider cases 2 or 3 since they would require $f_{\rm rm}^{\rm min}$ to be larger than the existing switch frequency $f_{\rm rm}^{{\rm end}, *}$.

On the other hand, to find the maximum frequency, we need to examine cases 2 and 3 in table \ref{table:1}. Let us begin with case 2, i.e., $f^{{\rm end}, *}_{\rm rm}$ exists and fulfills $f^{{\rm end}, *}_{\rm rm} < f_{\rm rm}^{\rm max}$. Furthermore, either $f^{{\rm start}, *}_{\rm rm}$ exists and meets the condition $f_{\rm rm}^{\rm max}<f^{{\rm start}, *}_{\rm rm}$ or $f^{{\rm start}, *}_{\rm rm}$ does not exist.  The criterion in equation \eqref{eq:rm_min_assumption} turns into $ x_m^{\rm eq} = x_m^0$
 or equivalently $a_{\rm eq}/a_0= 1$, which is never true.
 This means that either $f^{{\rm start}, *}_{\rm rm}$ exists, which implies that a possible maximum frequency must be larger than this frequency, or $f^{{\rm start}, *}_{\rm rm}$ does not exist, and there is no maximum frequency such that the spectrum extends to arbitrarily high frequencies. 

The last case is that $f^{{\rm start}, *}_{\rm rm}$ exists and $f_{\rm rm}^{\rm max}>f^{{\rm start}, *}_{\rm rm}$. The criterion \eqref{eq:RM_Max_Min} becomes 
 \begin{align}
 \varphi_3(\tilde{x}_m^{\rm ini}, \chi_r) = x_m^0 \,.
 \end{align}
Using the same argument as in the derivation of $f^{{\rm end}, *}_{\rm rm}$, we obtain, upon exchanging $\tilde{x}_m^{\rm ini} \leftrightarrow \tilde{x}_m^{\rm eq}$,
\begin{align}
    f_{\rm rm}^{\rm max} = \frac{f_m^0}{\left(\frac{\tilde{a}_m^{\rm ini}}{a_0}\right)^{3/2}(1+\chi_r)-1} \,. \label{eq:f_rm_max}
\end{align}
Therefore, a maximum can only exist if $\tilde{a}_m^{\rm ini} (1+\chi_r)^{2/3} > a_0$. The validity of our assumption $f_{\rm rm}^{\rm max}> f^{{\rm start}, *}_{\rm rm}$ can easily be proven by following an argument analogous to the one below \eqref{eq:f_rm_min}. As for the RR spectrum, one can heuristically argue that there is the possibility for a maximum frequency below which the spectrum strictly vanishes. Such a high-frequency cutoff must arise if none of the string loops has fully decayed until today. The ones that would decay first are those produced with the smallest length and the longest time to decay. These loops are those produced at $t_{\rm ini}$ and their length today sets the minimum length $l_{\rm rm}^{\rm min}$ of all loops, in particular all RM loops. Via $f_{\rm rm}^{\rm max}= 2/l_{\rm rm}^{\rm min}$, this minimum length sets the maximum frequency of the spectrum, which, if one follows the presented arguments, leads again to the frequency \eqref{eq:f_rm_max} we obtained by evaluating the Heaviside function in the spectrum. 

The relevant frequencies, their order, and the cases in which they occur can be schematically summarized as follows:
\begin{align}
   f_{\rm rm}^{\rm min} \begin{array}{l} 
         \Leftarrow \\
         <
    \end{array} f_{\rm rm}^{{\rm end}, *} \begin{array}{l}
         \Leftarrow \\
         <
    \end{array}f_{\rm rm}^{{\rm start}, *} \begin{array}{l}
         \Leftarrow  \\
         < 
    \end{array} f_{\rm rm}^{\rm max} \,.
\end{align}
The implication arrow indicates that, e.g., if $f_{\rm rm}^{\rm max}$ exists, then so does $f_{\rm rm}^{\rm start, *}$ and so on. The minimum frequency always exists, while the first switch frequency almost always exists, except for very large string tensions. The second switch frequency and the maximum frequency exist only in the case of very low $G\mu$ and very large $t_{\rm ini}$.
  
Equipped with the minimum and maximum frequencies, we can replace the Heaviside function in \eqref{eq:Orm1} with the more explicit expression
\begin{align}
    \left[1-\Theta(f-f_{\rm rm}^{\rm  max})\Theta(f_{\rm rm}^{\rm max})\right]\left[1-\Theta(f_{\rm rm}^{\rm min}-f)\Theta(f_{\rm rm}^{\rm min})\right] \, .
\end{align}

Having determined the most important frequencies, except for the peak frequency to which we will come later, we shall now discuss the shape of the spectrum. 
Let us begin with asymptotically low frequencies, at which we find from a small $x$ expansion of the spectral shape function \eqref{eq:RM_Shape_Function} that the spectrum can in first non-vanishing order be written as
\begin{align}
    h^2 \Omega_{\rm rm, low}^{(1)} = \frac{3\mathcal{A}_{\rm rm}}{4\sqrt{x_m^0}} \left((x_{\rm rm}^{\rm end})^2 - (x_{\rm rm}^{\rm start})^2\right) \, . \label{eq:rm_low_frequency}
\end{align}
Moreover, it is possible to use the low-frequency expansion
\begin{align}
    \varphi_3(x, \chi)\stackrel{x\to 0}{\longrightarrow} \sqrt{1+\chi}\,x^{3/2} \, . \label{eq:phi3_low_x_expansion}
\end{align}    
    Therefore, we obtain in case 1 of table \ref{table:1}
\begin{align}
    h^2 \Omega_{\rm rm, low}^{(1)} = \frac{3\mathcal{A}_{\rm rm}}{4(x_m^0)^{1/2}} \left((1+\chi_r) \left(\tilde{x}_m^{\rm eq}\right)^3 - \left(x_m^{\rm eq}\right)^2\right).
\end{align}
 If we are considering low frequencies but still $f\gg f_{\rm rm}^{\rm min}$, then the second term is negligible. Upon noting that furthermore $\chi_r\gg 1$, we can approximate 
\begin{align} \label{eq:Orm_low1}
    h^2 \Omega_{\rm rm, low}^{(1)} &=  \frac{3 \mathcal{A}_{\rm rm}}{4} \chi_r \left(\frac{\tilde{a}_m^{\rm eq}}{a_0}\right)^{1/4} \left(\frac{f}{\tilde{f}_m^{\rm eq}}\right)^{5/2}\simeq \\ \nonumber &\simeq 1.95\times 10^{-14} \left(\frac{C_r}{0.171}\right) \left(\frac{\Gamma}{50}\right) \left(\frac{G\mu}{10^{-10}}\right)^2 \left(\frac{f}{10^{-12} {\, \rm Hz}}\right)^{5/2}\, .
\end{align}
In case 2, we find straightforwardly
\begin{align} \label{eq:Orm_low2}
    h^2\Omega_{\rm rm, low}^{(1)} &= \frac{3\mathcal{A}_{\rm rm}}{4 (x_m^0)^{1/2}} \left((x_m^0)^2-(x_m^{\rm eq})^2\right) =  \frac{3\mathcal{A}_{\rm rm}}{4} \left(\frac{f}{f_m^0}\right)^{3/2} \left(1-\frac{a_{\rm eq}}{a_0}\right)\simeq  \\ \nonumber &\simeq 2.45\times 10^{-10}\left(\frac{C_r}{0.171}\right) \left(\frac{\Gamma}{50}\right) \left(\frac{G\mu}{10^{-10}}\right)^2 \left(\frac{f}{10^{-10} {\, \rm Hz}}\right)^{3/2}\, .
\end{align}
In case 3, we again have to use \eqref{eq:phi3_low_x_expansion} and find
\begin{align} 
    h^2\Omega_{\rm rm, low}^{(1)}(f) = \frac{3 \mathcal{A}_{\rm rm}}{4 (x_m^0)^{1/2}} \left((x_m^0)^2 -(1+\chi_r)(\tilde{x}_m^{\rm ini})^3\right) \, .
\end{align}
If we furthermore assume that $f\ll f_{\rm rm}^{\rm max}$, then the second term becomes negligible, and this expression simplifies to
\begin{align} \label{Orm_low3}
    h^2\Omega_{\rm rm, low}^{(1)}(f) &=   \frac{3 \mathcal{A}_{\rm rm}}{4}\left(\frac{f}{f_m^0} \right)^{3/2} \simeq \\ \nonumber&\simeq 2.45\times 10^{-18}\left(\frac{C_r}{0.171}\right) \left(\frac{\Gamma}{50}\right) \left(\frac{G\mu}{10^{-20}}\right)^2 \left(\frac{f}{10^{-2} {\, \rm Hz}}\right)^{3/2}\, , \label{eq:RM_lowF_powerlaw}
\end{align}
which is the same as the expression derived for case 2, up to a relative deviation in the prefactor of the order of $3\times 10^{-4}$ .  

For asymptotically large frequencies, the spectrum takes, at leading order, generally the form
\begin{align}
    h^2 \Omega_{\rm rm, high}^{(1)} = \frac{3 \mathcal{A}_{\rm rm}}{(x_m^0)^{1/2}} \left(\frac{1}{(x_{\rm rm}^{\rm start})^{1/2}}-\frac{1}{(x_{\rm rm}^{\rm end})^{1/2}}\right).  
\end{align}
We furthermore can use the behavior
\begin{align}
    \varphi_3(x, \chi)\stackrel{x\to \infty}{\longrightarrow} (1+\chi)^{1/3} \,x
\end{align}
and immediately see that $h^2 \Omega_{\rm rm, high}^{(1)}\left(f\right) \propto f^{-1}$ in all cases of table \ref{table:1}. 
In more detail, we obtain in the first case 
\begin{align}
 h^2 \Omega_{\rm rm, high}^{(1)} = \frac{3\mathcal{A}_{\rm rm}}{f} \left(\left(f_m^0 f_m^{\rm eq}\right)^{1/2} - \frac{\left(f_m^0 \tilde{f}_m^{\rm eq}\right)^{1/2}}{(1+\chi_r)^{1/6}}\right)\, . 
\end{align}
As an approximation, which becomes slightly rough for large string tensions, we can neglect the second term since $\chi_r\gg 1$ and find the parameter dependence
\begin{align}
     h^2 \Omega_{\rm rm, high}^{(1)} \simeq 1.10\times 10^{-16} \left(\frac{C_r}{0.171}\right)\left(\frac{50}{\Gamma}\right)^{3/2}\left(\frac{10^{-10}}{G\mu}\right)^{1/2} \left(\frac{1 {\, \rm Hz}}{f}\right) \, .\label{eq:Orm_high1}
\end{align}
For case 2, we have
\begin{align} \label{eq:Orm1_HF2}
 h^2 \Omega_{\rm rm, high}^{(1)} &= \frac{3\mathcal{A}_{\rm rm}}{(x_m^0)^{1/2}} \left(\frac{1}{(x_m^{\rm eq})^{1/2}} - \frac{1}{(x_m^0)^{1/2}}\right) = 3\mathcal{A}_{\rm rm}\frac{f_m^0}{f}\left(\left(\frac{a_0}{a_{\rm eq}}\right)^{1/4} -1 \right)\simeq \\ \nonumber &\simeq 9.54\times 10^{-17} \left(\frac{C_r}{0.171}\right)\left(\frac{50}{\Gamma}\right)^{3/2}\left(\frac{10^{-10}}{G\mu}\right)^{1/2} \left(\frac{1 {\, \rm Hz}}{f}\right)\, , 
\end{align}
while in the third case, we find
\begin{align}
 h^2 \Omega_{\rm rm, high}^{(1)} = 3\mathcal{A}_{\rm rm} \left( \frac{1}{(1+\chi_r)^{1/6}} \left( \frac{a_0}{\tilde{a}_m^{\rm ini}} \right)^{1/4} -1 \right) \frac{f_m^0}{f} \, . \label{eq:Orm1_HF3}
 \end{align}
For the reasonable part of parameter space, there is no large hierarchy between the two occurring terms, and the expression cannot be further simplified in a sensible way, so one cannot extract a simple power-law dependence on the model parameters.

From the power laws we derived for the high-frequency and the low-frequency behavior, we can acquire comparatively simple approximations for the peak frequency by determining the intersection between these asymptotic expressions: $h^2\Omega_{\rm rm, low}^{(1)}(f_{\rm rm}^{\rm peak})=h^2\Omega_{\rm rm, high}^{(1)}(f_{\rm rm}^{\rm peak})$. If the peak lies in a region where case 1 of table \ref{table:1} applies, one obtains
\begin{align}
    f_{\rm rm}^{\rm peak} = \frac{\tilde{f}_{\rm m}^{\rm eq}}{(1+\chi_r)^{1/3}}\left(4(1+\chi_r)^{1/6}\left(\frac{\tilde{a}_m^{\rm eq}}{a_{\rm eq}}\right)^{1/4}-4\right)^{2/7}. \label{eq:f_rm_peak_1}
\end{align}
If the peak lies in a region where the second case is valid, one finds
\begin{align}
    f_{\rm rm}^{\rm peak} = \gamma f_m^0 && \text{with}  && \gamma =  \left(4\frac{\left(\frac{a_0}{a_{\rm eq}}\right)^{1/4}-1}{1-\frac{a_{\rm eq}}{a_0}}\right)^{2/5} \,. \label{eq:f_rm_peak_2}
\end{align}
Finally, in the third case, the peak frequency lies at\footnote{This case can only be realized for a small range of $G\mu$ values if one chooses $t_{\rm ini} = t_{\rm fric}$, since in this case, the existence of this peak frequency also requires $f_{\rm rm}^{{\rm start}, *}$ to exist, which translates to the criterion $3H_m^0\alpha \xi_r t_{\rm Pl}/(2\Gamma) <\left(G\mu\right)^3 < \left(a_0/a_{\rm eq}\right)^{3/2} 3H_m^0\alpha \xi_r t_{\rm Pl}/(2\Gamma)$. For $\alpha\xi_r =0.1$ and $\Gamma = 50$, this means $5\times 10^{-22}\lesssim G\mu \lesssim 4\times 10^{-20}$.}
\begin{align}
    f_{\rm rm}^{\rm peak} = f_m^0 \left(\frac{4}{(1+\chi_r)^{1/6}}\left(\frac{a_0}{\tilde{a}_m^{\rm ini}}\right)^{1/4}-4\right)^{2/5}. \label{eq:f_rm_peak_3}
\end{align}
In the usually realized case at which the peak frequency satisfies the criteria of case 2 in table \ref{table:1}, the full RM spectrum can be evaluated at this peak frequency and one finds a relatively simple expression for the peak amplitude. This amplitude is proportional to $\mathcal{A}_{\rm rm}$ and the proportionality factor depends on $a_{\rm eq}/a_0$ only, i.e., it is independent of string-related parameters:
\begin{align}\label{eq:Orm_peak2}
    h^2 \Omega_{\rm rm, peak}^{(1)} = \frac{\mathcal{A}_{\rm rm}}{\gamma^{1/2}}\left(\frac{2+3\left(\frac{a_{\rm eq}}{a_0}\right)^{1/2}\gamma}{\left(1+\left(\frac{a_{\rm eq}}{a_0}\right)^{1/2}\gamma\right)^{3/2}} -\frac{2+3\gamma}{\left(1+\gamma\right)^{3/2}} \right)\simeq 0.37 \mathcal{A}_{\rm rm} \, .
\end{align}

Let us briefly summarize the features that can be found in the RM spectrum. Starting at low frequencies, it generally contains a minimum frequency $f_{\rm rm}^{\rm min}$ below which it vanishes. Still at low frequencies, but sufficiently far above this minimum cutoff, we also find a generically occurring $f^{5/2}$ power-law behavior of the spectrum, as can be seen in all plots \ref{fig:RM1} to \ref{fig:RM4}.  If, on the other hand, $\left(1+\chi_r\right)^{2/3}\tilde{a}_m^{\rm eq}>a_0$, as is the case for sufficiently low string tensions $G\mu \lesssim 10^{-9}$, this low-frequency behavior turns first into an $f^{3/2}$ power law above a frequency $f_{\rm rm}^{{\rm end}, *}$. Such a transition is present for the parameter values chosen for Figs.~\ref{fig:RM2} to \ref{fig:RM4}. At even lower string tensions or later initial times $t_{\rm ini}$, another switch frequency $f_{\rm rm}^{{\rm start}, *}$ occurs. At frequencies below this switch frequency, the lower integration boundary in Eq.~\eqref{eq:RM_First_Formula} is determined by the condition $t>t_{\rm eq}$. Above the switch frequency, the condition $t_*>t_{\rm ini}$ determines the lower integration boundary. While qualitatively different, the $f^{3/2}$ power law remains, up to a prefactor, the same, and can be observed in Figs.~\ref{fig:RM3} and \ref{fig:RM4}. Leaving the low-frequency range, the spectrum has two options. In the first case, it has no high-frequency cutoff, implying that it extends to arbitrarily large frequencies. Here, one finds an $f^{-1}$ decay at asymptotically large frequencies, as illustrated in Figs.~\ref{fig:RM1} to \ref{fig:RM3}. The spectrum peaks then, depending on whether the $f^{5/2}$, the first or the second $f^{3/2}$ power law was realized before reaching the peak, at the peak frequency \eqref{eq:f_rm_peak_1}, \eqref{eq:f_rm_peak_2}, or \eqref{eq:f_rm_peak_3}, respectively.
In the second case, realized at very low string tensions, and very late initial times $t_{\rm ini}$, or, more quantitatively, if $\tilde{a}^{\rm ini}_m\left(1+\chi_r\right)^{2/3}>a_0$, a maximum frequency $f_{\rm rm}^{\rm max}$ arises, above which the spectrum vanishes and which cuts off the high-frequency region. This behavior is shown in Fig.~\ref{fig:RM4} and discussed in detail in Ref.~\cite{Schmitz_2024}. {Note that the analytic expressions in Ref.~\cite{Sousa_2020} do not capture  the $f^{5/2}$ power law at low frequencies and, furthermore, a minimum frequency cutoff has to be imposed by hand.}

\subsection{MM loops}
\label{subsec:MM_Loops_Fundamental}
The last contribution to the fundamental spectrum is the gravitational radiation emitted from string loops produced and decaying during the matter-dominated era.
For this case, we have to use the loop number density \eqref{eq:matter} with \eqref{eq:Beta_Loop_Number_Density} in the general formula for the GW spectrum \eqref{eq:GW_Spectrum_Fundamental}, which yields
\begin{align}
    h^2 \Omega_{\rm mm}^{(1)} = h^2 \frac{16\pi}{3} &\left(\frac{G\mu}{H_0}\right)^2 \frac{\Gamma}{f} C_m \Theta(a_{\rm end}-a_{\rm start})\times \\ \nonumber &\times\int_{a_{\rm start}/a_0}^{a_{\rm end}/a_0} \frac{\left(\frac{a}{a_0}\right)^4 \pd \left(\frac{a}{a_0}\right)}{t^2_{m}(a) \left(\frac{2}{f}\frac{a}{a_0} + \Gamma G\mu t_{m}(a)\right)^{2} H_{m}(a)} \,,
\end{align}
where the expressions for the Hubble parameter $H_m$ and cosmic time $t_m$ are given in \eqref{eq:H_mat+t_mat}. As in the RM case, we change the integration variable to $x_m$ defined in \eqref{eq:def:x_m}, such that
\begin{align}
h^2 \Omega_{\rm mm}^{(1)} = 18\pi h^2 \Omega_m G\mu C_m \left(\frac{h_m^0}{f}\right)^2 \Theta(x_{\rm mm}^{\rm end} - x_{\rm mm}^{\rm start}) \int_{x_{\rm mm}^{\rm start}}^{x_{\rm mm}^{\rm end}} \frac{x_m^2}{(1+x_m)^{2}} \,\pd x_m \, ,
\end{align}
which can easily be evaluated as 
\begin{equation}
\label{eq:Omm1}
h^2\Omega_{\rm mm}^{(1)} = \mathcal{A}_{\rm mm} \Theta\left(x_{\rm mm}^{\rm end} - x_{\rm mm}^{\rm start}\right) \left[S_{\rm mm}^{(1)}(x_{\rm mm}^{\rm end}) - S_{\rm mm}^{(1)}(x_{\rm mm}^{\rm start})\right] \,.
\end{equation}
The resulting amplitude reads
\begin{equation}
\mathcal{A}_{\rm mm} = 18\pi\,C_m\,h^2\Omega_{m} \,G\mu \simeq 3.13\times 10^{-11} \left(\frac{C_m}{0.0387}\right)\left(\frac{G\mu}{10^{-10}}\right) \,,
\end{equation}
and the spectral shape function is of the form
\begin{equation}
S_{\rm mm}^{(1)}\left(x\right) =\frac{1}{(x_m^0)^2}\left( \frac{x^2+x-1}{1+x} - 2\,\ln\left(1+x\right)\right) \,. \label{eq:MM_Shape_function}
\end{equation}
{The fundamental MM spectrum was also computed in \cite{Sousa_2020}. It differs from our result due to the presence of the Heaviside function in Eq.~\eqref{eq:Omm1} and the values of the integration boundaries $x_{\rm mm}^{\rm start, end}$, at which the spectral shape functions are evaluated.} In our treatment, using again the notation defined in \eqref{eq:def:tilde_x_m_ini_eq} and \eqref{eq:def:x_m_0_eq}, one finds for the integration boundaries in the MM case
\begin{align}
x_{\rm mm}^{\rm start} = \max\big\{\chi_m^{-1},\varphi_3\left(\tilde{x}_m^{\rm eq},\chi_m\right)\big\} && \text{and} && 
x_{\rm mm}^{\rm end} = x_m^0 \,,
\end{align}
where we further introduced
\begin{align}
    \chi_m = \frac{\alpha \xi_m}{\Gamma G\mu}\,.
\end{align}
The boundary $x_{\rm mm}^{\rm start}$ derives directly from the Heaviside functions in \eqref{eq:matter}, i.e., it accounts for the conditions $t>t_*>t_{\rm eq}$, while the boundary $x_{\rm mm}^{\rm end}$ corresponds to the upper integration boundary in the general expression for the GW spectrum today \eqref{eq:GW_Spectrum_Fundamental}, which is merely $t<t_0$. 

\begin{figure}
\tiny
    \centering
    \begin{subfigure}{0.49\textwidth}
     \begin{overpic}[width = \textwidth]{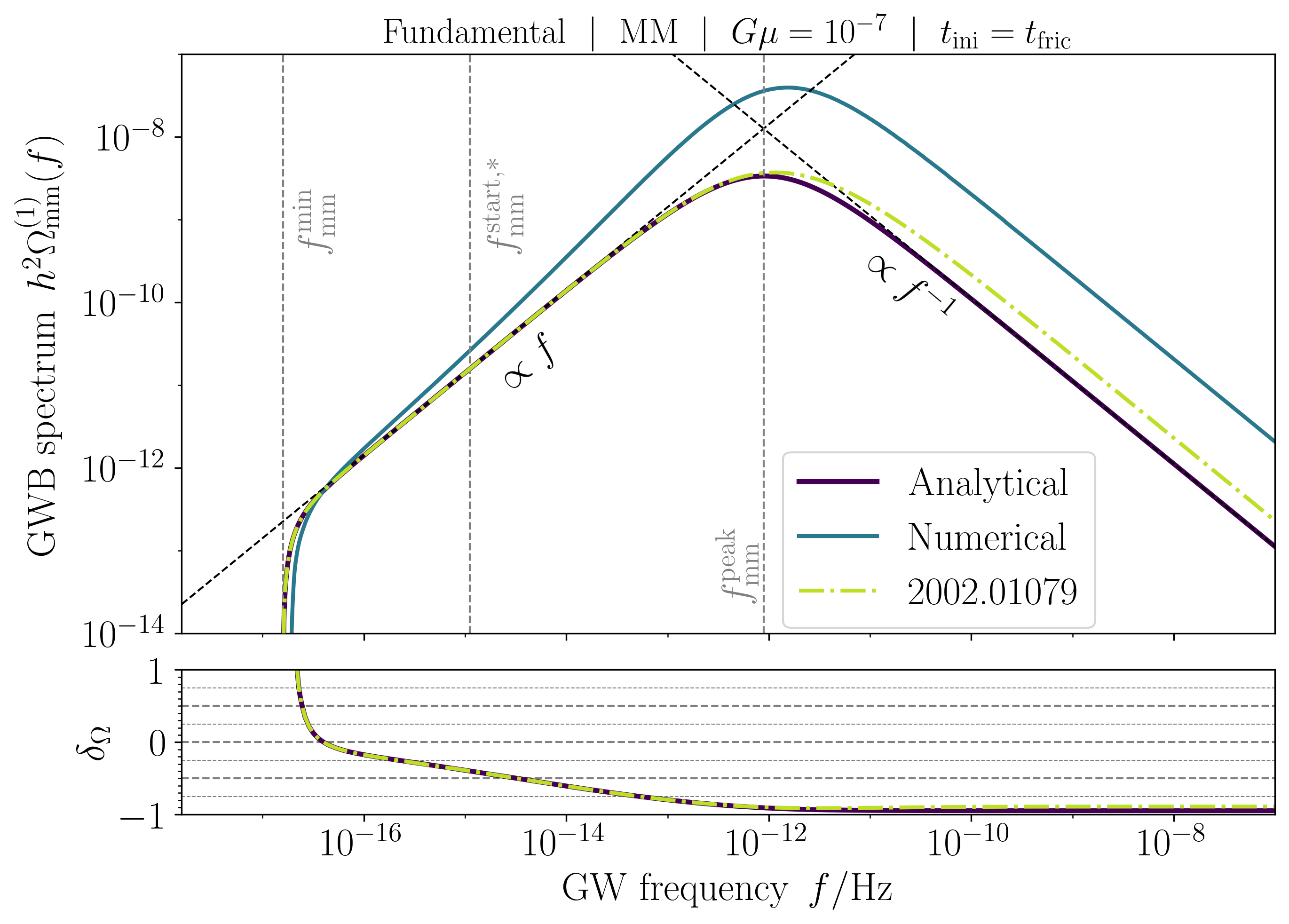}
    \put(18.5,50){\rotatebox{90}{\eqref{eq:f_mm_min}}}
    \put(33,50){\rotatebox{90}{\eqref{eq:fmmstarstart}}}
    \put(56,32){\rotatebox{90}{\eqref{f_mm_peak}}}
    \put(43,45){\rotatebox{40}{\eqref{eq:Omega_mm_low}}}
    \put(73,45.75){\rotatebox{-40}{\eqref{eq:Omega_mm_high}}}
\end{overpic}
\caption{ }
\label{fig:MM1}
    \end{subfigure}
    \begin{subfigure}{0.49\textwidth}
   \begin{overpic}[width = \textwidth]{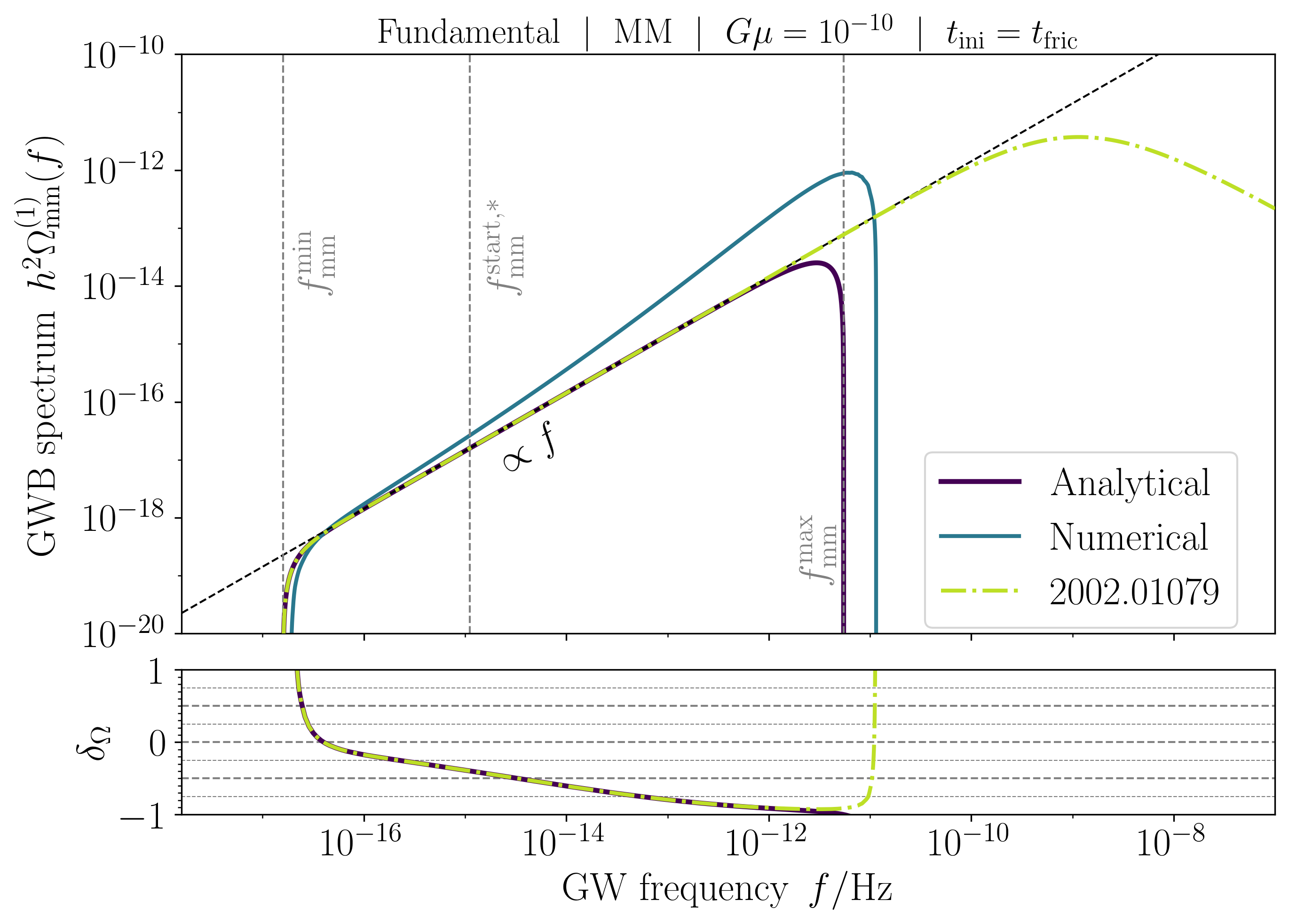}
       \put(18.5,47){\rotatebox{90}{\eqref{eq:f_mm_min}}}
    \put(33,47){\rotatebox{90}{\eqref{eq:fmmstarstart}}}
    \put(62,33){\rotatebox{90}{\eqref{def:f_mm_max}}}
    \put(44,38){\rotatebox{30}{\eqref{eq:Omega_mm_low}}}
\end{overpic}
\caption{  }
\label{fig:MM2}
    \end{subfigure}
    \caption{\footnotesize
    Fundamental GW spectra for the two qualitatively different cases of MM spectra. Panels and color codes are the same as in Fig.~\ref{fig:RR1}.}
\end{figure}

The spectrum obtained from these analytical considerations is depicted in Figs.~\ref{fig:MM1} and \ref{fig:MM2}, which we briefly want to discuss before turning to studying the features of this spectrum in detail. As can easily be seen from the plots, the agreement between the analytical prediction and the numerical result is much worse than it is for the RM and the RR cases. This is, however, not unexpected and can be explained by the rough approximation we used for the cosmic expansion history. In particular, our analytical treatment assumes the reduced correlation length $\xi$ to be constant during the radiation-dominated and the matter-dominated era. Recalling the numerical solution to the VOS model shown in Fig.~\ref{fig:VOS}, it is clear that, while this is a good approximation during radiation domination, it becomes poor for the matter-dominated era. 
In spite of this, our analytical approximation correctly reproduces the qualitative behavior of the spectrum and, in contrast to Ref.~\cite{Sousa_2020}, predicts the existence of a maximum frequency above which the spectrum vanishes. In absence of a maximum frequency, the results from Ref.~\cite{Sousa_2020} reproduce the numerical spectra slightly better. Overall, the MM spectrum, however, is less important for the full spectrum than the RM and the RR spectrum since it typically yields a subdominant contribution. {However, for values of $\alpha\xi_r$ much smaller than our benchmark value $0.1$, the MM contribution can actually become relevant.} In Section \ref{subsec:FullFundamentalSpectrum}, we will see how a simple modification of our analytical expressions can restore very good agreement between the analytical and the numerical complete fundamental spectrum.

\medskip\noindent
\textbf{Discussion of the spectrum:} Let us now discuss the characteristics of the MM spectrum and begin with the determination of the minimum and maximum frequencies. These are reached when the spectrum \eqref{eq:Omm1} vanishes, which happens if
\begin{align}
    x_{\rm mm}^{\rm start} = x_{\rm mm}^{\rm end} \, . \label{eq:MM_max_min_condition}
\end{align}
Depending on the values of $f_{\rm mm}^{\rm min}$ and $f_{\rm mm}^{\rm max}$, we have to distinguish between two cases for $x_{\rm mm}^{\rm start}$. The switch from one behavior to the other is located at the frequency which satisfies $\chi_{m}^{-1} = \varphi_3(\tilde{x}_m^{\rm eq}, \chi_m)$. Upon using \eqref{eq:def_phi3}, this can be seen to be equivalent to
\begin{align}
    f_{\rm mm}^{{\rm start}, *} = \frac{\tilde{f}_m^{\rm eq}}{\chi_m} \, . \label{eq:fmmstarstart}
\end{align}
We recall from the discussion of the RM spectrum that at very large frequencies, $\varphi_3(x, \chi)\propto x$. Hence, if $f$ is sufficiently large, then $\varphi_3(\tilde{x}_m^{\rm eq}, \chi_m)>\chi_m^{-1}$ and, therefore,
\begin{align}
    x_{\rm mm}^{\rm start} = \begin{cases} \chi_m^{-1} & \text{if } f<f_{\rm mm}^{{\rm start}, *} \\
    \varphi_3(\tilde{x}_m^{\rm eq}, \chi_m) & \text{else}
    \end{cases}.
\end{align}
Let us now come back to the minimum frequency and start from the assumption that $f_{\rm mm}^{\rm min}< f_{\rm mm}^{{\rm start}, *}$. Then \eqref{eq:MM_max_min_condition} is equivalent to $x_m^0 = \chi_m^{-1}$. This results in
\begin{align}
    f_{\rm mm}^{\rm min} = \frac{f_m^0}{\chi_m} \, . \label{eq:f_mm_min}
\end{align}
To verify our initial assumption, observe that $a_0 > \tilde{a}_m^{\rm eq}$. Thus $f_m^0 < \tilde{f}_m^{\rm eq}$, and correspondingly
$f_{\rm mm}^{\rm min} = f_m^0/\chi_m < \tilde{f}_m^{\rm eq}/\chi_m = f_{\rm mm}^{{\rm start}, *}$. Observe that the minimum frequency always occurs in the MM spectrum. It is associated with the radiation from the largest loops ever produced, which are those created today with length $l_{\rm mm}^{\rm max} = \alpha \xi_m t_0$. This maximum length gives rise to the minimum frequency $f_{\rm mm}^{\rm min} = 2/l_{\rm mm}^{\rm max}$, which is, moreover, the overall minimum frequency. 

Next, we turn to the maximum frequency and set out from the assumption that $f_{\rm mm}^{\rm max}>f_{\rm mm}^{{\rm start}, *}$. This means, the spectrum vanishes if $x_m^0 = \varphi_3(\tilde{x}_m^{\rm eq}, \chi_m)$. The equation is the same as the one we had to solve for $f^{{\rm start}, *}_{\rm rm}$ upon making the replacements $\tilde{x}_m^{\rm ini} \leftrightarrow \tilde{x}_m^{\rm eq}$, $x_m^{\rm eq} \leftrightarrow x_m^0$, and $\chi_r \leftrightarrow \chi_m$. Hence, we find 
\begin{align}
    f_{\rm mm}^{\rm max} = \frac{f_m^0}{(1+\chi_m)\left(\frac{\tilde{a}_m^{\rm eq}}{a_0}\right)^{3/2} - 1} \,. \label{def:f_mm_max}
\end{align}
While the switch and minimum frequencies always exist, a maximum is only present if $(1+\chi_m)^{2/3} \tilde{a}_m^{\rm eq} > a_0$. Heuristically, this is satisfied if and only if the earliest loops in matter domination born at time $t_{\rm eq}$ have not fully decayed until today. The maximum frequency then corresponds to the length $l_{\rm mm}^{\rm min}$ these shortest loops have nowadays due to the relation $f_{\rm mm}^{\rm max} = 2/l_{\rm mm}^{\rm min}$.
We still have to check whether our supposition that $f_{\rm mm}^{\rm max}>f_{\rm mm}^{{\rm start},*}$ was actually legitimate. This assumption can be brought into the equivalent form
\begin{align}
     1 < \frac{\left(\frac{\tilde{a}_m^{\rm eq}}{a_0}\right)^{1/2}\chi_m}{(1+\chi_m)\left(\frac{\tilde{a}_m^{\rm eq}}{a_0}\right)^{3/2} - 1} \,.\label{eq:mm_assumption}
\end{align}
Since $\left(\tilde{a}_m^{\rm eq}/a_0\right)^{1/2} < 1$, it follows that $\left(\tilde{a}_m^{\rm eq}/a_0\right)^{1/2}\chi_m > \left(\tilde{a}_m^{\rm eq}/a_0\right)^{1/2}(1+\chi_m) -1$ and besides
\begin{align}
    1 < \frac{(1+\chi_m) \left(\frac{\tilde{a}_m^{\rm eq}}{a_0}\right)^{1/2} - 1}{(1+\chi_m) \left(\frac{\tilde{a}_m^{\rm eq}}{a_0}\right)^{3/2} - 1} \,.
\end{align}
Combining these two results, we find that \eqref{eq:mm_assumption} is indeed satisfied. 

The determination of the extremal frequencies allows us to replace the Heaviside function occurring in \eqref{eq:Omm1} with the expression
\begin{align}
     \left[1-\Theta(f-f_{\rm mm}^{\rm max})\Theta(f_{\rm mm}^{\rm max})\right]\Theta(f-f_{\rm mm}^{\rm min}) \,.
\end{align}

After discussing the most important frequencies of the spectrum, we want to consider its asymptotic behavior. Let us begin with the low-frequency regime. The shape function \eqref{eq:MM_Shape_function} allows for the expansion 
\begin{align}
    S_{\rm mm}^{(1)}(x)\stackrel{x\to 0}{\longrightarrow} \frac{1}{(x_m^0)^2}\left(-1 + \frac{x^3}{3}\right). 
\end{align}
We first look at the case in which $f< f_{\rm mm}^{{\rm start}, *}$ and hence $x_{\rm mm}^{\rm start}= \chi_m^{-1}$. This means we cannot use the above expansion of the shape function for $S_{\rm mm}^{(1)}(x_{\rm mm}^{\rm start})$ but need to continue with the full form instead. We then find
\begin{align}
    h^2 \Omega_{\rm mm, low}^{(1)} = \mathcal{A}_{\rm mm}\left(-\frac{1}{(x_m^0)^2}+\frac{x_m^0}{3}- S_{\rm mm}^{(1)}(\chi_m^{-1})\right) \, .
\end{align}    
Still restricting to low frequencies, but assuming that we are sufficiently far above the minimum frequency $f\gg f_{\rm mm}^{\rm min}$, the spectrum further simplifies to 
\begin{align}
\label{eq:Omega_mm_low}
h^2 \Omega_{\rm mm, low}^{(1)} &= \frac{\mathcal{A}_{\rm mm}}{3}\,x_m^0 = \\ \nonumber &= 1.42\times 10^{-18} \left(\frac{C_m}{0.0387}\right)\left(\frac{\Gamma}{50}\right)\left(\frac{G\mu}{10^{-10}}\right)^2\left(\frac{f}{10^{-16} {\, \rm Hz}}\right) \,. 
\end{align}
For the second case, in which $f> f_{\rm mm}^{{\rm start}, *}$, we have $x_{\rm mm}^{\rm start} = \varphi_3(\tilde{x}_m^{\rm eq}, \chi_m)$. Expanding at low frequencies to leading order, we obtain $x_{\rm mm}^{\rm start} \to (1+\chi_m)^{1/2}(\tilde{x}_m^{\rm eq})^{3/2}$, and hence find the same expression as before
\begin{align}
    h^2 \Omega_{\rm mm, low}^{(1)} = \frac{\mathcal{A}_{\rm mm}}{3}\, x_m^0 \, .
\end{align}

For the high-frequency regime, we can, in principle, also distinguish the cases $f< f_{\rm mm}^{{\rm start}, *}$ and $f> f_{\rm mm}^{{\rm start}, *}$. The former can, however, never be realized since $f_{\rm mm}^{{\rm start}, *}/f_{\rm mm}^{\rm min} = \left(a_0/\tilde{a}_m^{\rm eq}\right)^{1/2}$, which means that the minimum frequency $f_{\rm mm}^{\rm min}$ is always only about two orders of magnitude smaller than $f_{\rm mm}^{{\rm start}, *}$. 
For the remaining case that $f> f_{\rm mm}^{{\rm start}, *}$, we find, upon using the expansion $x_{\rm mm}^{\rm start} = \varphi_3(\tilde{x}_m^{\rm eq}, \chi_m) \longrightarrow (1+\chi_m)^{1/3} \tilde{x}_m^{\rm eq}$ that
\begin{align}
\label{eq:Omega_mm_high}
    h^2 \Omega_{\rm mm, high}^{(1)} = \frac{\mathcal{A}_{\rm mm}}{x_m^0}\left(1- \left(1+\chi_m\right)^{1/3} \left(\frac{\tilde{a}_m^{\rm eq}}{a_0}\right)^{1/2}\right)\, .
\end{align}
Note that, sensibly, this expansion breaks down at the latest when the condition for the existence of the maximum frequency $f_{\rm mm}^{\rm max}$ in \eqref{def:f_mm_max} is satisfied. For sufficiently small $\chi_m$ and thus large enough string tensions, the first term on the right-hand side of \eqref{eq:Omega_mm_high} dominates, 
\begin{align}
     h^2 \Omega_{\rm mm, high}^{(1)} = \frac{\mathcal{A}_{\rm mm}}{x_m^0} \simeq 2.31\times 10^{-10} \left(\frac{C_m}{0.0387}\right) \left(\frac{\Gamma}{50}\right) \left(\frac{G\mu}{10^{-7}}\right)^2
     \left(\frac{10^{-10} {\, \rm Hz}}{f}\right) \,. 
\end{align}

With the results for the behavior of the spectrum at low and high frequencies \eqref{eq:Omega_mm_low} and \eqref{eq:Omega_mm_high} at hand, we can determine the peak frequency by requiring that both expressions yield the same value at this frequency. This yields
\begin{align} \label{f_mm_peak}
    f_{\rm mm}^{\rm peak} = f_m^0 \left(3-3\left(1+\chi_m\right)^{1/3}\left(\frac{\tilde{a}_m^{\rm eq}}{a_0}\right)^{1/2}\right)^{1/2}
\end{align}
and will, of course, break down if $f_{\rm mm}^{\rm peak} \sim f_{\rm mm}^{\rm min}$ or if $f_{\rm mm}^{\rm max}$ exists and thus the high-frequency part of the spectrum does not. The latter is actually the typical case for reasonably small values of the string tension. This being said, the MM contribution is normally, at frequencies around its peak {and for the standard value $\alpha \xi_r\simeq 0.1$}, strongly subdominant compared to the RM contribution, which makes the peak position and height less relevant from a phenomenological point of view. 

We conclude with a short summary of the features of the MM spectrum. It always exhibits a minimum frequency below which it is zero. Above, it rises according to a power law $\propto f$. The spectrum invariably contains a switch frequency, which leaves the power-law behavior unaffected, though. All these characteristics can be observed in both Figs.~\ref{fig:MM1} and \ref{fig:MM2}. For very large string tensions, the spectrum then reaches its peak and turns into an $f^{-1}$ power law, as one can see in Fig.~\ref{fig:MM1}. For lower string tensions, the spectrum exhibits a maximum frequency, is cut off, and a high-frequency power-law behavior is not present (cf.\ Fig.~\ref{fig:MM2}). As mentioned previously, the difference between our analytical calculation and the numerical results can be understood as a result of an improper treatment of the reduced correlation length $\xi$ as a constant $\xi_m$ during matter domination. This explains, in particular, two features, best illustrated in Fig.~\ref{fig:MM2}. Recall that the maximum frequency is associated with loops produced at $t_{\rm eq}$. At this time, we have $\xi(t_{\rm eq})\simeq 0.30$ in comparison to $\xi_m = 0.63$. This will typically shift the analytical maximum frequency to about half the value that is found numerically. Indeed, replacing $\xi_m$ in \eqref{def:f_mm_max} with $\xi(t_{\rm eq})$ reproduces the numerically determined maximum frequency correctly. Moreover, an increase in $\xi$ over time, as is found in the numerical solution, implies that, in contrast to the case where $\xi_m$ is constant, the transition from small to large initial loop lengths is faster than the usual $l_*\propto t$ growth. This, in turn, means that the spectrum decreases towards low frequencies faster than $\propto f$ as predicted analytically. Having fixed $\xi_m$ to a larger value than $\xi(t_{\rm eq})$, we also predict too little power emitted from small MM loops, thus underestimating the amplitude of the spectrum close to the maximum frequency. All these effects can be observed in Fig.~\ref{fig:MM2}.

\begin{figure}
\tiny
    \centering
    \begin{subfigure}{0.49\textwidth}
     \begin{overpic}[width = \textwidth]{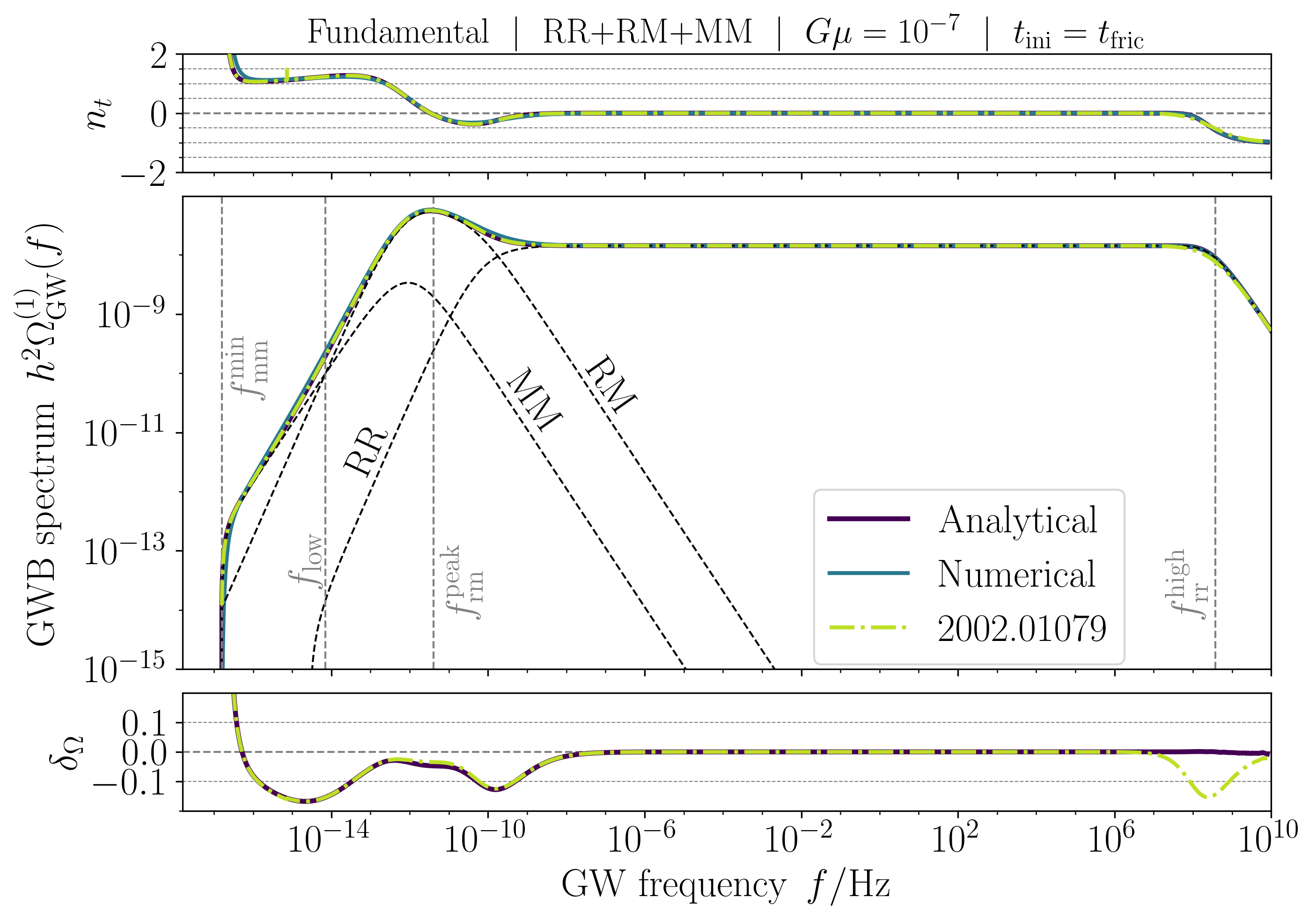}
    \put(17,45){\rotatebox{90}{\eqref{eq:f_mm_min}}}
    \put(21,45){\rotatebox{90}{\eqref{eq:LF_Transition_MM_to_RM}}}
    \put(33.25,29.5){\rotatebox{90}{\eqref{eq:f_rm_peak_2}}}
    \put(88.75,29){\rotatebox{90}{\eqref{eq:f_rr_high}}}
\end{overpic}
\caption{ }
\label{fig:Full1}
    \end{subfigure}
    \begin{subfigure}{0.49\textwidth}
    \begin{overpic}[width = \textwidth]{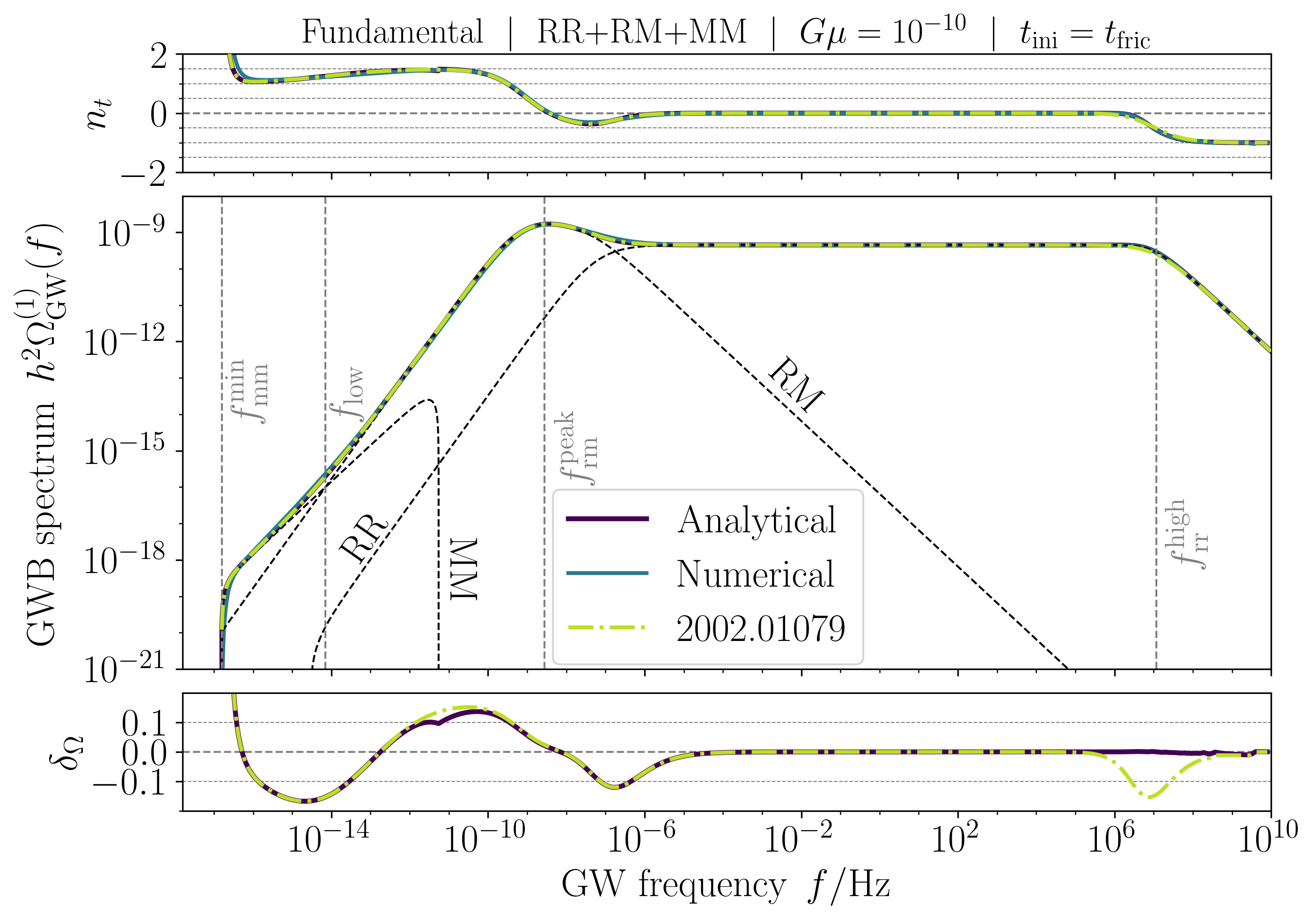}
    \put(17,44){\rotatebox{90}{\eqref{eq:f_mm_min}}}
    \put(24.75,44){\rotatebox{90}{\eqref{eq:LF_Transition_MM_to_RM}}}
    \put(37.75,31){\rotatebox{90}{\eqref{eq:f_rm_peak_2}}}
    \put(88.75,33){\rotatebox{90}{\eqref{eq:f_rr_high}}}
\end{overpic}
\caption{  }
\label{fig:Full2}
    \end{subfigure}

\bigskip
    
    \begin{subfigure}{0.49\textwidth}
    \begin{overpic}[width = \textwidth]{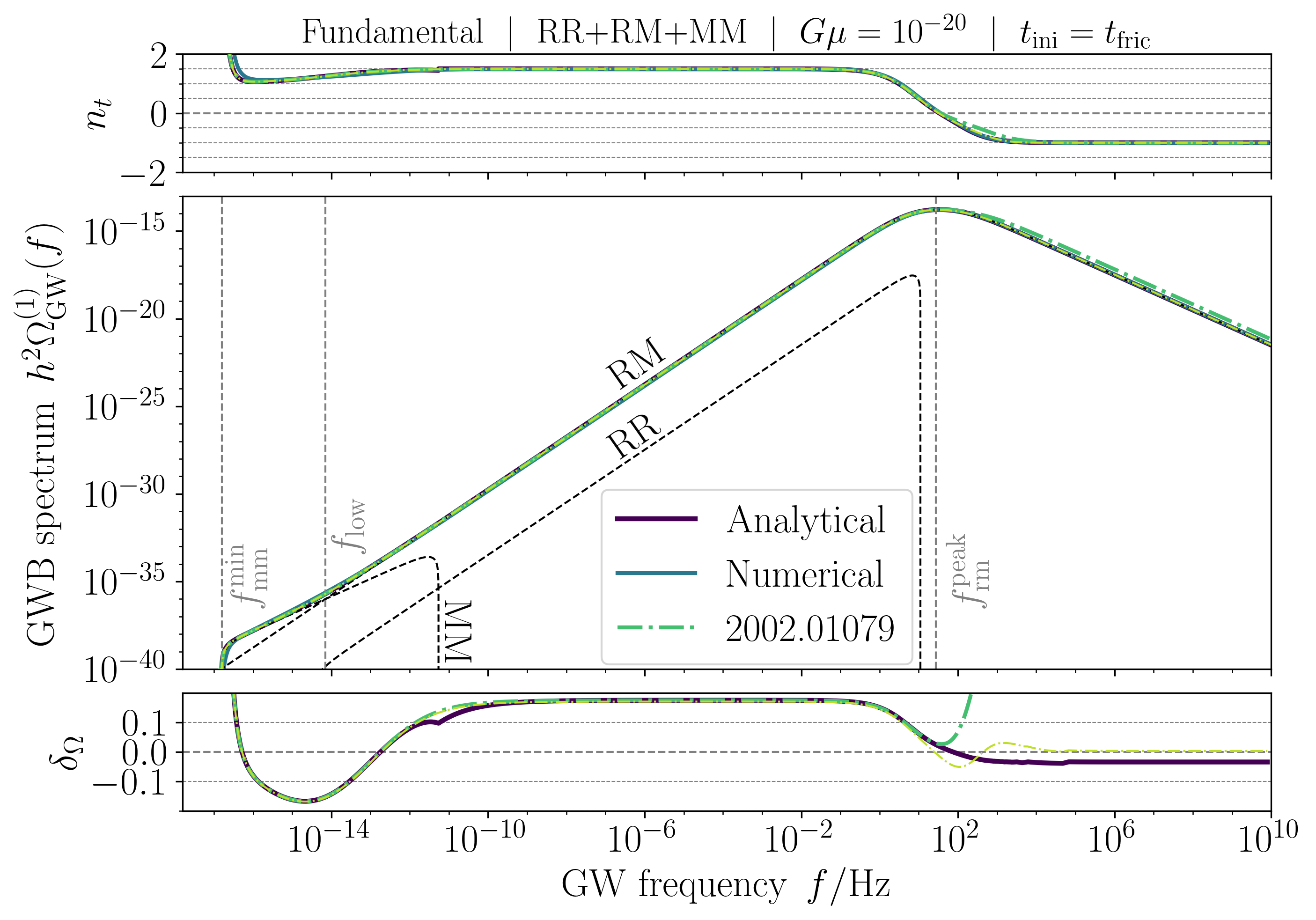}
        \put(17,30){\rotatebox{90}{\eqref{eq:f_mm_min}}}
    \put(25,33.5){\rotatebox{90}{\eqref{eq:LF_Transition_MM_to_RM}}}
    \put(72,31){\rotatebox{90}{\eqref{eq:f_rm_peak_3}}}
\end{overpic}
\caption{  }
\label{fig:Full3}
    \end{subfigure}
    \begin{subfigure}{0.49\textwidth}
   \begin{overpic}[width = \textwidth]{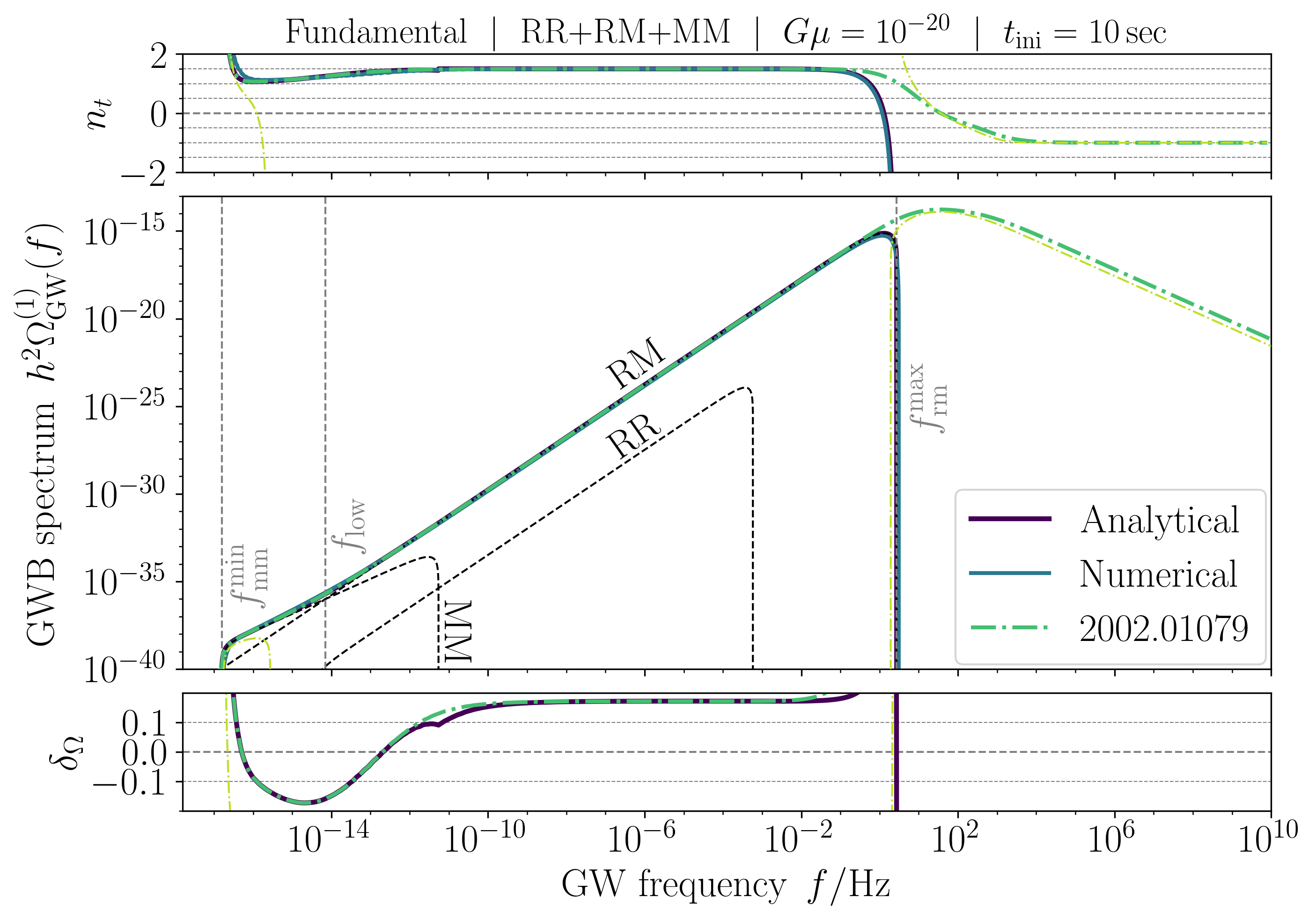}
   \put(17,30){\rotatebox{90}{\eqref{eq:f_mm_min}}}
    \put(25,33.5){\rotatebox{90}{\eqref{eq:LF_Transition_MM_to_RM}}}
\put(69,43.5){\rotatebox{90}{\eqref{eq:f_rm_max}}}
\end{overpic}
\caption{  }
\label{fig:Full4}
    \end{subfigure}
    \caption{\footnotesize
     Complete fundamental GW spectrum for four qualitatively different cases. {Our analytical RM spectrum is plotted with the replacement $x_{\rm rm}^{\rm end}\to x_m^0$ and the modified amplitude in \eqref{eq:modifiedamp} as discussed in Sec.~\ref{subsec:FullFundamentalSpectrum}}.
 Main and lower panels and color codes are the same as in Fig.~\ref{fig:RR1}. For the analytical result derived in Ref.~{\cite{Sousa_2020}}, we show {in panels (c) and (d)} two lines: A thin line for the RR+RM+MM spectrum {(light green)} and a thick line for the RM+MM contribution {(dark green)}. {We do this because their RR spectrum turns negative, leading to large deviations in the full spectrum, while the RM+MM contributions can still provide a good fit for parts of the frequency range.} Additionally, the upper panels show the spectral index $n_t$ (cf.\ \eqref{eq:spectralindex}) of the numerical and the two analytical spectra. {For panel (d), we remark that while the RM and MM spectra of \cite{Sousa_2020} are independent of $t_{\rm ini}$, the RR spectrum was derived there assuming $t_{\rm ini}=t_{\rm Pl}/G\mu^2$. We have modified this result to make it applicable to other values of $t_{\rm ini}$ as well. Following the discussion in \cite{Sousa_2020, Blanco-Pillado:2024aca}, we also cut their RR and RM spectra off at the frequency $f_{\rm rr}^{\rm min}$.}}
     \label{fig:Full}
\end{figure}

\subsection{Complete fundamental spectrum} \label{subsec:FullFundamentalSpectrum}

In principle, summing the fundamental RR, RM, and MM spectra will give rise to the complete fundamental spectrum. Before we turn to this sum, recall that both our analytical and numerical results up to now are based on the VOS model, assuming a constant effective number of DOFs. On the numerical side, however, we account for the full time dependence of all time-dependent quantities, such as $a(t)$, $\xi(t)$, and  $v_\infty(t)$, based on our numerical solution of the Friedmann and VOS equations and the standard energy composition of our Universe in terms of radiation, matter, and dark energy (keeping the effective numbers of DOFs fixed at their present-day values). In our analytical calculation, on the other hand, we introduce an artificial splitting of the cosmic expansion into an early pure-radiation era and a late pure-matter era, which renders our analytical calculation manageable, but which at the same time also introduces a slight error. We emphasize that, if it were not for this error, our analytical and numerical results would agree with each other perfectly because they would represent the same computation. 

Fortunately, there are simple changes that can be applied to find better agreement with the numerical calculation. These corrections enter the RM spectrum only. {First, we make the replacement $x_{\rm rm}^{\rm end} \to x_{m}^{0}$.}\footnote{{Note that the frequency range in which the RM case 1 of table \ref{table:1} applies is, thereby, removed completely.}} {This, however, also removes the minimum frequency in the RM spectrum, which would then extend to arbitrarily low frequencies. There should be an overall low-frequency cutoff to the spectrum, associated with the GWs produced by the largest string loops, i.e., the loops produced today. Correspondingly, we also need to cut the RM spectrum off at this universal minimum frequency, which is nothing but the low-frequency cutoff of the MM spectrum $f_{\rm mm}^{\rm min}$.} In addition, we assume pure radiation domination before matter--radiation equality to evaluate $t_{\rm eq}$ in the prefactor $\mathcal{A}_{\rm rm}$ such that $t_{\rm eq}^{-3/2}\left(a_{\rm eq}/a_0\right)^3 = \left(2H_r^0\right)^{3/2}$ and the amplitude \eqref{eq:A_rm} becomes
\begin{align}
    \mathcal{A}_{\rm rm} = \sqrt{\frac{128}{3}} \pi C_r h^2 \Omega_m \left(\frac{{4} \Omega_r}{\Omega_m}\right)^{3/4} \left(\frac{G\mu}{\Gamma}\right)^{1/2} \, . \label{eq:modifiedamp}
\end{align}

Some benchmark spectra, including the above corrections and
covering different cases, are shown in Figs.~\ref{fig:Full1} to \ref{fig:Full4}. In an additional panel, we show for each of these plots also the spectral index defined as
\begin{align}
    n_t(f) = \frac{\pd \ln\left(h^2\Omega(f)\right)}{\pd \ln\left(f/{\rm Hz}\right)} \, . \label{eq:spectralindex}
\end{align}
All spectra have a minimum frequency in common, which is given by $f_{\rm mm}^{\rm min}$ in \eqref{eq:f_mm_min}. The spectrum roughly rises then first with an $f^{1}$ power law given by expression \eqref{eq:Omega_mm_low}, which stems from the MM contribution. Afterwards, it smoothly transitions into an $f^{3/2}$ power law that is described in \eqref{eq:RM_lowF_powerlaw} and derives from the RM contribution. This behavior is visible in all plots. The two power laws intersect at a frequency 
\begin{align} \label{eq:LF_Transition_MM_to_RM}
f_{\rm low} = \left(\frac{4\mathcal{A}_{\rm mm}}{9\mathcal{A}_{\rm rm}}\right)^2 f_m^0
\end{align}
such that the two different low-frequency power laws are assumed well below and well above this frequency. Note that $f_{\rm low}$ is, just as $f_{\rm mm}^{\rm min}$, independent of $\Gamma$ and $G\mu$. The range in which the $f^{1}$ behavior is relevant, therefore, stays the same. $f_{\rm rm}^{\rm peak}$ increases, however, with decreasing $\Gamma$ and $G\mu$. The region in which the $f^{3/2}$ power law is important, hence, becomes larger for smaller string tensions.
We can observe this in Figs.~\ref{fig:Full1} to \ref{fig:Full4}.
If we consider late initial times $t_{\rm ini}$ and low string tensions, an overall cut-off $f_{\rm rm}^{\rm max}$ given in \eqref{eq:f_rm_max} in the spectrum may arise. In this case, the low-frequency power law suddenly drops to zero towards larger frequencies. This can be observed in Fig.~\ref{fig:Full4}. Otherwise, the spectrum will extend to larger frequencies. It peaks then at a frequency $f_{\rm rm}^{\rm peak}$, either given by \eqref{eq:f_rm_peak_2} or \eqref{eq:f_rm_peak_3}. This is shown in Figs.~\ref{fig:Full1} to \ref{fig:Full3}. Without the $f_{\rm rm}^{\rm max}$ cutoff, but
still at low string tensions, we might find a maximum frequency $f_{\rm rr}^{\rm max}$ (cf.\ \eqref{eq:f_rr_max}) in the RR contribution. In this case, the RR spectrum will be cut off, and the overall spectrum will be dominated by the RM spectrum. This means that, after reaching the peak of the spectrum, we will find an $f^{-1}$ decay, which is given by \eqref{eq:Orm1_HF3}.\footnote{The high-frequency behavior for case $2$ is not relevant here since, if $f_{\rm rr}^{\rm max}$ exists, $f_{\rm rm}^{{\rm start},*}$ will as well. Case $2$ will then not apply to asymptotically large frequencies.} We can see such a spectrum in Fig.~\ref{fig:Full3}. At larger string tensions, the RR contribution will not be cut off, and instead, we find a plateau with height $A_{\rm rr}$ given in \eqref{eq:Orr1_plateau}. At a frequency $f_{\rm rr}^{\rm high}$ given in \eqref{eq:f_rr_high}, the plateau ends and turns into an $f^{-1}$-power law, whose precise form is described by \eqref{eq:Omega_mm_high}. This high-frequency decay as well as the plateau are visible in Figs.~\ref{fig:Full1} and \ref{fig:Full2}.
\begin{figure}
\tiny
    \centering
    \begin{subfigure}{0.49\textwidth}
     \begin{overpic}[width = \textwidth]{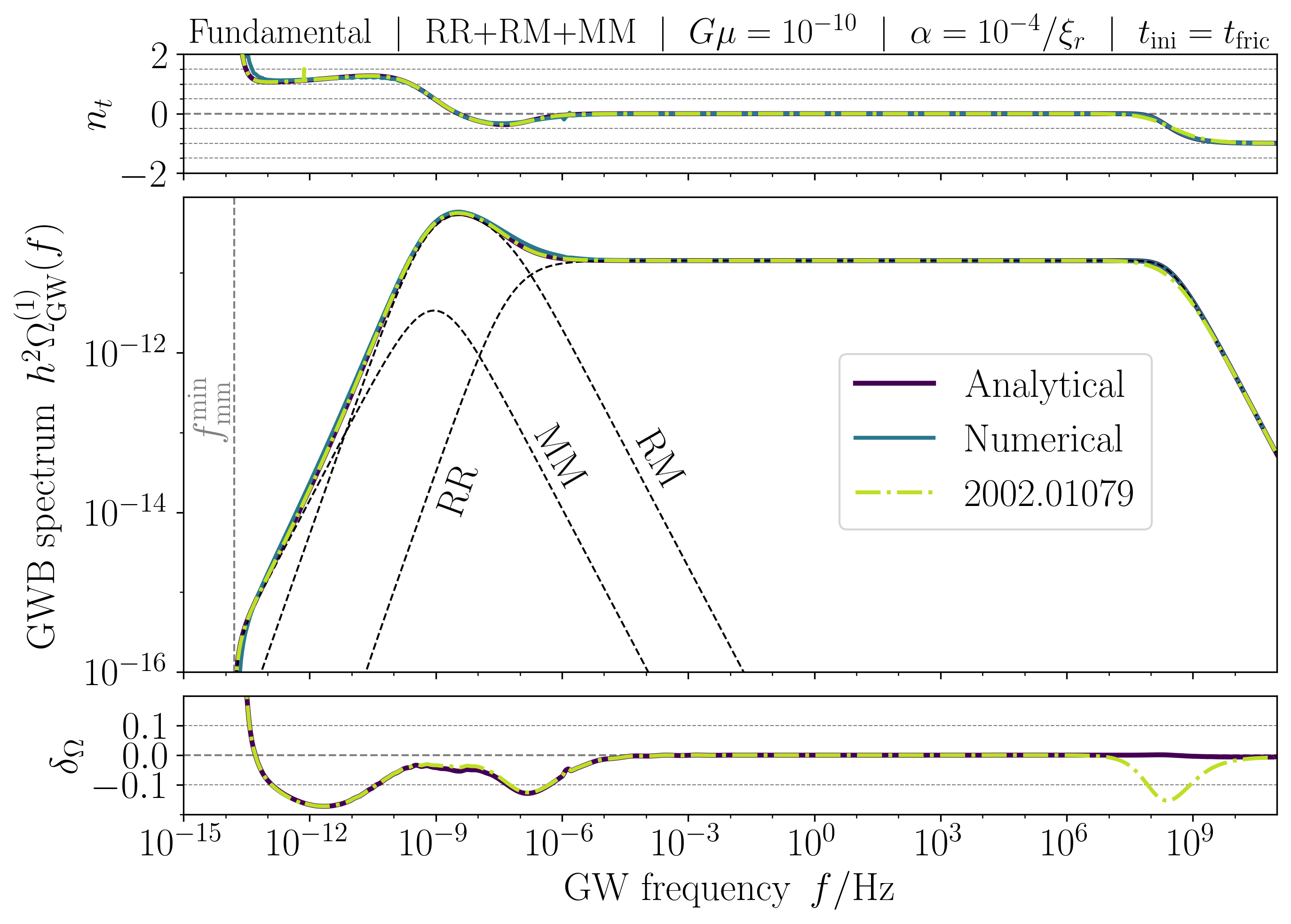}
    \put(15,44){\rotatebox{90}{\eqref{eq:f_mm_min}}}
\end{overpic}
\caption{ }
\label{fig:alpha-4}
    \end{subfigure}
    \begin{subfigure}{0.49\textwidth}
   \begin{overpic}[width = \textwidth]{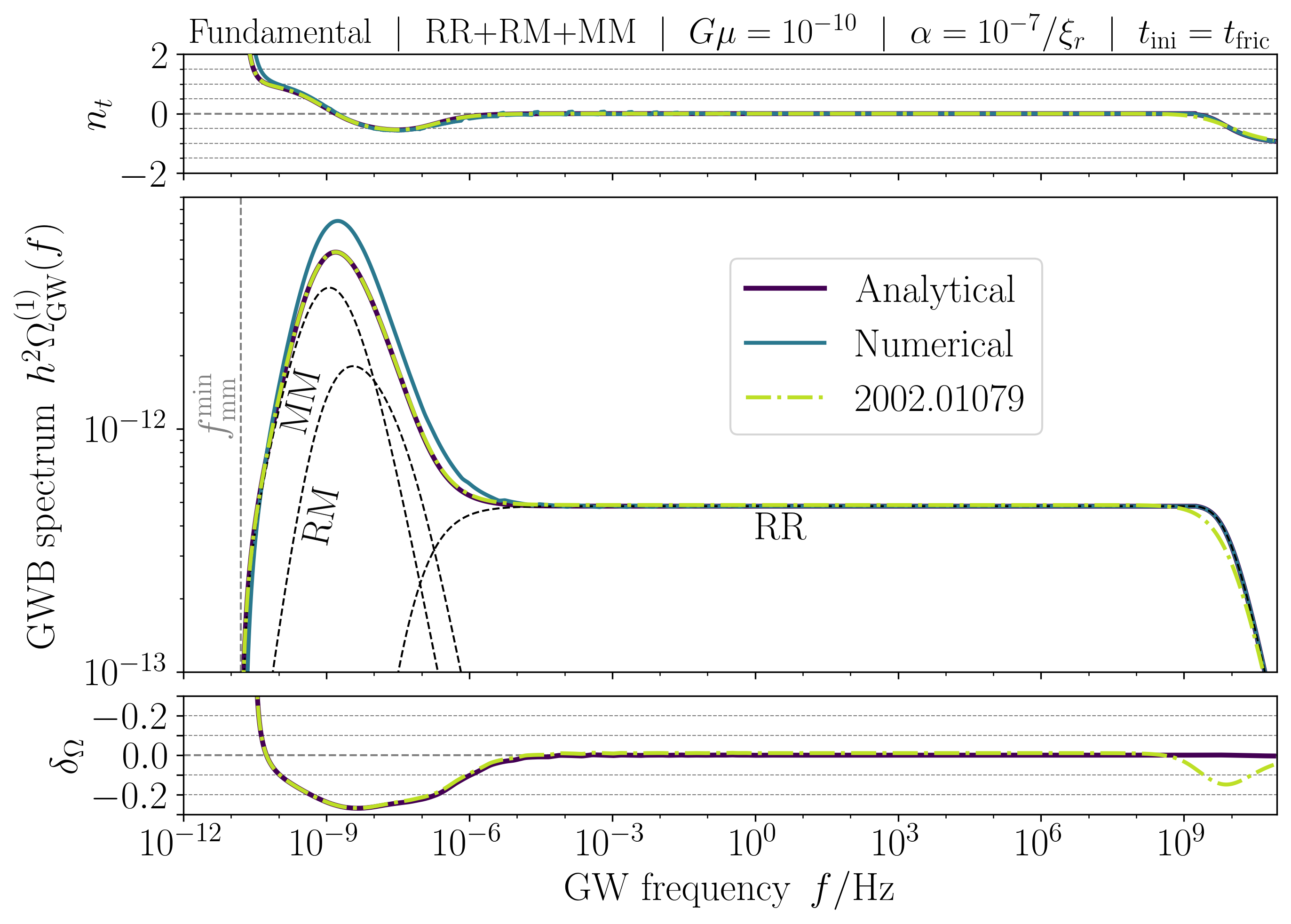}
       \put(15.5,44){\rotatebox{90}{\eqref{eq:f_mm_min}}}
\end{overpic}
\caption{  }
\label{fig:alpha-7}
    \end{subfigure}
    \caption{\footnotesize
    {Same as in Fig.~\ref{fig:Full}, but for small values of $\alpha$.}}
\end{figure}

{In addition to the spectra evaluated for the benchmark value $\alpha\xi_r = 0.1$, we also show how the spectra behave for much smaller values $\alpha \xi_r = 10^{-4}$ and $10^{-7}$ in Figs.~\ref{fig:alpha-4} and \ref{fig:alpha-7}. In contrast to Fig.~\ref{fig:Full2}, we can see that the MM contribution does not exhibit a maximum frequency. This is easily understood since the much smaller initial loop length requires, at the same decay efficiency determined by $\Gamma G\mu$, a much shorter duration to completely decay. Apart from the overall lower amplitude of the spectra, the MM contribution becomes, in comparison to the RM contribution, increasingly relevant for small values of $\alpha$. While in Fig.~\ref{fig:alpha-4} the RM contribution still dominates the spectrum at its peak, for the parameters chosen in Fig.~\ref{fig:alpha-7}, the MM contribution is dominant at the peak frequency. As discussed in Sec.~\ref{subsec:MM_Loops_Fundamental}, our analytical expressions provide in the MM case a worse fit to our numerical spectra than in the RR and RM cases. As the peak region in Fig.~\ref{fig:alpha-7} is dominated by the MM contribution, this leads to a larger error in overall spectrum, which stays, however, below $30\%$. The same increased error occurs in the analytical expressions of Ref.~\cite{Sousa_2020}.}

\section{Total spectrum}
\label{sec:TotalSpectrum}

So far, we were only concerned with the spectrum from the first harmonic, $\Omega^{(1)}_{\rm GW}$. The observable result is, however, not the contribution from a single mode, but the sum of all harmonics. As pointed out in Sec.~\ref{subsec:Gravitational_Wave_Spectrum}, higher-mode contributions have the convenient property that they may be expressed in terms of the fundamental spectrum. In the following, we want to make use of this relation and derive an explicit analytical approximation to the total spectrum based on our analytical treatment of the fundamental spectrum. 

Our starting point is the sum in \eqref{eq:Sum_Full_Spectrum}, which we reproduce here for convenience,\footnote{For reasons explained in Section \ref{subsec:Gravitational_Wave_Spectrum}, we ignore in this treatment, without introducing any inaccuracies, that the earliest GW emission time will be independent of the mode number, but can depend on frequency. We treat the earliest emission time $t_{\rm ini}$ here merely as another free model parameter.}
\begin{align}\label{eq:summed_spectrum}
\Omega_{\rm GW}\left(f\right) =\frac{1}{H_{n_{\rm max}}^{(q)}}\sum_{k=1}^{n_{\rm max}} \frac{1}{k^q}\, \Omega_{\rm GW}^{(1)} \left(\frac{f}{k}\right) \,. 
\end{align}
It appears that, even in the limit $n_{\rm max}\to \infty$, this sum cannot be further simplified. Instead, we proceed by approximating the sum as follows: Denoting the summands by $\Phi(k)$, we can generalize the function $\Phi$ from mode number $k\in \mathbb{N}$ to $k\in \mathbb{R}_{\geq 1}$. For arbitrary $m, {n_{\rm max}}\in \mathbb{N}$ and monotonically decreasing $\Phi$, one can easily show that 
\begin{align}
  \sum_{k=1}^m \Phi(k) +\int_{m+1}^{{n_{\rm max}}+1}\Phi(k) \,\pd k  \leq \sum_{k=1}^{n_{\rm max}} \Phi(k) \leq  \sum_{k=1}^m \Phi(k) + \int_{m}^{n_{\rm max}} \Phi(k)\,\pd k \,.\label{Upperandlowerbound} 
\end{align}
Choosing a larger value of $m$ will generally improve the approximation. If $\Phi$ is monotonically increasing, the upper and lower estimates of the sum are interchanged. Hereafter, we will only write down the explicit expressions for the upper bound. The lower bound is then obtained by the trivial replacement $(n_{\rm max},\, m)\mapsto (n_{\rm max}+1,\, m+1)$ in the integrated part. 

For the following considerations, it is important that $q>1$. Accordingly, if $\Omega^{(1)}_{\rm GW}(f)$ decreases in $f$ nowhere faster than $f^{-1}$, the function $\Phi(k) \propto  \Omega^{(1)}_{\rm GW}\left(f/k\right)/k^q$ will be monotonically decreasing and \eqref{Upperandlowerbound} can be applied without any changes. As a matter of fact, the $f^{-1}$ decay is, in the absence of a maximum frequency, the fastest decrease we find for the fundamental RR, RM, and MM spectra, occurring at asymptotically large frequencies. Even though this is no longer true in the presence of a maximum frequency, the above formula still provides a good approximation of the actual spectrum.

{
It should be noted that $\Phi(k)$ represents the analytical functions for the spectra derived in previous sections. Since these expressions may deviate from the numerical results, the band defined by the upper and lower bounds does not necessarily contain the numerically computed total spectrum. Moreover, if one is primarily interested in a single representative value of the spectrum rather than the full band, we recommend using the geometric mean of the upper and lower bounds as an estimate.}

\subsection{General features}
To avoid a repetition of reasoning, let us start with a discussion of the ubiquitous features of the GW spectra. For this, consider again the sum over all modes \eqref{eq:summed_spectrum}. We can visualize this sum as adding up $n_{\rm max}$ copies of the fundamental spectrum, which are, with increasing mode number $k$, shifted towards higher frequencies and smaller amplitudes. It is immediately clear that the minimum frequency of the total spectrum is set by the lowest mode and consequently the same as for the fundamental mode $f_{\rm min}^{\rm tot} = f_{\rm min}$ {\cite{Blanco-Pillado:2024aca}}. The maximum frequency is, on the other hand, determined by the largest contributing mode number $f_{\rm max}^{\rm tot} = n_{\rm max} f_{\rm max}$. It is important to note at this point that while the total minimum frequency is a sensible prediction, the maximum frequency of the total spectrum is artificially introduced by arbitrarily choosing a finite $n_{\rm max}$ and, accordingly, unphysical. Similarly, the spectrum should, in general, only be considered reliable at frequencies that satisfy $f\ll f_{\rm max}^{\rm tot}$.

Apart from minimum and maximum frequencies, all three fundamental spectra approach power-law behaviors in the asymptotic regions, $h^2\Omega^{(1)}_{\rm GW}(f) \longrightarrow C f^{p}$. In these cases, the sum over harmonics can be evaluated explicitly, namely as
\begin{align} \label{eq:General_Features}
   h^2 \Omega_{\rm GW}(f) = \frac{C}{H_{n_{\rm max}}^{(q)}} \left[\zeta (q+p,1)-\zeta (q+p,n_{\rm max}+1)\right] f^{p} \, ,
\end{align}
where $\zeta$ is the Hurwitz zeta function. Note that this is not true in the case of the UHF regime, though, since the different harmonics do not depend only on $f/k$.

\subsection{RR loops}
\label{subsec:Full_RR_Loops}
We begin the discussion of the total spectrum once again with the gravitational radiation from loops decaying during radiation domination. For the fundamental mode of the spectrum, we found the expression \eqref{eq:Orr1}. To analytically carry out the integration over the mode number, we need to approximate the term containing $\varphi_2\left(x_r^{\rm ini},\chi_r\right)$. For $x_{\rm r}^{\rm ini}  \gg 1$, we can again use the expansion $\varphi_2(x_r^{\rm ini}, \chi_r) \to \sqrt{1+\chi_r}\,x_r^{\rm ini}$, and it turns out that this expression leads to a very good approximation also at intermediate frequencies. This may not come as a surprise, recalling from Sec.~\ref{subsec:RR_Loops_Fundamental} that at those frequencies, the term depending on $x_r^{\rm eq}$ is strongly dominant. 
In the parameter regime in which $\varphi_2$ is relevant and simultaneously $x_r^{\rm ini}\ll 1$, we need the expansion
$ \varphi_2(x_r^{\rm ini}, \chi_r) \to (1 + \chi_r) \left(x_r^{\rm ini}\right)^2 - (1 + \chi_r)^2 \left(x_r^{\rm ini}\right)^4$. This is particularly important when $f_{\rm rr}^{\rm max}$ exists. Since the shape function deriving from the low-frequency expansion is also only relevant in this regime, we present here only the expanded spectrum, which is additionally easier to evaluate numerically. Combining all of these considerations, we find as an upper limit for the total RR spectrum the formula
\begin{align}
    h^2 \Omega_{\rm rr}^{m, {\rm upper}}(f) =  &\frac{1}{H_{n_{\rm max}}^{(q)}}\sum_{k=1}^m \frac{1}{k^q} h^2\Omega_{\rm rr}^{(1)}\left(\frac{f}{k}\right) +\\ &+ \frac{\mathcal{A}_{\rm rr}}{H_{n_{\rm max}}^{(q)}} \left\{S_{\rm rr,1}(k, x_r^{\rm eq})\big\vert_{k_{\rm rr}^{\rm min}}^{k_{\rm rr}^{\rm max}} - S_{\rm rr,2}(k, \sqrt{1+\chi_r}x_r^{\rm ini})\big\vert_{k_{\rm rr}^{\rm min}}^{k_{\rm rr}^{\rm start}} - S_{\rm rr,3}\left(k\right)\big\vert_{k_{\rm rr}^{\rm start}}^{k_{\rm rr}^{\rm max}} \right\} \,, \nonumber
\end{align}
where we introduced the shape functions
\begin{align}
    S_{\rm rr,1}(k, x) &= -\frac{{_2F_1}\left[\frac{3}{2}, 1-q, 2-q, -\frac{k}{x}\right]}{(q-1)k^{q-1} } \, , \\
    S_{\rm rr, 2}(k, x) &= \begin{cases} 
   \frac{x^3}{k^{q+2}} \left(-\frac{1}{q+2}+\frac{3}{ q+4}\frac{x^2}{k^2}-\frac{6}{q+6}\frac{x^4}{k^4}\right) & \text{if } x\ll 1 \\
    S_{\rm rr, 1}(k, x) & \text{otherwise} 
    \end{cases} \, ,\\
    S_{\rm rr,3}(k) &= -\frac{S_{\rm rr}^{(1)}(\chi_r^{-1})}{(q-1)k^{q-1}} \, .
\end{align}
The frequency-dependent mode numbers that appear above are given by
\begin{subequations}
\begin{align}
   k_{\rm rr}^{\rm min}&= \max\left\{m, f/f_{\rm rr}^{\rm max}\right\}, \\ k_{\rm rr}^{\rm max} &= \max\left\{\min\left\{n_{\rm max}, f/f_{\rm rr}^{\rm min}\right\},k_{\rm rr}^{\rm min}\right\},\\  k_{\rm rr}^{\rm start} &= \min\left\{\max\left\{k_{\rm rr}^{\rm min}, f/f_{\rm rr}^{{\rm start}, *}\right\}, k_{\rm rr}^{\rm max}\right\}.
\end{align}
\end{subequations}
The mode numbers $k_{\rm rr}^{\rm min/max}$ derive, upon the replacement $f\to f/k$, directly from \eqref{eq:Theta-RR}. In a similar fashion, $k_{\rm rr}^{{\rm start}}$ corresponds to $f_{\rm rr}^{{\rm start},*}$ derived in \eqref{eq:f_rr_star_start}. {In the limit $n_{\rm max}\to \infty$, the expression for the maximum mode number reduces to
\begin{align}
    k_{\rm rr}^{\rm max} \stackrel{n_{\rm max}\to \infty}{\longrightarrow} \max\left\{f/f_{\rm rr}^{\rm min} , k_{\rm rr}^{\rm min}\right\} \, .
\end{align}
}
{This effective limit was taken in \cite{Blanco-Pillado:2024aca} in an analogous way.}

At low frequencies ($x\lesssim 10^{-3}$), the evaluation of the hypergeometric function can become numerically difficult. An expansion around $x=0$ reveals a leading-order term which is independent of the mode number and, therefore, cancels out. Removing it, one may use to a very good approximation
\begin{align}
S_{\rm rr,1}(k,x) \to \frac{2  \Gamma (2-q)}{(q-1) (2 q+1) \Gamma (1-q)} \frac{x^{3/2}}{k^{1/2+q}} \, .
\end{align}
If the spectrum exhibits no maximum, a good value to start using the low-frequency expression in $S_{\rm rr, 2}$ is below $x_r^{\rm ini}\sim 10^{-8}$. If $f_{\rm rr}^{\rm max}>0$, it is advisable to use this expression only.

Having found a complete expression for the total RR spectrum, we continue with a discussion of its features. While the descriptions of the low-frequency, plateau, and high-frequency regions can be extracted via \eqref{eq:General_Features} directly from the fundamental spectrum, for large $n_{\rm max}$, an intermediate region between the plateau and high-frequency region develops. We will now derive the corresponding power law. For the region below $f_{\rm rr}^{\rm high}$ in \eqref{eq:f_rr_high}, the fundamental spectrum is well described by the plateau $h^2 \Omega_{\rm rr, plateau}^{(1)}=\mathcal{A}_{\rm rr}$ in \eqref{eq:Orr1_plateau}. For $f\gg f_{\rm rr}^{\rm high}$, the spectrum approaches a high-frequency power law of the form  $h^2 \Omega_{\rm rr, high}^{(1)}(f) = \mathcal{A}_{\rm rr} f_{\rm rr}^{\rm high}/f$, which we found in \eqref{eq:Orr1_high}.
Correspondingly, we can write for the total spectrum (for $1<q<2$)
\begin{align}
    h^2 \Omega_{\rm rr}(f) =& \frac{1}{H_{n_{\rm max}}^{(q)}} \sum_{k=1}^{n_{\rm max}} \frac{1}{k^q}h^2 \Omega_{\rm rr}^{(1)}\left(\frac{f}{k}\right) \simeq \\ \simeq &  \frac{1}{H_{n_{\rm max}}^{(q)}}\left(\sum_{k=1}^{\lfloor k_{\rm high}(f)\rfloor} \frac{ \mathcal{A}_{\rm rr}}{k^{q-1}}\frac{f_{\rm rr}^{\rm high}}{f} +\sum_{k = \lceil k_{\rm high}(f)\rceil }^{n_{\rm max}} \frac{\mathcal{A}_{\rm rr}}{k^q}\right)=\nonumber \\ =& \frac{\mathcal{A}_{\rm rr}}{H_{n_{\rm max}}^{(q)}}\left(\frac{f_{\rm rr}^{\rm high}}{f}\zeta\left(q-1, k\right)\Big\vert_{k=\lfloor k_{\rm high}(f) \rfloor +1}^{k=1} + \zeta\left(q, k\right)\Big\vert_{k={n_{\rm max}} +1}^{k=\lceil k_{\rm high}(f)\rceil} \right) 
\end{align}
with $k_{\rm high}(f)=f/f_{\rm rr}^{\rm high}$. The spectrum can only develop an intermediate power law if there is a region in which $f_{\rm rr}^{\rm high} \ll f \ll n_{\rm max} f_{\rm rr}^{\rm high}$. Assuming that this is realized and picking a frequency within this region, we have $k_{\rm high}(f)\gg 1$. Expanding around this value and approximating $\lfloor k_{\rm high}(f)\rfloor \simeq \lceil k_{\rm high}(f)\rceil \simeq k_{\rm high}(f) $ and $k_{\rm high}(f) +1 \simeq k_{\rm high}(f)$, we find
\begin{align}
    h^2\Omega_{\rm rr}(f) &= \frac{\mathcal{A}_{\rm rr}}{H_{n_{\rm max}}^{(q)}}\left(\frac{f_{\rm rr}^{\rm high}}{f}\zeta\left(q-1, 1\right) + \left(\frac{1}{2-q}+\frac{1}{q-1}\right)\left(\frac{f_{\rm rr}^{\rm high}}{f}\right)^{q-1} - \zeta\left(q, n_{\rm max} +1 \right)\right) \sim \nonumber\\&\sim \frac{\mathcal{A}_{\rm rr}}{H_{n_{\rm max}}^{(q)}}\frac{1}{(2-q)(q-1)} \left(\frac{f_{\rm rr}^{\rm high}}{f}\right)^{q-1}.
\end{align}
Here, we used that the first term is smaller than the second term by a factor of $(f_{\rm rr}^{\rm high}/f)^{2-q}$. The last term is suppressed since $\zeta\left(q, n_{\rm max} +1 \right)\simeq n^{1-q}_{\rm max}/(q-1)$ and, by assumption, $k_{\rm high}(f) \ll n_{\rm max}$. Note, however, that this approximation breaks down as $q\to2$. 
In summary, the following power laws can be found in the spectrum in the absence of a maximum frequency:
\begin{align} \label{eq:RR_Total_Power_Laws}
    h^2 \Omega_{\rm rr}(f)\simeq \frac{\mathcal{A}_{\rm rr}}{H_{n_{\rm max}}^{(q)}} \begin{cases} \zeta\left(q+\frac{3}{2}, k\right)\big\vert_{k=n_{\rm max}+1}^{k=1} \left(\frac{f}{f_r^{\rm eq}}\right)^{3/2} & \text{if } f_{\rm rr}^{\rm min} \ll f \ll f_{\rm rr}^{\rm low} \\ H_{n_{\rm max}}^{(q)}
    & \text{if } f_{\rm rr}^{\rm low} \ll f \ll f_{\rm rr}^{\rm high}
    \\ \frac{1}{(2-q)(q-1)} \left(\frac{f_{\rm rr}^{\rm high}}{f}\right)^{q-1}
    & \text{if } f_{\rm rr}^{\rm high} \ll f \ll n_{\rm max} f_{\rm rr}^{\rm high}\\
    \zeta\left(q-1, k\right)\big\vert_{k=n_{\rm max}+1}^{k=1} \left(\frac{f_{\rm rr}^{\rm high}}{f}\right) & \text{if } n_{\rm max}f_{\rm rr}^{\rm high} \ll f
    \end{cases} \, .
\end{align}

\begin{figure}
\tiny
    \centering
    \begin{subfigure}{0.49\textwidth}
     \begin{overpic}[width = \textwidth]{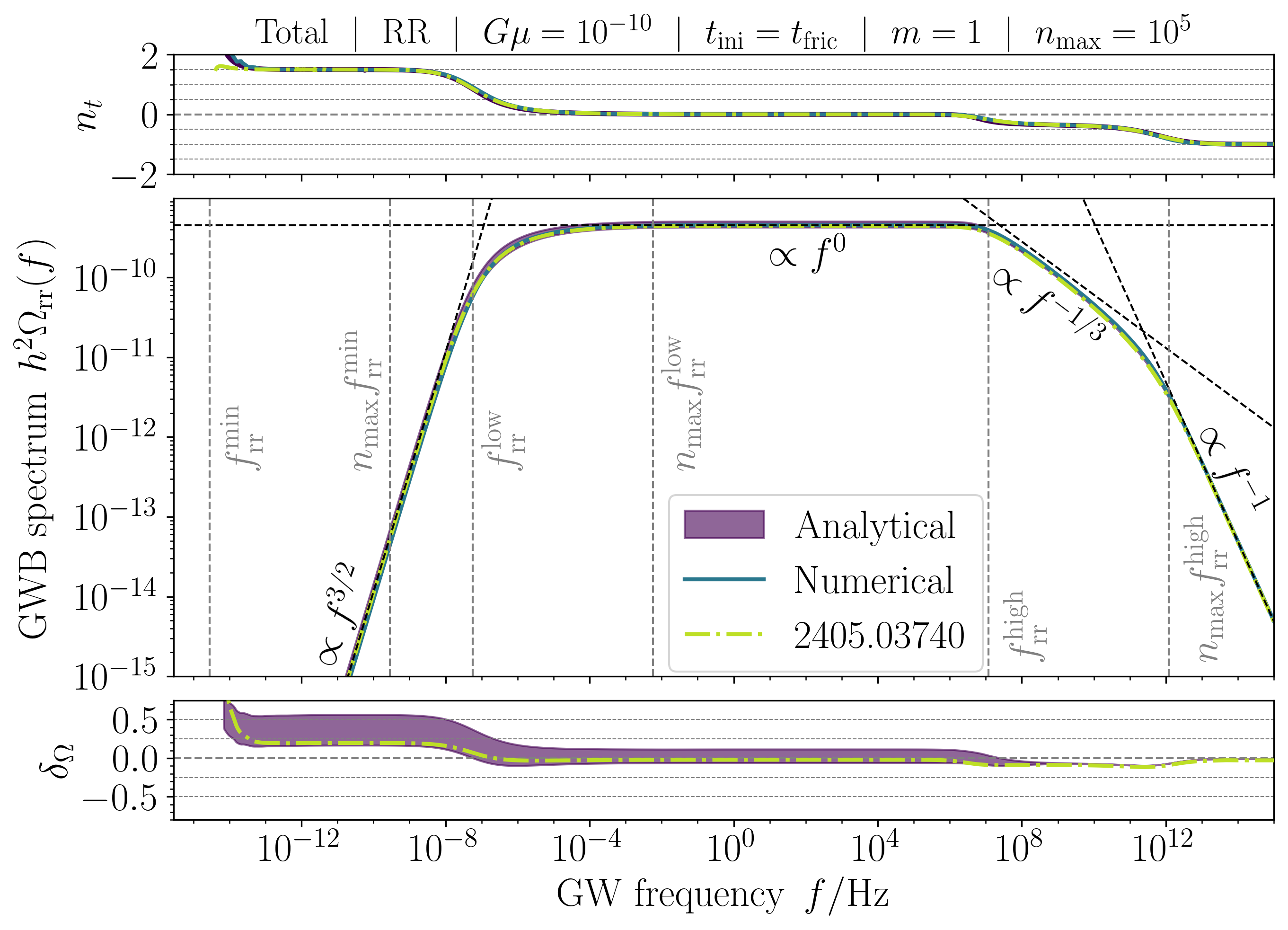}
     \put(17.5,41.5){\rotatebox{90}{\eqref{eq_f_rr_min}}}
    \put(38,41.5){\rotatebox{90}{\eqref{eq:f_rr_low}}}
    \put(78.5,27){\rotatebox{90}{\eqref{eq:f_rr_high}}}
     \put(57.5,40){\rotatebox{0}{power laws:}}
     \put(57.5,36.5){\rotatebox{0}{\eqref{eq:RR_Total_Power_Laws}}}
\end{overpic}
\caption{ }
\label{fig:RR_Total_1}
    \end{subfigure}
    \begin{subfigure}{0.49\textwidth}
  \begin{overpic}[width = \textwidth]{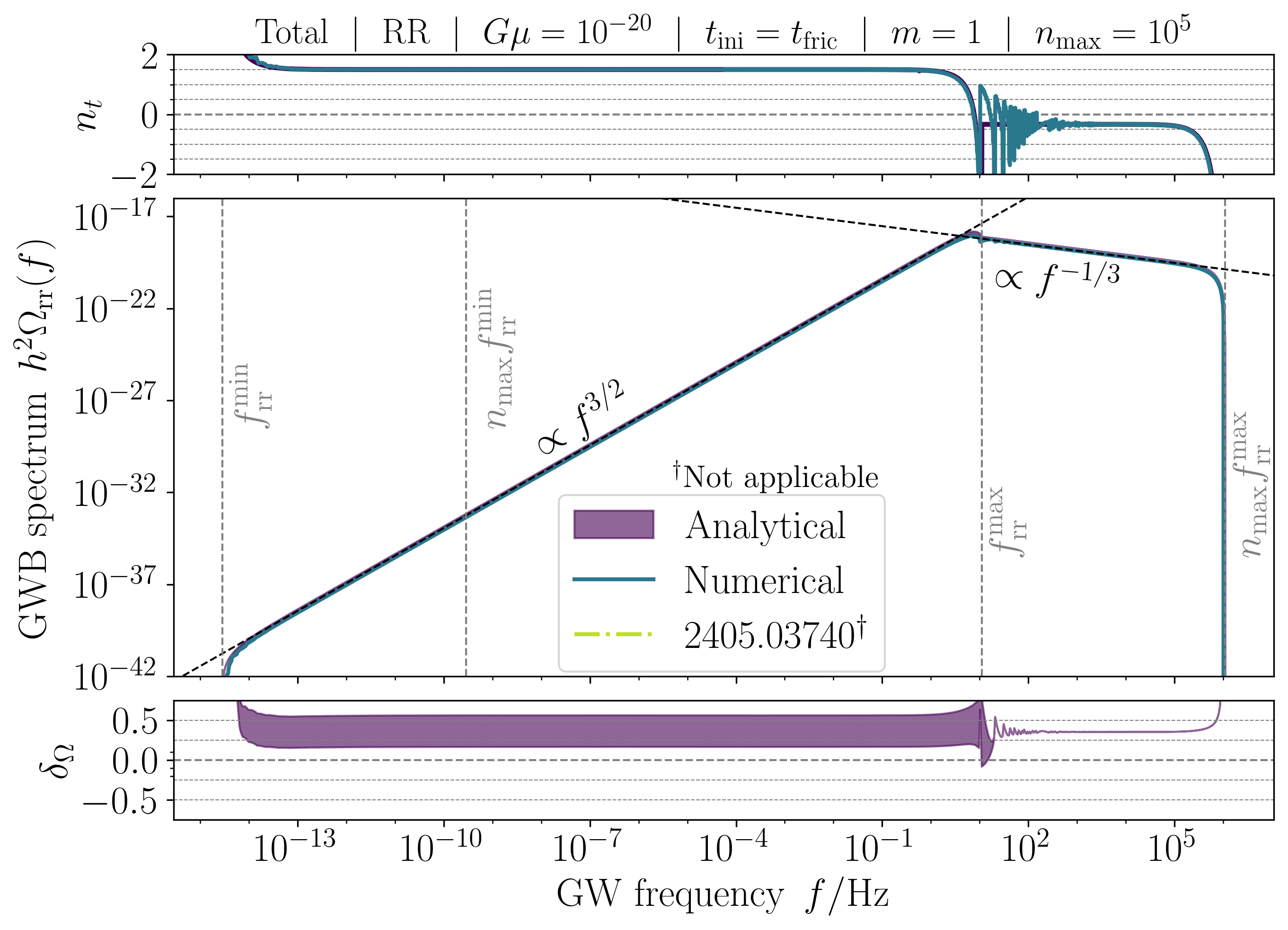}
  \put(18.25,45){\rotatebox{90}{\eqref{eq_f_rr_min}}}
    \put(77,35){\rotatebox{90}{\eqref{eq:f_rr_max}}}
        \put(80,54.5){\rotatebox{-7}{\eqref{eq:RR_HF_Power_LAW_nmax}}}
        \put(49,41.5){\rotatebox{29.5}{\eqref{eq:Orr1_low_simplified}, \eqref{eq:General_Features}}}
\end{overpic}
\caption{  }
\label{fig:RR_Total_2}
    \end{subfigure}
    \caption{\footnotesize
    Total GW spectrum for RR loops and the two qualitatively different cases. 
     Main panels: Result from numerical calculations with a fixed effective number of DOFs (teal, cf.~Section \ref{subsec:Numerical}), analytical result derived in this paper with $m=1$ and $n_{\rm max}=10^5$ ({deep purple band, showing values between lower and upper bounds given in Eq.~\eqref{Upperandlowerbound}}), and analytical result derived in Ref.~\cite{Blanco-Pillado:2024aca} (light green, dash-dotted; {in \cite{Blanco-Pillado:2024aca}, $n_{\rm max}$ is referred to as $N$}). {Note that the parameter $m$ appears only in our analytical template.}    
     Grey dashed lines show characteristic frequencies of the spectrum, and black dashed lines show different power-law behaviors of the spectrum.
     Upper panels: Spectral index $n_t$ (cf. \eqref{eq:spectralindex}) of the three spectra.
     Lower panels: Relative deviation $\delta_\Omega$ (cf. \eqref{eq:DeltaOmega}) of the two analytical results from the numerical result.}\label{fig:RR_Total}
\end{figure}

A corresponding benchmark spectrum is shown in Fig.~\ref{fig:RR_Total_1}, which displays our numerical and analytical results alongside the analytical spectrum derived in Ref.~\cite{Blanco-Pillado:2024aca} on the basis of Ref.~\cite{Sousa_2020}. As evident from this plot, all three results are in good agreement. Observe that, as in the case of the fundamental spectrum (cf.\ Fig.~\ref{fig:RR1}), our analytical result slightly overestimates the actual spectrum in the low-frequency tail. Due to the approximation we applied to $\varphi_2$, our spectrum has the same deviation from the numerical results as Ref.~\cite{Blanco-Pillado:2024aca} does, when transitioning from the plateau to the high-frequency region. Note that we cut off the spectrum of Ref.~\cite{Blanco-Pillado:2024aca} at very low frequencies close to $f_{\rm rr}^{\rm min}$ as it becomes numerically ill-behaved there.

In the presence of a maximum frequency, the behavior is different. The spectrum consists only of the low-frequency power law \eqref{eq:Orr1_low_simplified} and has a sharp cutoff at the maximum frequency. We can, therefore, approximate the fundamental spectrum by $h^2\Omega_{\rm rr}^{(1)} \simeq \mathcal{A}_{\rm rr} \left(f/f_r^{\rm eq}\right)^{3/2} \Theta\left(f_{\rm rr}^{\rm max} - f\right)$. 
Hence, we obtain for the total spectrum
\begin{align}
    h^2\Omega_{\rm rr}(f) &\simeq \frac{1}{H_{n_{\rm max}}^{(q)}} \sum_{k = \lceil f/f_{\rm rr}^{\rm max}\rceil}^{n_{\rm max}} \frac{\mathcal{A}_{\rm rr}}{k^{q-3/2}}\left(\frac{f}{f_r^{\rm eq}}\right)^{3/2} = \\ \nonumber &= \frac{\mathcal{A}_{\rm rr}}{H_{n_{\rm max}}^{(q)}} \left(\frac{f}{f_r^{\rm eq}}\right)^{3/2} \zeta\left(q+3/2, k\right)\Big\vert_{k=\lceil f/f_{\rm rr}^{\rm max}\rceil}^{k= n_{\rm max} + 1} \, .
\end{align}
For $f_{\rm rr}^{\rm max}\ll f \ll n_{\rm max} f_{\rm rr}^{\rm max}$, we can furthermore expand the $\zeta$ function, which results in  
\begin{align} \label{eq:RR_HF_Power_LAW_nmax}
    h^2\Omega_{\rm rr}(f) \simeq \frac{\mathcal{A}_{\rm rr}}{\left(q+\frac{1}{2}\right)H_{n_{\rm max}}^{(q)}} \left(\frac{f}{f_{r}^{\rm eq}}\right)^{3/2} \left(\frac{f}{f_{\rm rr}^{\rm max}}\right)^{-q-1/2} \propto f^{1-q} \, .
\end{align}
Note that this overestimates the amplitude of the power law by a factor of $\sim 3$ since, at $f_{\rm rr}^{\rm max}$, the spectrum already deviates from the low-frequency power law.  

A benchmark scenario exhibiting the described features is depicted in Fig.~\ref{fig:RR_Total_2},  in which we compare our analytical and numerical results. Our formulae can reproduce the shape of the numerical spectrum well, but slightly overestimate the amplitude, as was already the case for the fundamental spectrum. Additionally, we also see that the low-frequency and high-frequency power laws describe the spectrum adequately. For the latter power law, we included the mentioned correction factor of $3$. {Observe that the numerical spectrum exhibits an oscillatory feature, which is most clearly visible in the spectral index $n_t$. This is of completely analogous origin as the oscillatory feature of low-scale strings discussed in \cite{Schmitz_2024}. While our analytical expression correctly reproduces the overall power-law behaviour, it cannot resolve this feature, as is evident from the oscillations in $\delta_\Omega$.} The result of Ref.~\cite{Blanco-Pillado:2024aca} cannot be applied in this range of parameter space as it is based on the expression for the fundamental spectrum in  
Ref.~\cite{Sousa_2020}, which becomes negative in this case.

\medskip\noindent
\textbf{UHF regime:} For the UHF regime, we can again directly draw conclusions from the results we obtained for the fundamental mode. We simply need to adapt the initial time $t_{\rm ini}$ as described around \eqref{eq:a_min2}. 
We have to be careful, though, when calculating the prefactor for the high-frequency power law. From the $f^{-2}$ dependence of \eqref{eq:RR_UHF}, one might expect that the different harmonics decay like $k^{-q-2}$. In reality, they decrease like $k^{-q-1}$ since one factor $f^{-1}$ derives from the lower integration boundary, which is independent of $k$. Keeping this in mind and applying \eqref{eq:General_Features}, one finds that the $f^{-1}$ high-frequency power law transitions at a frequency \eqref{eq:UHF_start_frequency} to an UHF power law of the form
\begin{align}\label{eq:RR_UHF_full}
      h^2 \Omega_{\rm rr, high}^{\delta} &\simeq \frac{3\mathcal{A}_{\rm rr}}{2}\left(\frac{G\mu}{\chi_r}\right)^{1/2} \frac{h_r^0}{t_{\rm Pl} f^2} \, .
\end{align}

\begin{figure}
\centering
\begin{subfigure}{.5\textwidth}
  \centering
\begin{overpic}[width = \textwidth]{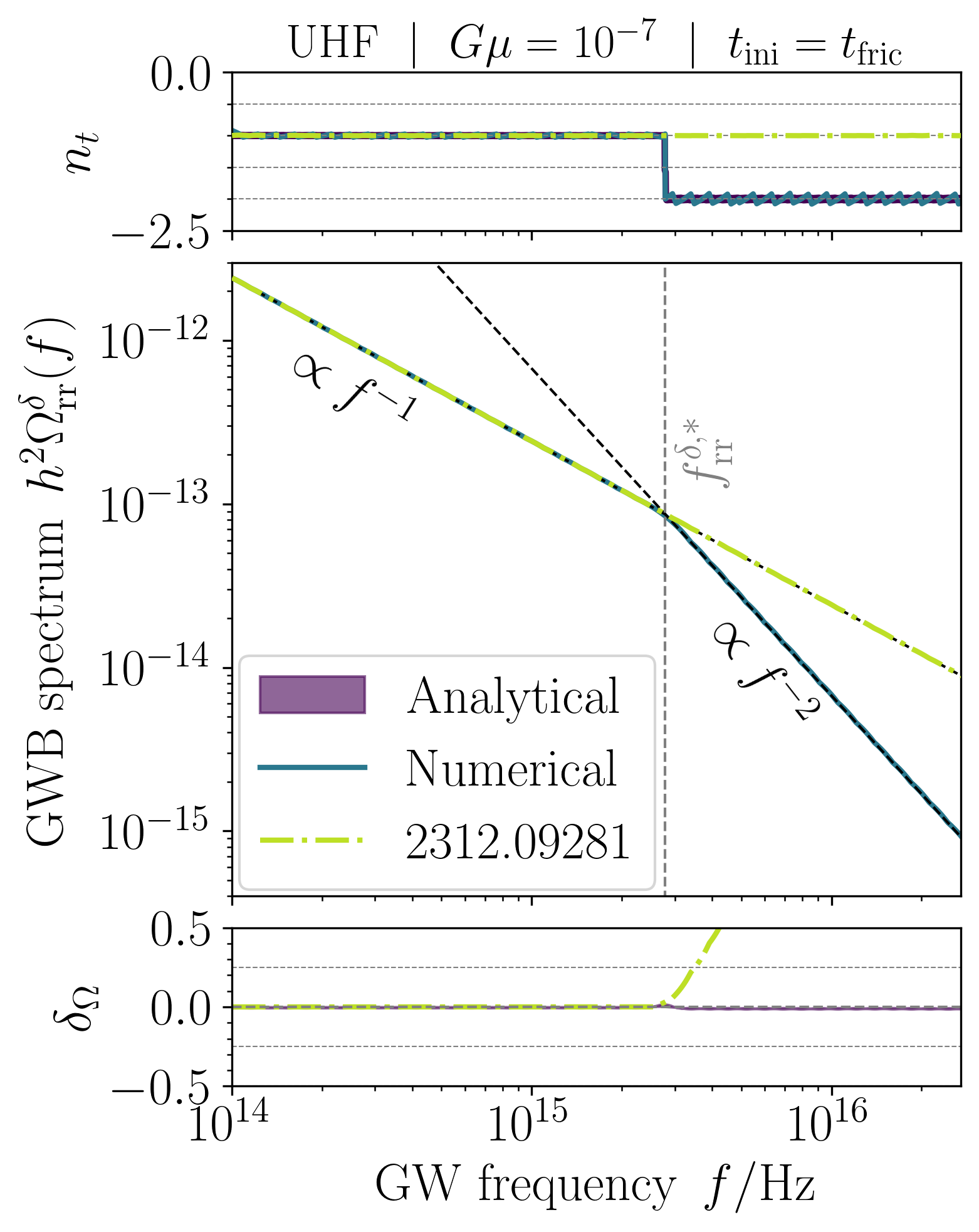}
\put(24,76.5){\rotatebox{-28.75}{\eqref{eq:Orr1_high}, \eqref{eq:General_Features}}}
\put(55.75,67){\rotatebox{90}{\eqref{eq:f_finite_width}}}
\put(64.75,40){\rotatebox{-47}{\eqref{eq:RR_UHF_full}}}
\end{overpic}
\end{subfigure}%
\begin{subfigure}{.5\textwidth}
  \centering
\begin{overpic}[width = \textwidth]{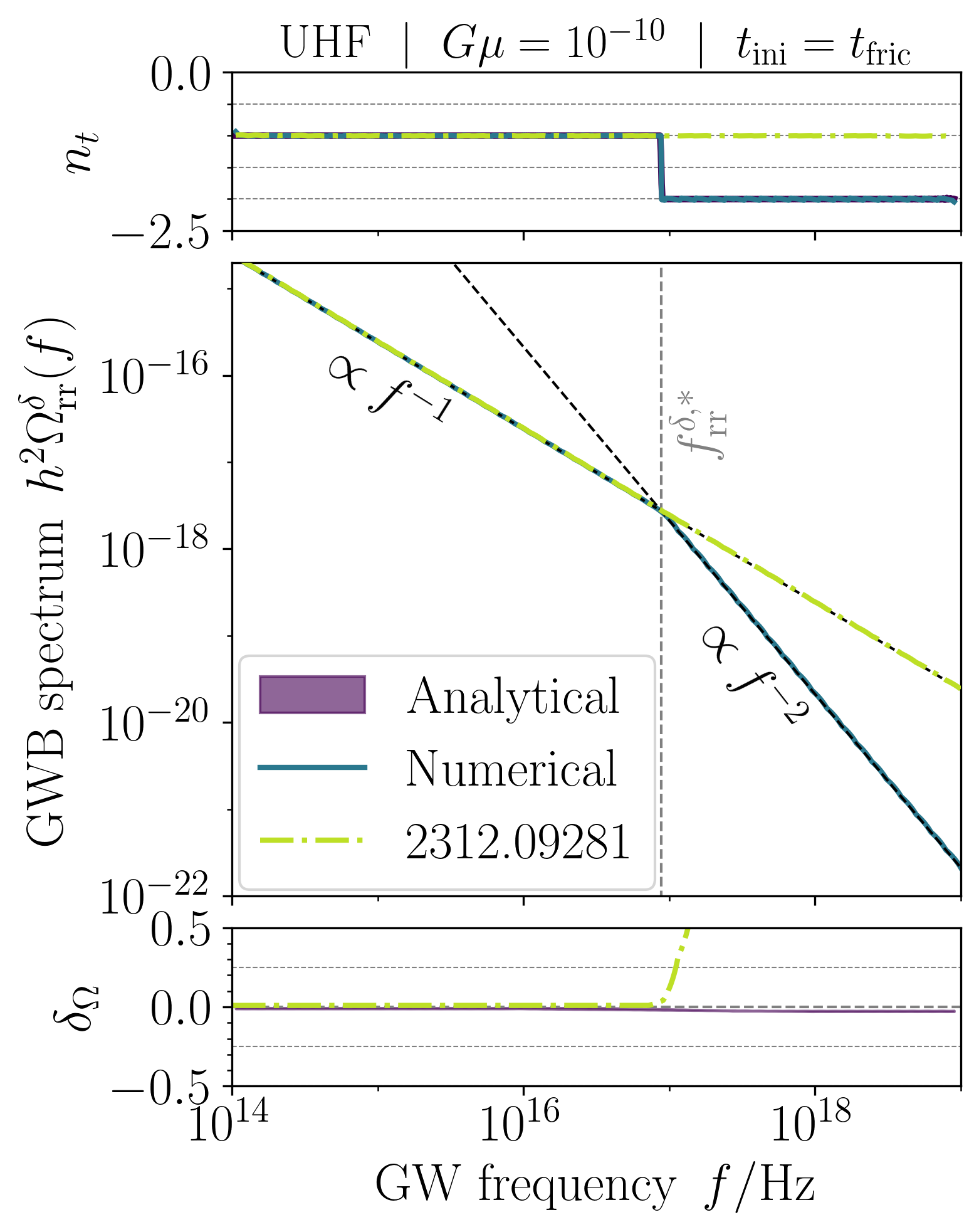}
\put(25.5,76.5){\rotatebox{-30.5}{\eqref{eq:Orr1_high}, \eqref{eq:General_Features}}}
\put(55.25,68.5){\rotatebox{90}{\eqref{eq:f_finite_width}}}
\put(63.5,39.5){\rotatebox{-49}
{\eqref{eq:RR_UHF_full}}}
\end{overpic}
\end{subfigure}
\caption{\footnotesize UHF regime of the total GW spectrum for the RR case, string tensions of $G\mu = 10^{-7}$ (left) and $G\mu = 10^{-10}$ (right), and initial times $t_{\rm ini}=t_{\rm fric}$. Main panels: Our exact numerical result based on the VOS model accounting for the full time dependence of all relevant quantities and for a fixed effective number of DOFs (teal), our analytical result derived in this paper for $m=1$ and $n_{\rm max} = 10^3$ ({deep purple band, showing values between lower and upper bound given in Eq.~\eqref{Upperandlowerbound}}) and the numerical result following from Ref.~\cite{Servant:2023tua}. The grey dashed line shows the frequency at which the UHF effects become relevant. The power laws describing the spectrum below and above this frequency are shown as black dashed lines. Upper panels: Spectral index (cf.\ \eqref{eq:spectralindex}) of the three spectra. 
Lower panels: Relative deviation (cf.~Eq.~\eqref{eq:DeltaOmega}) of the analytical result from the numerical result. The visible deviations are due to numerical noise.}
\label{fig:RR_UHF_full}
\end{figure}

In Fig.~\ref{fig:RR_UHF_full}, we can see the transition from the $f^{-1}$ to the $f^{-2}$ power law for two benchmark scenarios. We compare our analytical result to our numerical result as well as to the result of Ref.~\cite{Servant:2023tua}. For the initial time, we chose, as before, the friction cutoff $t_{\rm ini} = t_{\rm fric}$.  
As one can see, our analytical and numerical results agree perfectly; visible deviations are only due to numerical inaccuracies. The derived power laws shown in the figure give also an excellent fit to the spectrum. When comparing this result to the fundamental spectrum, one may observe that, in contrast to other features, the transition from high frequencies to ultrahigh frequencies is not washed out by the mode summation. This is because of the mode number independence of the cutoff that we found. This is in contrast to the result of Ref.~\cite{Servant:2023tua} which discusses the same effect in Appendix A. In this article, the authors allow for GW emission from a $k$--mode, as soon as a loop of appropriate size had a chance to be formed for the first time. They neglect, however, the fact that loops at later times shrink again to smaller size.  In our treatment, we ensured that the relevant condition on the GW wavelength is always fulfilled. In particular, the correction to the initial time found in Ref.~\cite{Servant:2023tua} is independent of the frequency but only depends on the mode number. For the depicted benchmark scenarios, this would only become relevant for mode numbers $k\gtrsim 10^6$. Since we consider in the plot only mode numbers as large as $10^3$, the correction has no effect here and generally underestimates the actual correction due to the UHF effect.

\begin{figure}
\tiny
    \centering
    \begin{subfigure}{0.49\textwidth}
     \begin{overpic}[width = \textwidth]{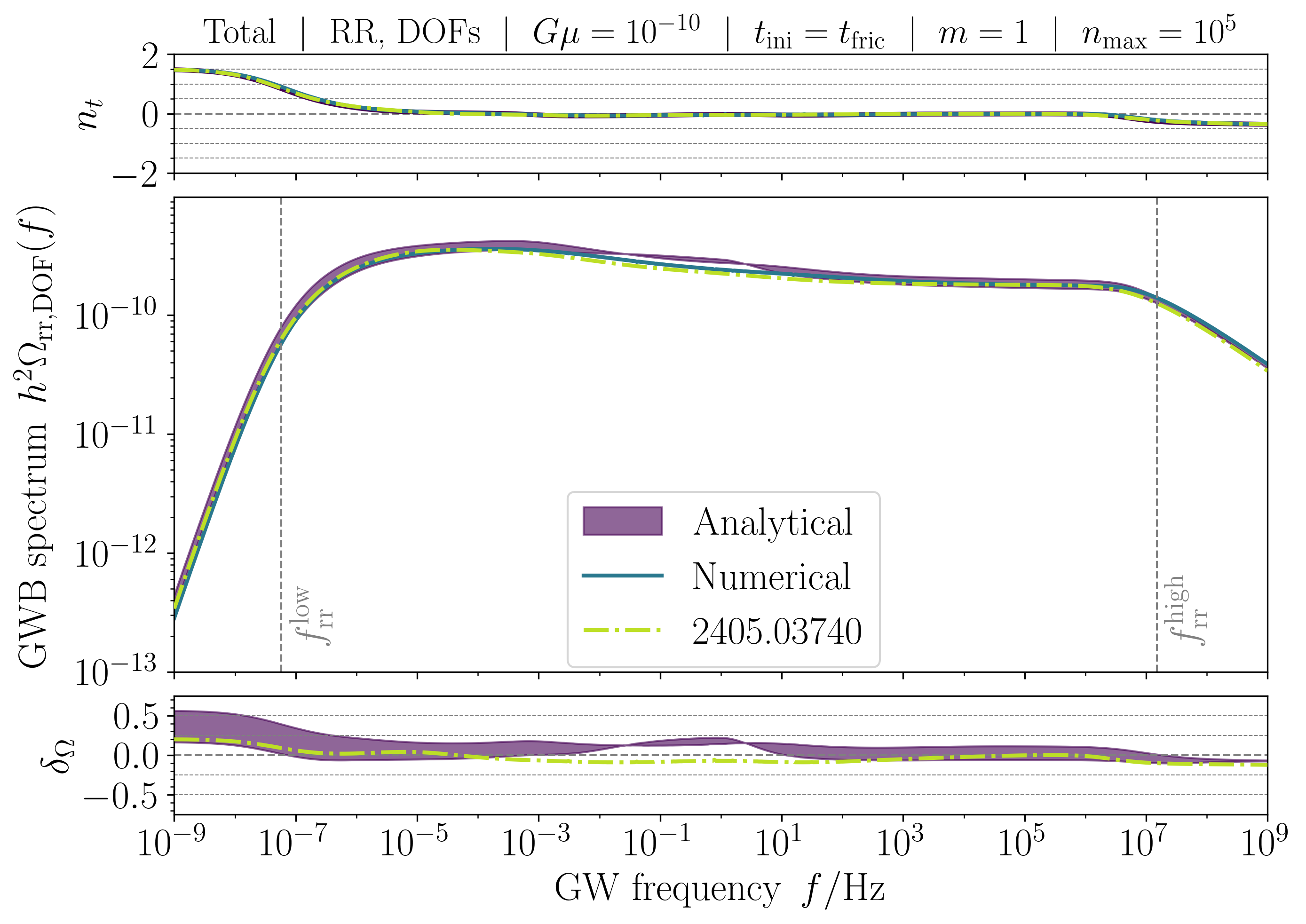}
\put(22.75, 28){\rotatebox{90}{\eqref{eq:f_rr_low}}}
\put(89.75, 28){\rotatebox{90}{\eqref{eq:f_rr_high}}}
\end{overpic}
\caption{ }
\label{fig:Varying_DOFs1}
    \end{subfigure}
    \begin{subfigure}{0.49\textwidth}
   \begin{overpic}[width = \textwidth]{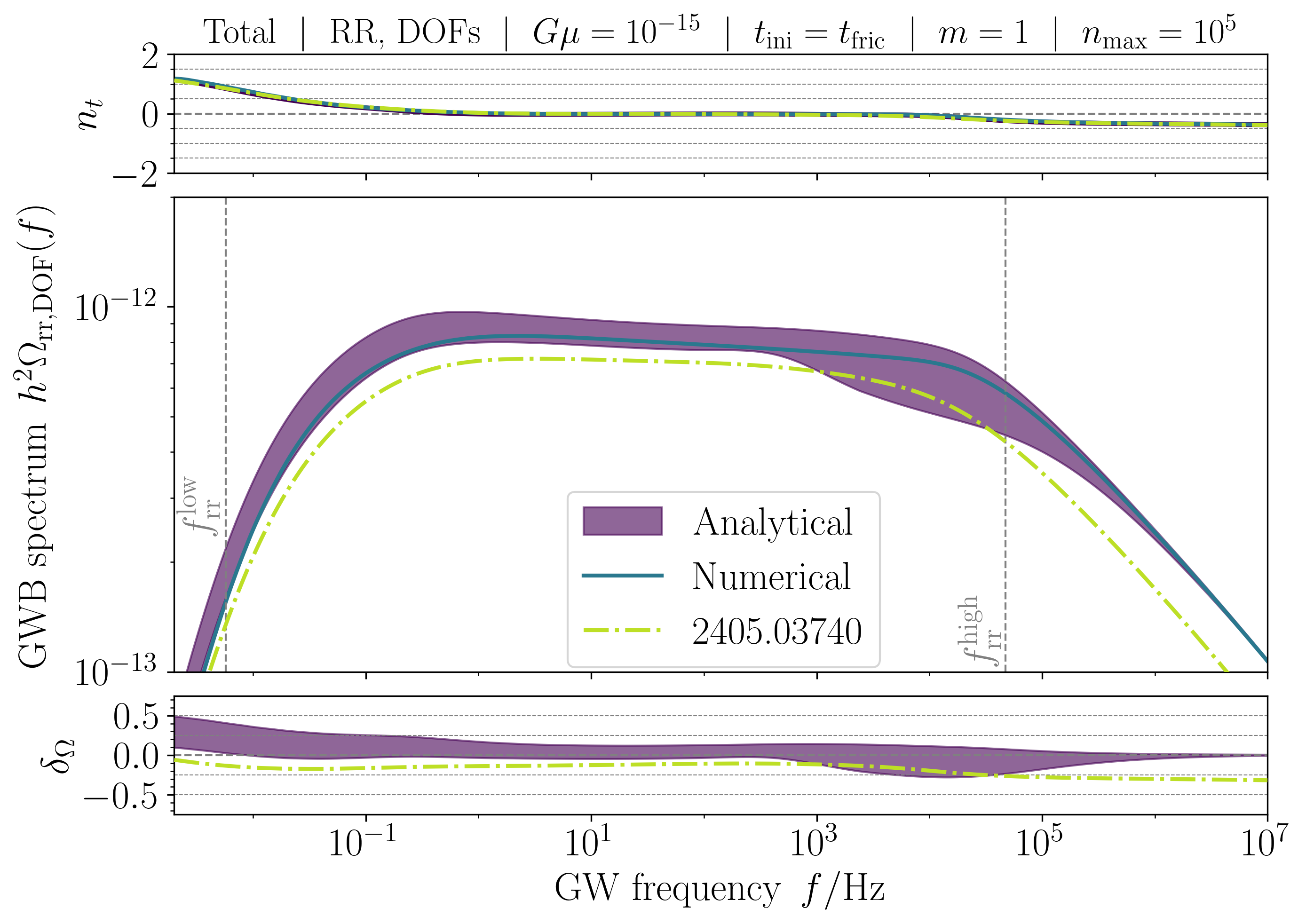}
\put(14, 35){\rotatebox{90}{\eqref{eq:f_rr_low}}}
\put(73.5, 26){\rotatebox{90}{\eqref{eq:f_rr_high}}}
\end{overpic}
\caption{  }
\label{fig:Varying_DOFs2}
    \end{subfigure}
     \caption{\footnotesize
   Total GW spectra for RR loops, accounting for changes in the effective number of DOFs. The benchmark points show the spectra for tensions of $G\mu = 10^{-10}$ (left panel) and $G\mu = 10^{-15}$ (right panel), and an initial time $t_{\rm ini} = t_{\rm fric}$. 
The panels and color code are the same as in Fig.~\ref{fig:RR_Total}. This time, our numerical comparison spectrum based on the VOS model is accounting for the full time dependence of all relevant quantities and for a time-dependent effective number of DOFs (cf.~Section \ref{subsec:Numerical}).}
    \label{fig:Varying_DOFs_Total_1}
\end{figure}

\subsection{Varying degrees of freedom}
\label{subsec:FullVaryingDOFs}

The previous derivation of the total RR spectrum still assumed a constant effective number of relativistic DOFs. To include variations in this number during radiation domination, we need to adapt the discussion of the previous Section and refer to the slightly more complex fundamental spectrum in Section \ref{subsec:VaryingDOFs}.
Directly working with this fundamental spectrum and the results we derived in the previous Section can be problematic, since the additional splitting of the radiation dominated era will introduce multiple regions in which our expansions of $\varphi_2$ can become invalid. We are saved by the fact that the fundamental RR spectrum is, in absence of a maximum frequency, very well approximated if we use the replacement $\varphi_2(x,\chi)\to \left(1+\chi_r\right)^{1/2}x$. 
This gives rise to a fundamental spectrum that can be integrated without any further approximations. In presence of a maximum frequency in the RR spectrum, there is no plateau region and the RR contribution will be strongly subdominant to the RM contribution. Therefore, if a maximum in the RR spectrum occurs, we can comfortably refer to the discussion in the previous Section. Let us now turn to the case in which there is no such maximum frequency and investigate the effect of changing DOFs. To use the above approximation consistently, we need to account for the fact that the switch, minimum, and maximum frequencies will be affected as well. Let us summarize these. With the notation used in Section \ref{subsec:VaryingDOFs}, we have as a result of our approximation, indicated by hats, for case $A$
\begin{align}
    \hat{x}_{A,i}^{\rm start} = \max\left\{\chi_r^{-1}, \sqrt{1+\chi_r}x_i^{(i)}\right\}, && \hat{x}_{A,i}^{\rm end} = x_i^{(i+1)}
\end{align}
which tells us for the minimum and switch frequency that
\begin{align}
 \hat{f}_{A,i}^{\rm min} = \frac{f_i^{(i+1)}}{\chi_r}, && \hat{f}_{A,i}^{{\rm start}, *}=\frac{f_i^{(i)}}{\chi_r\sqrt{1+\chi_r}} \, .
\end{align}
The spectrum will no longer contain a maximum frequency but is only non-vanishing if
\begin{align}
    \frac{a_{(i+1)}}{a_{(i)}}>\sqrt{1+\chi_r} \, .
\end{align}
The simplifications are much stronger in case $B$. Introducing the frequencies
\begin{align}
    \hat{f}_{B,ij}^{\rm start} = \min\left\{f_i^{(i)}, f_i^{(j)}/\sqrt{1+\chi_r}\right\} && \text{and} && \hat{f}_{B, ij}^{\rm end} = \max\left\{f_i^{(i+1)}, f_i^{(j+1)}/\sqrt{1+\chi_r}\right\}, 
\end{align}
we can write
\begin{align}
    \hat{x}_{B,ij}^{\rm start} = f/\hat{f}_{B,ij}^{\rm start}, && \hat{x}_{B,ij}^{\rm end} = f/\hat{f}_{B,ij}^{\rm end},
\end{align}
and we can also immediately see that the spectrum is only nonvanishing if 
\begin{align}
    \hat{f}_{B,ij}^{\rm start}>\hat{f}_{B,ij}^{\rm end} \, .
\end{align}
From this structure, it is clear that neither switch frequencies nor minimum or maximum frequencies exist in the approximated spectra. To make sure that the spectrum still does not extend to arbitrarily low frequencies, we cut it off at $f_{\rm rr}^{\rm min}$. 

Applying \eqref{Upperandlowerbound} to the spectra that arise by using these simplifications in combination with the fundamental spectrum \eqref{eq:RR_Varying_DOFs_Fundamental_Spectrum}, we find
\begin{align}
    h^2 &\Omega_{\rm rr,DOF}^{m, \text{upper}} = \frac{1}{H^{(q)}_{n_{\rm max}}} \sum_{k=1}^m \frac{1}{k^q} h^2 \Omega_{\rm rr, DOF}^{(1)}\left(\frac{f}{k}\right) +  \\\nonumber &+ \frac{\mathcal{A_{\rm rr}}}{H^{(q)}_{n_{\rm max}}} \sum_{i=0}^{N-1}\Theta_i^A \mathcal{G}_i \left\{S_{\rm rr,1}\left(k, x_i^{(i+1)}\right)\Big\vert_{m}^{k_{A,i}^{\rm max}} - S_{\rm rr,1}\left(k, \sqrt{1+\chi_r}x_i^{(i)}\right)\Big\vert_{m}^{k_{A,i}^{\rm start}} - S_{\rm rr,3}\left(k\right)\Big\vert_{k_{A,i}^{\rm start}}^{k_{A,i}^{\rm max}} \right\}+\\\nonumber &+\frac{\mathcal{A}_{\rm rr}}{H_{n_{\rm max}}^{(q)}} \sum_{i=1}^{N-1}\sum_{j=0}^{i-1}\Theta_{ij}^B \mathcal{G}_i^{1/4}\mathcal{G}_j^{3/4} \biggl\{ S_{{\rm rr}, 1}\left(k,\hat{x}_{B,ij}^{\rm end}\right)\Big\vert_{m}^{k_{B,ij}^{\rm max}} - S_{{\rm rr},1}\left(k,\hat{x}_{B,ij}^{\rm start}\right)\Big\vert_{m}^{k_{B,ij}^{\rm max}} \biggr\} \, .\nonumber
\end{align}
To evaluate the above expression, we need the following frequency-dependent mode numbers
\begin{subequations}
\begin{align}
{k_{A,i}^{\rm max} = \max\left\{\min\left\{n_{\rm max}, f/f_{A,i}^{\rm min}\right\}, m\right\} \stackrel{n_{\rm max}\to \infty}{\longrightarrow} \max\left\{ f/f_{A,i}^{\rm min}, m\right\} \, ,} \\
    k_{A,i}^{\rm start} = \min\left\{ \max\left\{m, f/\hat{f}_{A,i}^{{\rm start},*}\right\}, k_{A,i}^{\rm max}\right\} \, ,\\
    {k_{B,ij}^{\rm max} = \max\left\{\min\left\{n_{\rm max}, f/f_{\rm rr}^{\rm min}\right\}, m\right\} \stackrel{n_{\rm max}\to \infty}{\longrightarrow} \max\left\{ f/f_{\rm rr}^{\rm min}, m\right\}
    \, , }
\end{align}
\end{subequations}
as well as the Heaviside functions
\begin{align}
    \Theta_i^A = \Theta\left(\frac{a_{(i+1)}}{a_{(i)}}-\sqrt{1+\chi_r}\right) && \text{and} && \Theta_{ij}^B = \Theta\left(\hat{f}^{\rm start}_{B, ij} - \hat{f}^{\rm end}_{B, ij}\right) \, . 
\end{align}
In spite of the lack of a closed analytical expression for the time evolution of the effective number of DOFs, the above formulae provide a great approximation of our numerical spectrum; see Fig.~\ref{fig:Varying_DOFs_Total_1}. We observe that the deviations only begin to exceed the $25\,\%$ level in the low-frequency tail of the spectrum. This is unobservable and, hence, unproblematic since, in this part of the spectrum, the RR contribution is strongly subdominant to the RM contribution. 

There is a mentionable difference between our result and the one in Ref.~\cite{Blanco-Pillado:2024aca}. The latter accounts for changes in the DOFs by fitting the analytical SGWB to the numerically calculated one. For the result here, we used no information about the GW spectrum to choose our values of $a_{(i)}$ and $\mathcal{G}_i$, since they directly derive from the evolution of $g_s$ and $g_\rho$. Furthermore, Ref.~\cite{Blanco-Pillado:2024aca} does not distinguish between effects during loop production time and GW emission time, i.e., between $\mathcal{G}_i$ and $\mathcal{G}_j$.

\subsection{RM and MM loops}
\label{subsec:Full_RM_MM_Loops}
We continue with a discussion of all loops decaying during matter domination. While on a technical level, we still distinguish between RM and MM loops, due to the corrections described in Section \ref{subsec:FullFundamentalSpectrum}, it is no longer sensible to consider the resulting GW spectra separately.   

For the case of RM loops, we found that the fundamental spectrum is described by \eqref{eq:Orm1}. In order to carry out the integrals in the estimate \eqref{Upperandlowerbound} analytically, it is necessary to approximate $\varphi_3(\tilde{x}_m^{\rm ini}, \chi_r)$. For large frequencies $\tilde{x}_m^{\rm ini}\gg 1$ we can approximate $\varphi_3 (\tilde{x}_m^{\rm ini}, \chi_r) \to (1+\chi_r)^{1/3} \tilde{x}_{m}^{\rm ini}$, while for low frequencies $\tilde{x}_m^{\rm ini} \ll 1$, we have $\varphi_3 (\tilde{x}_m^{\rm ini}, \chi_r) \to \sqrt{1+ \chi_r}(\tilde{x}_m^{\rm ini})^{3/2}$. With this, we are able to find the following expressions for the spectrum:
\begin{align}
   h^2 &\Omega_{\rm rm}^{m, \text{upper}}(f)=\frac{1}{H_{n_{\rm max}}^{(q)}}\sum_{k=1}^m \frac{1}{k^q} h^2 \Omega_{\rm rm}^{(1)}\left(\frac{f}{k}\right) + \\ \nonumber &+\frac{\mathcal{A}_{\rm rm}}{H_{n_{\rm max}}^{(q)}}\left\{ S_{\rm rm, 1}\left(k, (1+\chi_r)^{1/3}\tilde{x}_m^{\rm ini}\right)\big\vert_{k_{\rm rm}^{\rm min}}^{k_{\rm rm}^{\rm start}} + S_{\rm rm, 2}\left(k, x_{m}^{\rm eq}\right)\big\vert_{k_{\rm rm}^{\rm start}}^{k_{\rm rm}^{\rm max}} - S_{\rm rm, 2}\left(k, x_m^0\right)\big\vert_{k_{\rm rm}^{\rm min}}^{k_{\rm rm}^{\rm max}}\right\} \,,
\end{align}
where we introduced the shape functions
\begin{align}\label{eq:full_RM_shape_1}
S_{\rm rm, 1}(k,x) &= \begin{cases}  \frac{3}{2(3+2q)k^{3/2+q}\sqrt{x_m^0}} x^3 + F_{\rm RM}(k) & \text{if } x\ll 1 \\
S_{\rm rm, 2}(k, x) & \text{else} 
\end{cases},\\
    S_{\rm rm, 2}(k,x) &= \frac{2k^{2-q}+2 (2 q-1) (k+x) \left(-x\right)^{1-q} \, _2F_1\left[\frac{1}{2},q-1;\frac{3}{2};\frac{k+x}{x}\right]}{\sqrt{x_m^0 (k+x)}} \, ,
\end{align}
as well as
\begin{subequations}
\begin{align}
   k_{\rm rm}^{\rm min}= \max\left\{m, f/f_{\rm rm}^{\rm max}\right\}, \\ {k_{\rm rm}^{\rm max} =\max\left\{\min\left\{n_{\rm max}, f/f_{\rm mm, min}\right\}, k_{\rm rm}^{\rm min}\right\} \stackrel{n_{\rm max}\to \infty}{\longrightarrow} \max\left\{f/f_{\rm mm, min}, k_{\rm rm}^{\rm min}\right\} \, ,}\\  k_{\rm rm}^{\rm start} = \min\left\{\max\left\{k_{\rm rm}^{\rm min}, f/f_{\rm rm}^{{\rm start}, *}\right\}, k_{\rm rm}^{\rm max}\right\} \, .
\end{align}
\end{subequations}
We also introduced the function $F_{\rm RM}(k)$, which only depends on the mode number but not on the frequency and does not need to be specified further {as it cancels out of any final expression for the spectrum}. 
At very high (low) frequencies, the evaluation of the hypergeometric functions can become numerically unstable. Therefore, it is beneficial to use series expansions. While working with the full hypergeometric function can be cumbersome, instead we can directly expand $S_{\rm rm}^{(1)}(x)$ for large (small)  $x$ and integrate over the mode number afterwards. In this way, we obtain
\begin{align}
    S_{\rm rm, 2}\left(k,x\right) \stackrel{x\to \infty}{\longrightarrow} \frac{3 k^{2-q}}{\left(2-q\right) \sqrt{x_m^0 x}}  \,, \qquad 
    S_{\rm rm, 2}\left(k,x\right) \stackrel{x\to 0}{\longrightarrow} \frac{3k^{-1/2-q} x^2}{\left(2+4q\right) \sqrt{x_m^0}} +F_{\rm RM}(k) \, .
\end{align}
Note that the above low-frequency expansion should only be applied simultaneously with the low-frequency expansion in \eqref{eq:full_RM_shape_1}. Only in this case, the function $F_{\rm RM}(k)$ is guaranteed to cancel out and can therefore be neglected. If there is no maximum frequency, a good choice to change from the low-frequency expansion to the full spectrum is at $x_{m}^0\sim 10^{-5}$. For the high-frequency expansion, we found $x_m^0 \sim 10^{10}$ to be a good value. If a maximum frequency exists, the low-frequency expansion can be applied everywhere. 

Most power laws arising in the spectrum derive again via \eqref{eq:General_Features} directly from the power laws present in the fundamental spectrum. However, when transitioning from the peak to the high-frequency region, a new power law can develop. Below the peak frequency \eqref{eq:f_rm_peak_2} or \eqref{eq:f_rm_peak_3}, the spectrum rises approximately like $A(f/f_m^0)^{3/2}$ with the amplitude $A$ given in \eqref{eq:RM_lowF_powerlaw}. Above, it decays like $B\,f_m^0/f$, and the amplitude $B$ can be found in \eqref{eq:Orm1_HF2} or \eqref{eq:Orm1_HF3}. 
The total spectrum can then be approximated by
\begin{align}
    h^2\Omega_{\rm rm}\left(f\right)\simeq \frac{1}{H_{n_{\rm max}}^{(q)}} \sum_{k=1}^{\lfloor k_{\rm rm}^{\rm peak}\rfloor} \frac{B}{k^{q-1}} \frac{f_m^0}{f} + \frac{1}{H_{n_{\rm max}}^{(q)}}\sum_{k= \lceil k_{\rm rm}^{\rm peak} \rceil}^{n_{\rm max}} \frac{A}{k^{q+3/2}} \left(\frac{f}{f_m^0}\right)^{3/2},
\end{align}
where we introduced $k_{\rm rm}^{\rm peak}(f)=f/f_{\rm rm}^{\rm peak}$. The above sums can be carried out explicitly in terms of Hurwitz zeta functions. Since we are interested in the regime where $f_{\rm rm}^{\rm peak} \ll f \ll n_{\rm max} f_{\rm rm}^{\rm peak}$ and thus $1\ll k_{\rm rm}^{\rm peak}(f) \ll n_{\rm max}$, we can expand the arising terms and find in total
\begin{align}
    h^2\Omega_{\rm rm}(f) \simeq \frac{f^{1-q}}{H_{n_{\rm max}}^{(q)}} \left( \frac{B}{2-q} \frac{f_m^0}{\left(f_{\rm rm}^{\rm peak}\right)^{2-q}} + \frac{A}{q+1/2} \frac{\left(f_{\rm rm}^{\rm peak}\right)^{q+1/2}}{\left(f_m^0\right)^{3/2}}\right). 
\end{align}
Within the reasonable parameter range, there is no clear hierarchy between the two remaining terms and we have to take both into account. Due to our approximation, we have overestimated the amplitude of the intermediate power law for the same reason as in the RR case. We can, however, correct for this quite well with a factor of $\sim 4$. 

The above reasoning can, of course, only be sensibly applied if there is no maximum frequency in the spectrum. If there is a maximum frequency, the fundamental spectrum will be roughly described by the low-frequency power $A(f/f_m^0)^{3/2}$ below the maximum frequency, and it vanishes above the cutoff. This means we can simply take the above expression for the total spectrum, set the high-frequency amplitude $B = 0$, and replace $f_{\rm rm}^{\rm peak}$ by $f_{\rm rm}^{\rm max}$. Let us use the explicit expression $A=3\mathcal{A}_{\rm rm}/4$ to find
\begin{align}
    h^2 \Omega_{\rm rm}(f) \simeq \frac{f^{1-q}}{H_{n_{\rm max}}^{(q)}} \frac{3 \mathcal{A}_{\rm rm}}{4q+2} \frac{\left(f_{\rm rm}^{\rm max}\right)^{q+1/2}}{\left(f_m^0\right)^{3/2}} \, .
\end{align}
 
Turning to MM loops, the fundamental spectrum is described by \eqref{eq:Omm1}. To carry out the integrals in \eqref{Upperandlowerbound}, we need to approximate $\varphi_3\left(\tilde{x}_m^{\rm eq}, \chi_m\right)$. Namely, for $\tilde{x}_m^{\rm eq}\gg 1$, we can approximate $\varphi_3\left(\tilde{x}_m^{\rm eq}\right) \to \left(1+\chi_m\right)^{1/3} \tilde{x}_m^{\rm eq}$ and for low frequencies $\varphi_3 \left(\tilde{x}_m^{\rm eq}, \chi_m\right)\to \sqrt{1+\chi_m} \left(\tilde{x}_m^{\rm eq}\right)^{3/2}$. This allows us to obtain
\begin{align} \label{eq:OmmFull}
    h^2 &\Omega_{\rm mm}^{m , \text{upper}} (f)=\frac{1}{H_{n_{\rm max}}^{(q)}}\sum_{k=1}^m \frac{1}{k^q} h^2 \Omega_{\rm mm}^{(1)}\left(\frac{f}{k}\right)+\\&\nonumber+\frac{\mathcal{A}_{\rm mm}}{H_{n_{\rm max}}^{(q)}}\left\{S_{\rm mm,1}\left(k, x_m^0\right)\big\vert_{k_{\rm mm}^{\rm min}}^{k_{\rm mm}^{\rm max}} - S_{\rm mm,3}\left(k, (1+\chi_m)^{1/3}\tilde{x}_m^{\rm eq}\right)\big\vert_{k_{\rm mm}^{\rm min}}^{k_{\rm mm}^{\rm start}} - S_{\rm mm, 2}(k)\big\vert_{k_{\rm mm}^{\rm start}}^{{k_{\rm mm}^{\rm max}}} \right\} 
\end{align}
where the shape functions read
\begin{align}
    S_{\rm mm, 1}(k, x) =& \left(\frac{x}{x_m^0}\right)^2 \frac{k^{1-q} \left(\, _2F_1\left(1,1-q;2-q;-\frac{k}{x}\right)-2 \left(\frac{k}{x}\right)^2 \ln \left(1+\frac{x}{k}\right)-\left(\frac{k}{x}-1\right)^2\right)}{3-q}\stackrel{x\ll 1}{\simeq} \nonumber\\ &\simeq \left(\frac{x}{x_m^0}\right)^2\left(-\frac{x}{3q k^q}+\frac{x^2}{2(1+q)k^{1+q}}\right)+F_1(x)+F_{\rm MM}(k),\\
    S_{\rm mm, 2}(k)=&
    \frac{k^{3-q} S_{\rm mm}^{(1)}\left(\chi_m^{-1}\right)}{3-q} \stackrel{\chi_m \gg 1}{\simeq} \frac{k^{3-q}}{3-q}\left(\frac{1}{x_m^0}\right)^2\left(\frac{\chi_m^{-3}}{3}-\frac{\chi_m^{-4}}{2}-1\right) \nonumber=\\&=\frac{k^{3-q}}{3-q}\left(\frac{1}{x_m^0}\right)^2\left(\frac{\chi_m^{-3}}{3}-\frac{\chi_m^{-4}}{2}\right)+F_{\rm MM}(k), 
    \end{align}
\begin{align}
    \nonumber
    \tilde{S}_{\rm mm, 3}(k,x ) =&  \frac{k^{3-q}}{(q-3) \left(x_m^0\right)^2} \Biggl\{1-{_2}F_1\left(1,2-\frac{2 q}{3};3-\frac{2 q}{3};-\left(\frac{k}{x}\right)^{3/2}\right) + 2 \ln \left(1+\left(\frac{x}{k}\right)^{3/2}\right)-\\& -\frac{2}{2q-3}\left(\frac{x}{k}\right)^{3/2} \left(3  {_2}F_1\left(1,\frac{2 q}{3}-1;\frac{2 q}{3};-\left(\frac{x}{k}\right)^{3/2}\right)+q-3\right)\Biggr\}
    \stackrel{x\ll 1}{\simeq} \nonumber \\&\simeq - \frac{2x^{9/2}}{3(3+2q)k^{3/2+q}(x_m^0)^2}+F_2(x)+F_{\rm MM}(k),\\
    S_{\rm mm, 3}(k,x) &= \begin{cases} 
    \tilde{S}_{\rm mm, 3}(k, x) & \text{if } x\ll 1\\
        S_{\rm mm, 1}(k,x) & \text{if } x\gg 1 
    \end{cases} \label{eq:Smm3}
\end{align}
The different terms are evaluated at the mode numbers
\begin{subequations}
\begin{align}
   k_{\rm mm}^{\rm min}= \max\left\{m, f/f_{\rm mm}^{\rm max}\right\} \, , \\ {k_{\rm mm}^{\rm max} = \max\left\{\min\left\{n_{\rm max}, f/f_{\rm mm}^{\rm min}\right\},k_{\rm mm}^{\rm min}\right\} 
   \stackrel{n_{\rm max}\to \infty}{\longrightarrow}\max\left\{ f/f_{\rm mm}^{\rm min},k_{\rm mm}^{\rm min}\right\} 
   \, ,}\\  k_{\rm mm}^{\rm start} = \min\left\{\max\left\{k_{\rm mm}^{\rm min}, f/f_{\rm mm}^{{\rm start},*}\right\}, k_{\rm mm}^{\rm max}\right\} \, .
\end{align}
\end{subequations}
A few comments are in order. First, the functions $F_1(x), F_2(x)$ are irrelevant since they cancel out immediately. {They have no physical meaning as they do not enter the observable spectrum.} Furthermore, we find it useful to distinguish four frequency regimes. At very low frequencies (we used $x_m^0<10^{-2}$), we use in \eqref{eq:OmmFull} the low-frequency expression of \eqref{eq:Smm3} together with the expanded versions of all the shape functions. In this case, they all contain at leading order a term $F_{\rm MM}(k)$, which cancels out in the complete spectrum. {It is, therefore, not explicitly specified.} For larger but still low frequencies (we applied this to $10^{-2}<x_m^0<10^3$), we still used the low-frequency expression of \eqref{eq:Smm3} but with the whole form of the shape functions, except for $S_{{\rm mm}, 2}$, for which we expanded in $\chi_m\gg 1$. Since this will be true at all frequencies, we use this expression for $S_{{\rm mm}, 2}$ everywhere. Note that, this time, we cannot neglect $F_{\rm MM}(k)$ since we do not expand the other shape functions and no cancellation will occur. At higher frequencies (we used $10^3<x_m^0 <10^6$), we switch to the high-frequency form of \eqref{eq:Smm3}, using the complete expressions for the shape functions and at very high frequencies ($x_m^0>10^6$), we expanded  $S_{{\rm mm},1}(k,x)$ to avoid problems with the numerical evaluation:
\begin{align}
    S_{\rm mm, 1}(k,x)\stackrel{x\gg 1}{\simeq} \frac{k^{2-q}}{(x_m^0)^2}\left(\frac{x}{2-q}-\frac{2k}{(3-q)^2} - \frac{2k}{3-q}\ln\left(\frac{x}{k}\right) - \frac{3k^2}{(4-q)x}\right).
\end{align}
Furthermore, note that in cases in which no maximum frequency exists, not only a low-frequency or a high-frequency regime exists, but also an intermediate regime. It turns out, though, that in this case, the entire spectrum is rather well described only using the high-frequency expression \eqref{eq:Smm3}. 
Since the MM spectrum is in all physically relevant cases, except at very low frequencies, subleading to the RM spectrum, the only relevant power law occurs in the very low-frequency regime. This power law derives with \eqref{eq:General_Features} directly from the low-frequency expression for the fundamental spectrum given in equation \eqref{eq:Omega_mm_low}. 

\begin{figure}
\tiny
    \centering
    \begin{subfigure}{0.49\textwidth}
     \begin{overpic}[width = \textwidth]{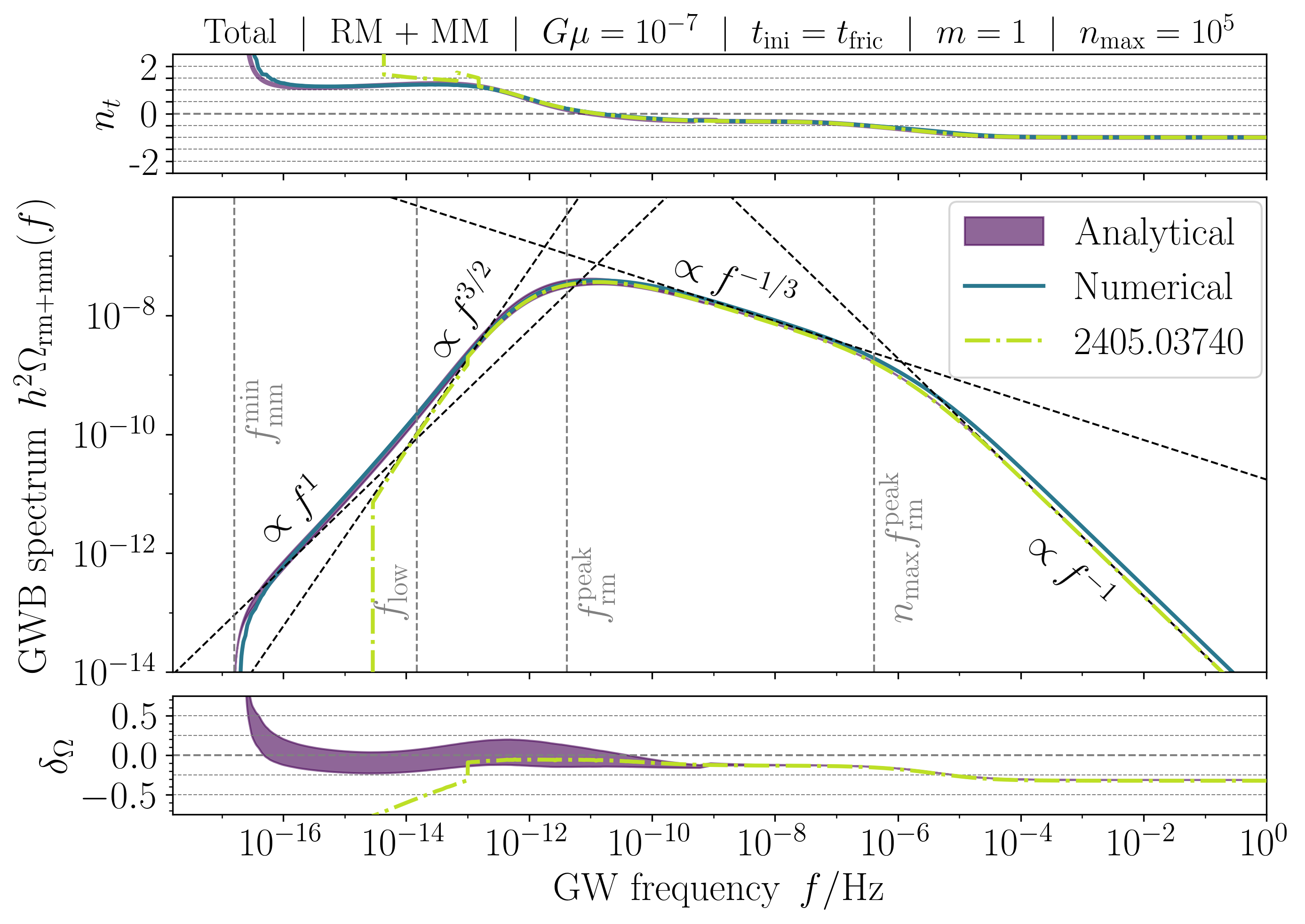}
     \put(19,42.5){\rotatebox{90}{\eqref{eq:f_mm_min}}}
     \put(32.75,20.25){\rotatebox{90}{\eqref{eq:LF_Transition_MM_to_RM}}}
     \put(44.4,30){\rotatebox{90}{\eqref{eq:f_rm_peak_2}}}
          \put(49,30){\rotatebox{0}{power laws:}}
     \put(49,26.5){\rotatebox{0}{\eqref{eq:RM_Total_PLs_no_max}}}
\end{overpic}
\caption{ }
\label{fig:RM+MM_Total_1}
    \end{subfigure}
    \begin{subfigure}{0.49\textwidth}
    \begin{overpic}[width = \textwidth]{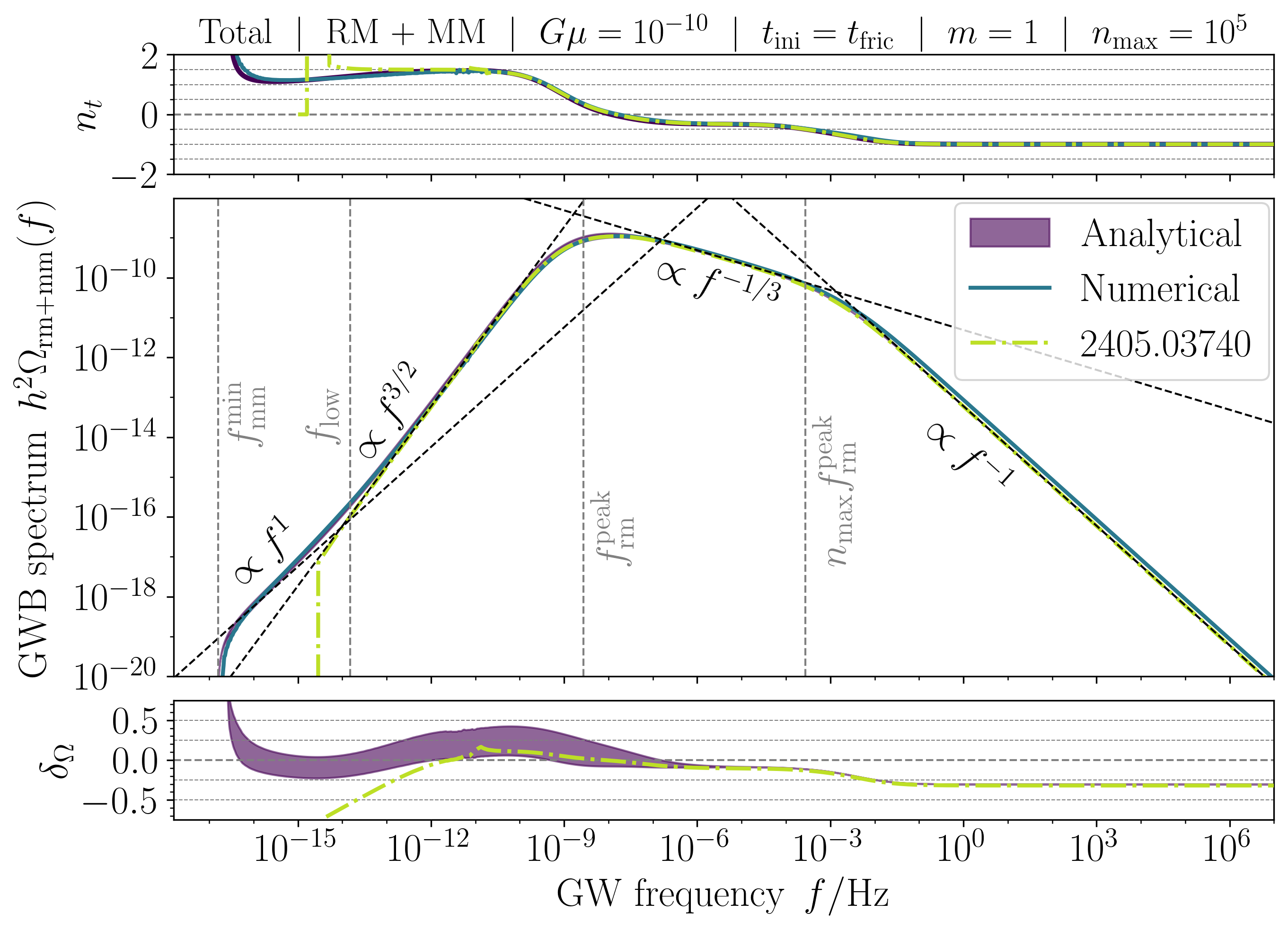}
    \put(18, 43.5){\rotatebox{90}{\eqref{eq:f_mm_min}}}
     \put(24, 43.5){\rotatebox{90}{\eqref{eq:LF_Transition_MM_to_RM}}}
     \put(46.5, 35){\rotatebox{90}{\eqref{eq:f_rm_peak_2}}}
               \put(27.5,25){\rotatebox{0}{power laws:}}
     \put(27.5,21.5){\rotatebox{0}{\eqref{eq:RM_Total_PLs_no_max}}}
\end{overpic}
\caption{  }
\label{fig:RM+MM_Total_2}
    \end{subfigure}

\bigskip
    
    \begin{subfigure}{0.49\textwidth}
  \begin{overpic}[width = \textwidth]{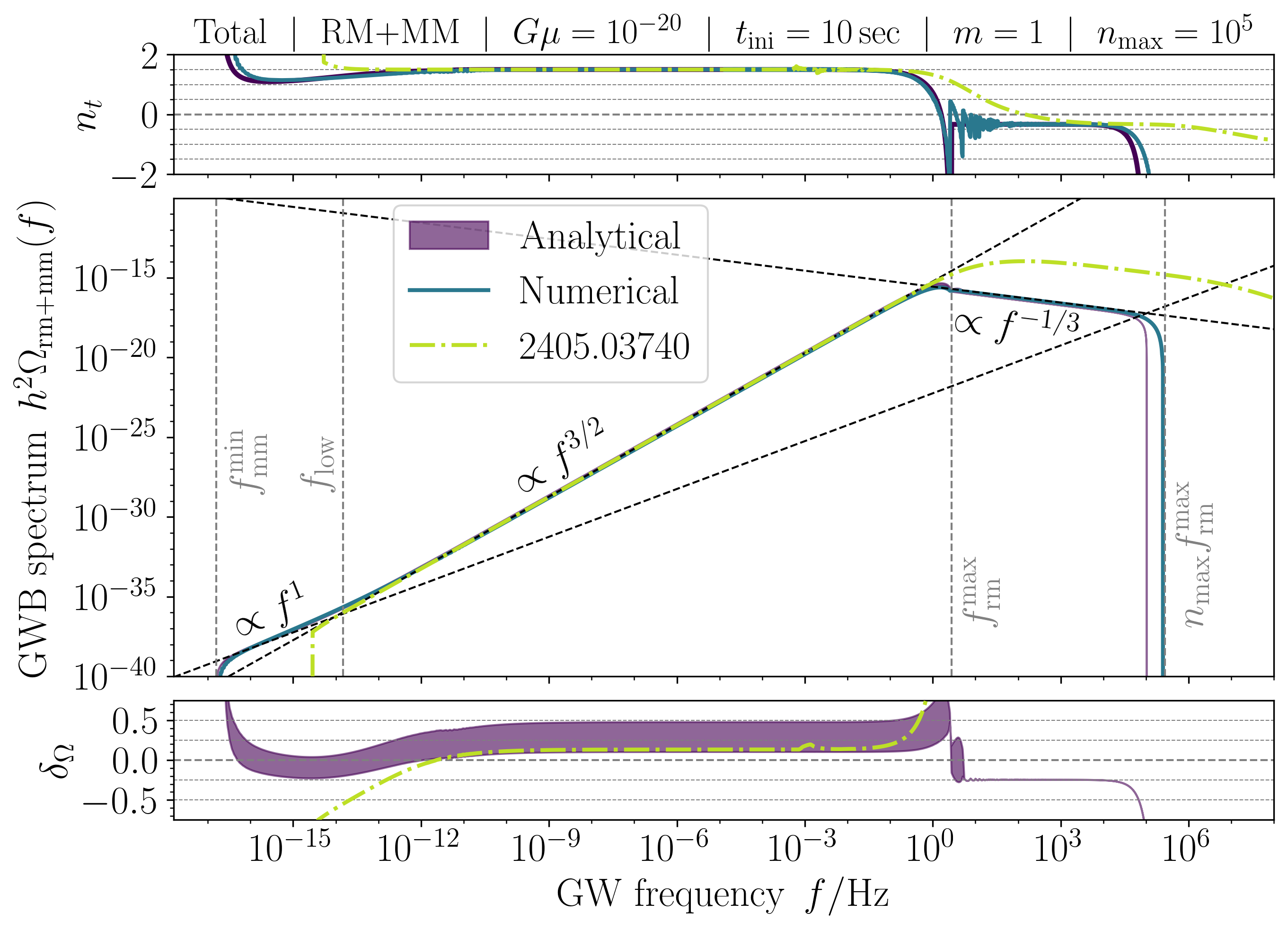}
    \put(17.75, 40.5){\rotatebox{90}{\eqref{eq:f_mm_min}}}
     \put(23.5, 40.5){\rotatebox{90}{\eqref{eq:LF_Transition_MM_to_RM}}}
     \put(75, 30.5){\rotatebox{90}{\eqref{eq:f_rm_max}}}
               \put(49,28){\rotatebox{0}{power laws:}}
     \put(49,24.5){\rotatebox{0}{\eqref{eq:RM_Total_PLs_max}}}
\end{overpic}
\caption{  }
\label{fig:RM+MM_Total_3}
    \end{subfigure}
    \caption{\footnotesize
    Three qualitatively different total GW spectra for RM and MM loops. 
     The panels and color code are the same as in Fig.~\ref{fig:RR_Total}. We cut off the MM contribution to the spectra from Ref.~\cite{Blanco-Pillado:2024aca} given by green dash-dotted lines at low frequencies (at $f = 10^{-13}\, {\rm Hz}, \, 10^{-11}\, {\rm Hz}, \, 10^{-3}\, {\rm Hz}$ for $G\mu = 10^{-7}, \, 10^{-10}, \, 10^{-20}$, respectively), as for lower frequencies, our numerical evaluation of this contribution is ill-behaved. {For panel (c), note also that the RM and MM spectra provided in Ref.~\cite{Blanco-Pillado:2024aca} are independent of $t_{\rm ini}$.}}
\end{figure}

The resulting spectra are shown in Figs.~\ref{fig:RM+MM_Total_1} to \ref{fig:RM+MM_Total_3} and cover the qualitatively different cases. In Fig.~\ref{fig:RM+MM_Total_1}, neither the RM nor the MM spectrum exhibits a maximum frequency, while for the parameter values in Fig.~\ref{fig:RM+MM_Total_2}, the MM spectrum does. Since this maximum frequency affects the MM spectrum only at rather large frequencies where it is strongly subleading to the RM contribution, the combined RM+MM spectra look qualitatively the same. For these cases, the spectrum can, in summary, be approximated by power laws of the form
\begin{align}\label{eq:RM_Total_PLs_no_max}
     h^2 \Omega_{\rm rm + mm}(f)\simeq \frac{1}{H_{n_{\rm max}}^{(q)}} \begin{cases}
        \frac{\mathcal{A}_{\rm mm}}{3} \zeta\left(q+1, k\right)\big\vert_{k=n_{\rm max}+1}^{k=1} \frac{f}{f_m^0} & \text{if } f_{\rm mm}^{\rm min} \ll f  \ll f_{\rm low}, \\ 
        \frac{3\mathcal{A}_{\rm rm}}{4} \zeta\left(q+3/2, k\right)\big\vert_{k=n_{\rm max}+1}^{k=1} \left(\frac{f}{f_m^0}\right)^{3/2} & \text{if } f_{\rm low} \ll f \ll f_{\rm rm}^{\rm peak}, \\
        \left(\frac{B}{2-q} \frac{f_m^0}{\left(f_{\rm rm}^{\rm peak}\right)^{2-q}} + \frac{A}{q+1/2} \frac{\left(f_{\rm rm}^{\rm peak}\right)^{q+1/2}}{\left(f_m^0\right)^{3/2}}\right) \frac{f^{1-q}}{4} & \text{if } f_{\rm rm}^{\rm peak} \ll f \ll n_{\rm max} f_{\rm rm}^{\rm peak}, \\
        B \zeta\left(q-1,n\right)\big\vert_{k=n_{\rm max} +1}^{k=1}\frac{f_m^0}{f} & \text{if } n_{\rm max} f_{\rm rm}^{\rm peak} \ll f,  
        \end{cases}
\end{align}
Here, $A=3\mathcal{A}_{\rm rm}/4$ and
\begin{align}
B=3\mathcal{A}_{\rm rm} \begin{cases} \left(\frac{a_0}{a_{\rm eq}}\right)^{1/4}-1\\
\frac{1}{\left(1+\chi_r\right)^{1/6}} \left(\frac{a_0}{\tilde{a}_m^{\rm ini}}\right)^{1/4}-1
\end{cases}
\end{align}
with the case distinction given in \eqref{eq:Orm1_HF2} and \eqref{eq:Orm1_HF3}. These power laws are depicted in the mentioned figures and, as can be seen, match the numerical spectrum well. Apart from this, it is clearly visible that at large frequencies, the numerically calculated spectrum is, in these cases, well approximated by our analytical calculations as well as by the result found in Ref.~\cite{Blanco-Pillado:2024aca}. {At low frequencies, our numerical evaluation of the MM contribution of Ref.~\cite{Blanco-Pillado:2024aca} becomes unstable. For these frequencies, we therefore show only the RM contribution in our plots.} Our expression matches our numerical spectrum at low frequencies for $m=1$ with an accuracy of $\lesssim 20\%$.  

A qualitatively different case occurs if not only the MM but also the RM spectrum exhibits a maximum frequency \eqref{eq:f_rm_max} and is depicted in Fig.~\ref{fig:RM+MM_Total_3}. In this case, the spectrum can be approximated by power laws of the form
\begin{align}\label{eq:RM_Total_PLs_max}
    h^2\Omega_{\rm rm+mm}(f)\simeq \frac{1}{H_{n_{\rm max}}^{(q)}} \begin{cases}
    \frac{\mathcal{A}_{\rm mm}}{3} \zeta\left(q+1, k\right)\big\vert_{k=n_{\rm max}+1}^{k=1} \frac{f}{f_m^0} & \text{if } f_{\rm mm}^{\rm min} \ll f  \ll f_{\rm low}, \\
     \frac{3\mathcal{A}_{\rm rm}}{4} \zeta\left(q+3/2, k\right)\big\vert_{k=n_{\rm max}+1}^{k=1} \left(\frac{f}{f_m^0}\right)^{3/2} & \text{if } f_{\rm low} \ll f \ll f_{\rm rm}^{\rm max}, \\
     \frac{3 \mathcal{A}_{\rm rm}}{4q+2} \frac{\left(f_{\rm rm}^{\rm max}\right)^{q+1/2}}{\left(f_m^0\right)^{3/2}} f^{1-q} & \text{if } f_{\rm rm}^{\rm max} \ll f \ll n_{\rm max} f_{\rm rm}^{\rm max}
    \end{cases}
\end{align}
From Fig.~\ref{fig:RM+MM_Total_3}, we can clearly see that the different power laws capture the result of the numerical computation formidably. Similarly, our full analytical result agrees with the numerical one very well, except for the frequency range close to $n_{\rm max}f_{\rm rm}^{\rm max}$. This is not unexpected, since $f_{\rm rm}^{\rm max}$ is determined from $\varphi_3\left(\tilde{x}_m^{\rm ini}, \chi_r\right)=x_m^0$. For our total spectrum, we used, however, the low-frequency expansion of $\varphi_3$ and we need to apply it in regions where this approximation starts to break down. Fortunately, as previously discussed, for frequencies close to $n_{\rm max}f_{\rm rm}^{\rm max}$, the spectrum becomes unphysical anyway since $n_{\rm max}$ is arbitrarily chosen and in principle, we should consider $n_{\rm max}\to \infty$ as long as the finite string width effects discussed in sections \ref{subsec:Gravitational_Wave_Spectrum}, \ref{subsec:RR_Loops_Fundamental}, and \ref{subsec:Full_RR_Loops} are taken into consideration. More problematically, for very small values in the denominator of $f_{\rm rm}^{\rm max}$,\footnote{In the language of Ref.~\cite{Schmitz_2024}, this corresponds to the parameter range very close to $T_{\rm cut}$.} the exact position of this maximum frequency is extremely sensitive to the exact parameter choice. Deviations due to our approximation can become large in this case. This is, of course, only the case for a very small parameter region and the power-law expressions give still excellent agreement with the numerical spectrum. Note that the result of Ref.~\cite{Blanco-Pillado:2024aca} is more or less incapable of capturing the maximum frequency at all, which is inherited from the fact that it derives from a fundamental spectrum that does not exhibit a maximum frequency. 

\begin{figure}
\tiny
    \centering
    \begin{subfigure}{0.49\textwidth}
     \begin{overpic}[width = \textwidth]{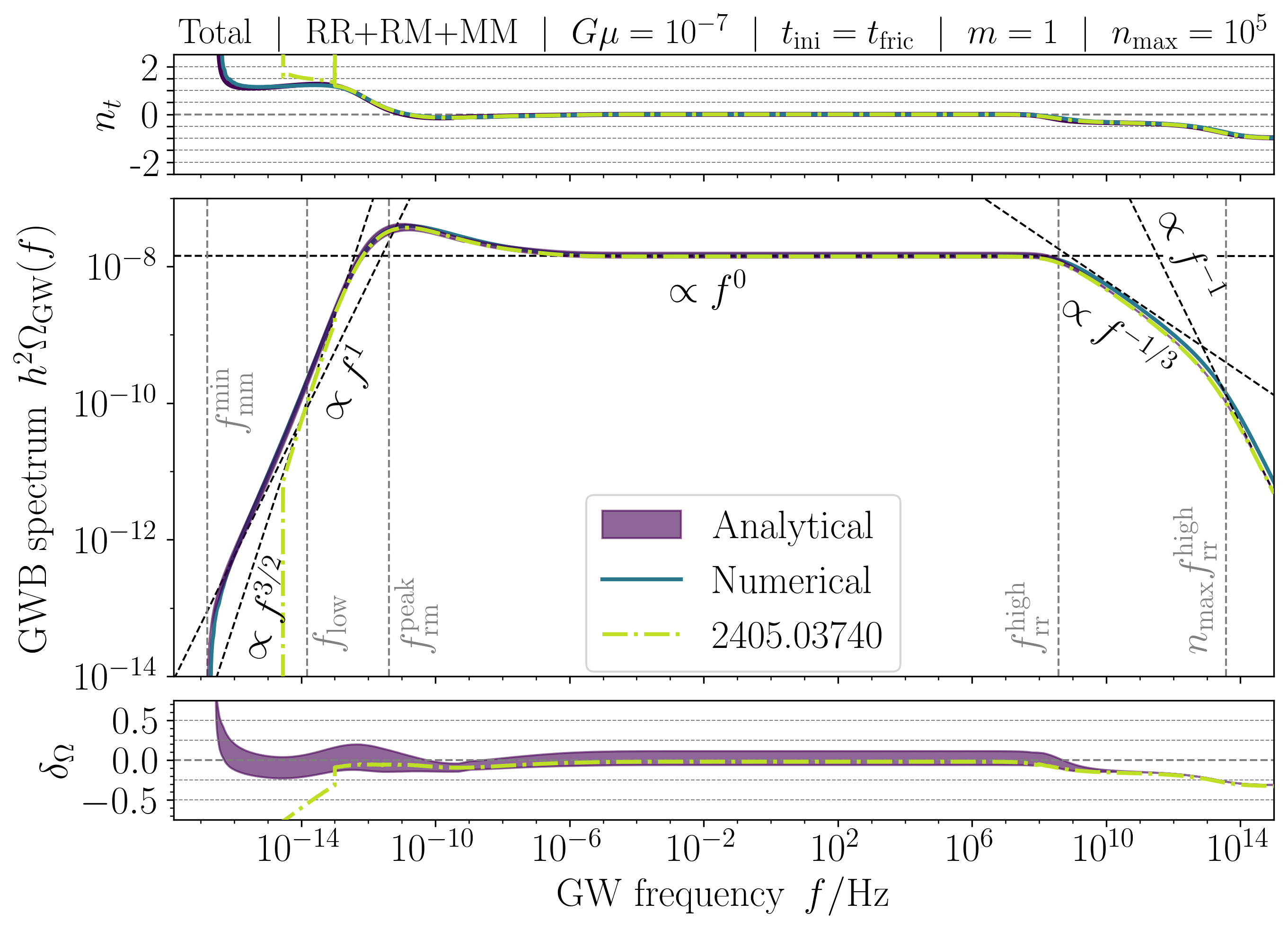}
\end{overpic}
\caption{ }
\label{fig:Full_Total_1}
    \end{subfigure}
    \begin{subfigure}{0.49\textwidth}
    \begin{overpic}[width = \textwidth]{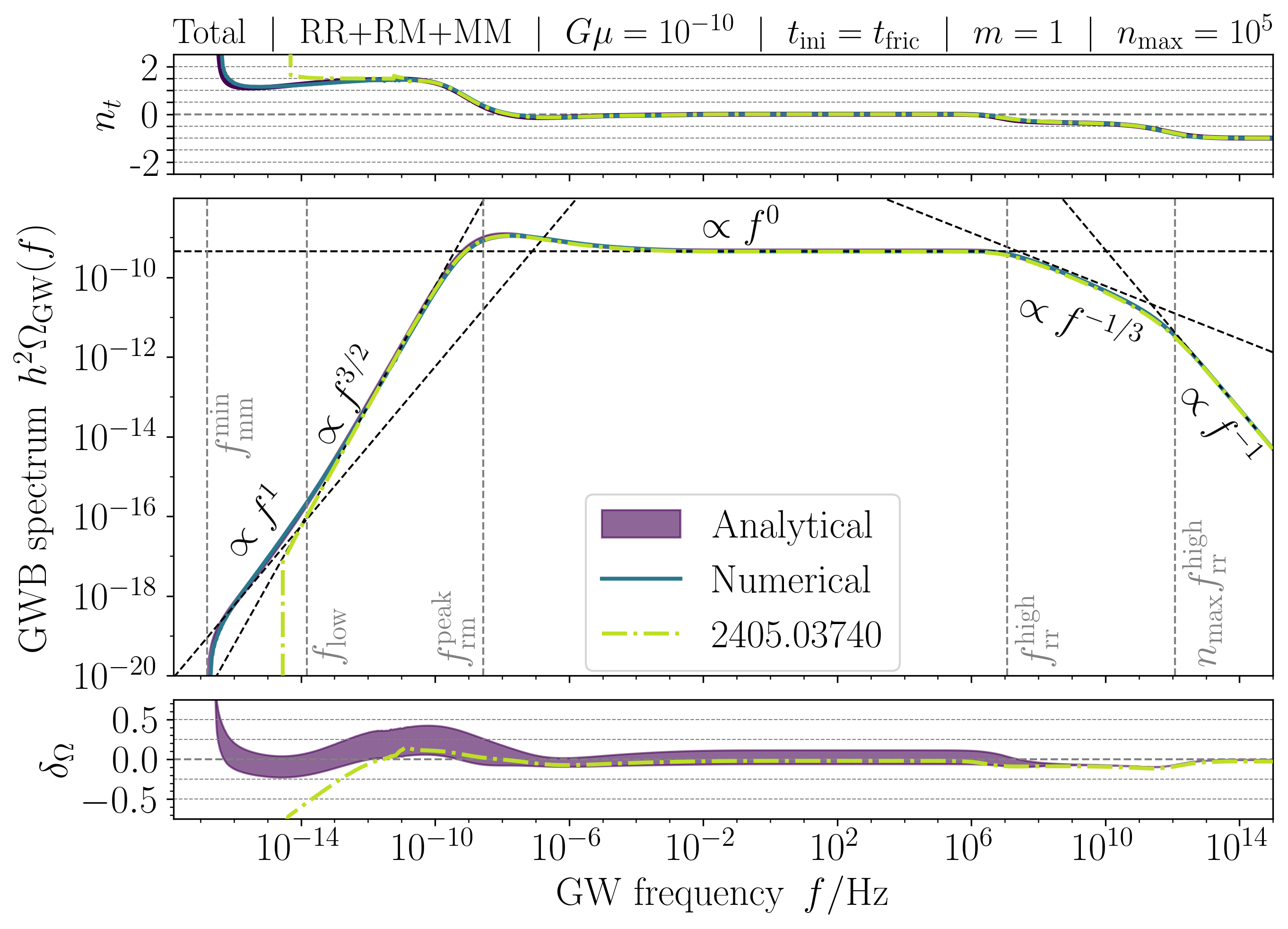}
\end{overpic}
\caption{  }
\label{fig:Full_Total_2}
    \end{subfigure}

\bigskip
    
    \begin{subfigure}{0.49\textwidth}
    \begin{overpic}[width = \textwidth]{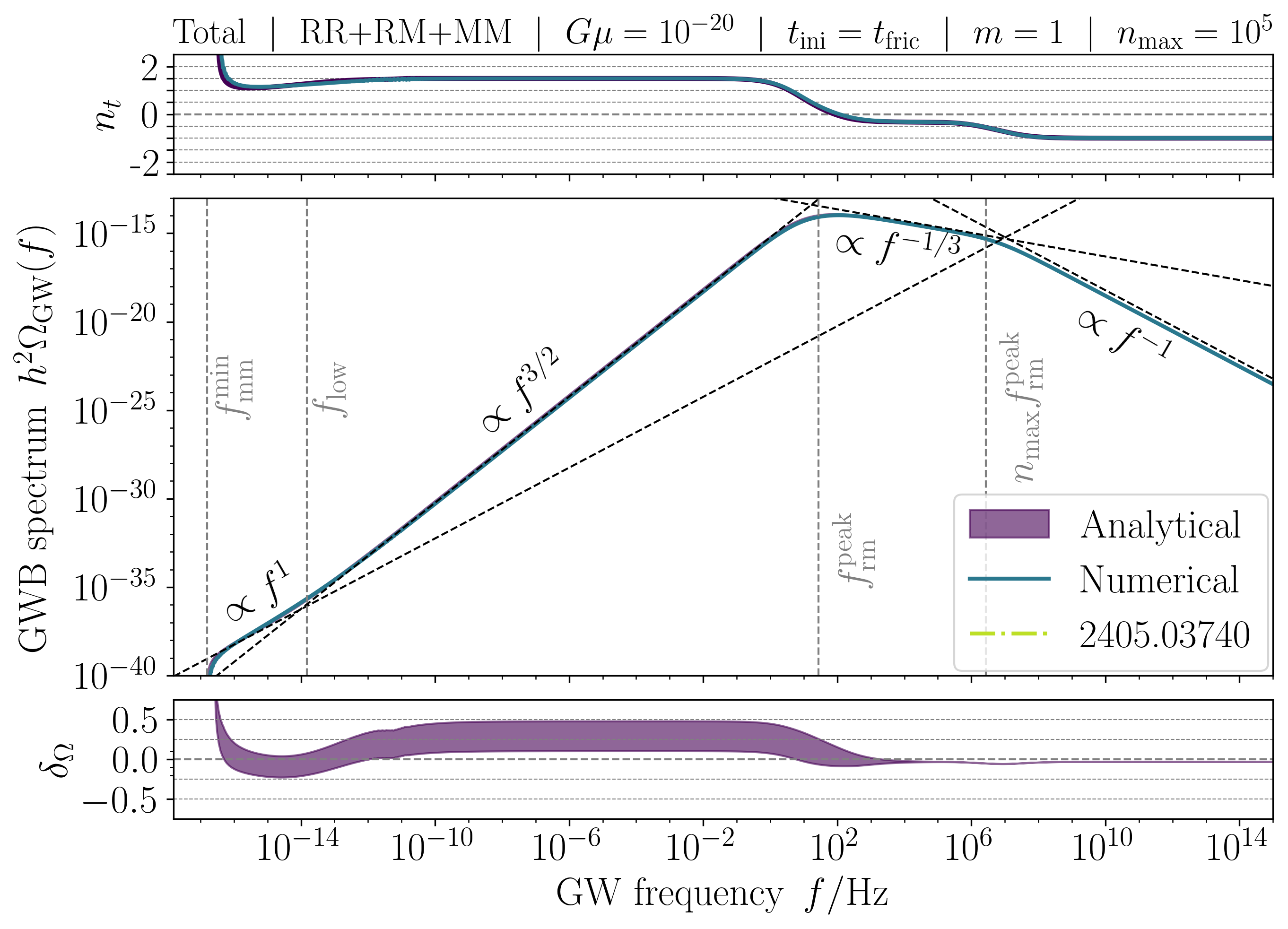}
\end{overpic}
\caption{  }
\label{fig:Full_Total_3}
    \end{subfigure}
    \begin{subfigure}{0.49\textwidth}
   \begin{overpic}[width = \textwidth]{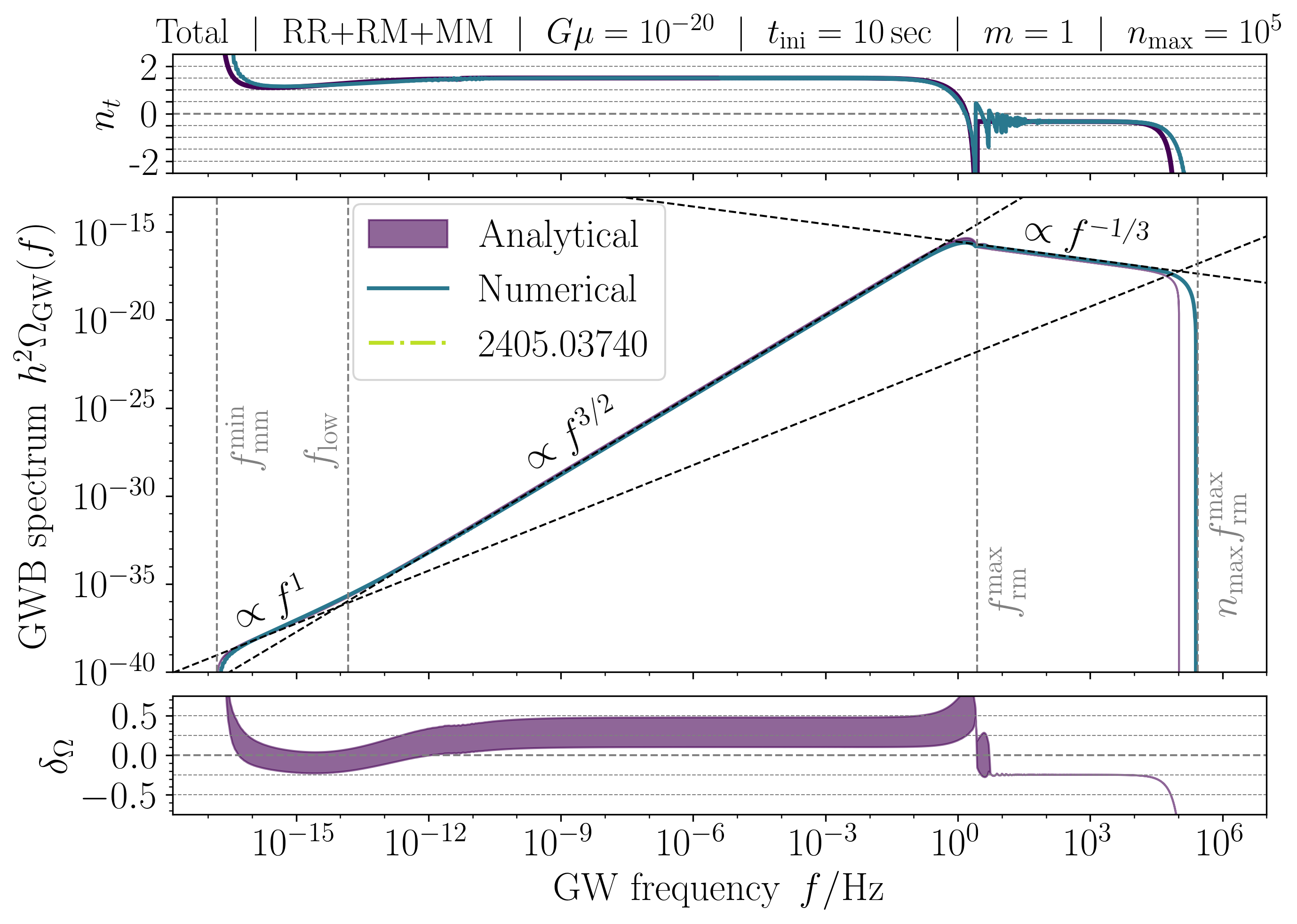}
\end{overpic}
\caption{  }
\label{fig:Full_Total_4}
    \end{subfigure}
    \caption{\footnotesize
     Total GW spectrum with all loop contributions for four qualitatively different cases. 
     Main panels: Our exact numerical result based on the VOS model accounting for the full time dependence of all relevant quantities and for a fixed effective number of DOFs (teal, cf.~Sec.~\ref{subsec:Numerical}), our analytical result with $m=1$ and $n_{\rm max}=10^5$ ({deep purple band, showing values between lower and upper bound given in Eq.~\eqref{Upperandlowerbound}}), and the analytical result derived in Ref.~\cite{Blanco-Pillado:2024aca} (light green, dash-dotted; {in \cite{Blanco-Pillado:2024aca}, $n_{\rm max}$ is referred to as $N$}). {Note that the parameter $m$ appears only in our analytical template.}    
    {We cut off the MM contribution to the spectra from Ref.~\cite{Blanco-Pillado:2024aca} at low frequencies (below $f = 10^{-13}\, {\rm Hz}, \, 10^{-11}\, {\rm Hz}, \, 10^{-3}\, {\rm Hz}$ for $G\mu = 10^{-7}, \, 10^{-10}, \, 10^{-20}$, respectively), as our numerical evaluation of this contribution is ill-behaved.} Grey dashed lines show characteristic frequencies, and black dashed lines indicate different power-law behaviors.
     Upper panels: Spectral index (cf. \eqref{eq:spectralindex}) of the three spectra.
     Lower panels: Relative deviation (cf. \eqref{eq:DeltaOmega}) of the two analytical results from the numerical result.}
     \label{fig:Full_Total}
\end{figure}

\begin{figure}
\tiny
    \centering
    \begin{subfigure}{0.49\textwidth}
     \begin{overpic}[width = \textwidth]{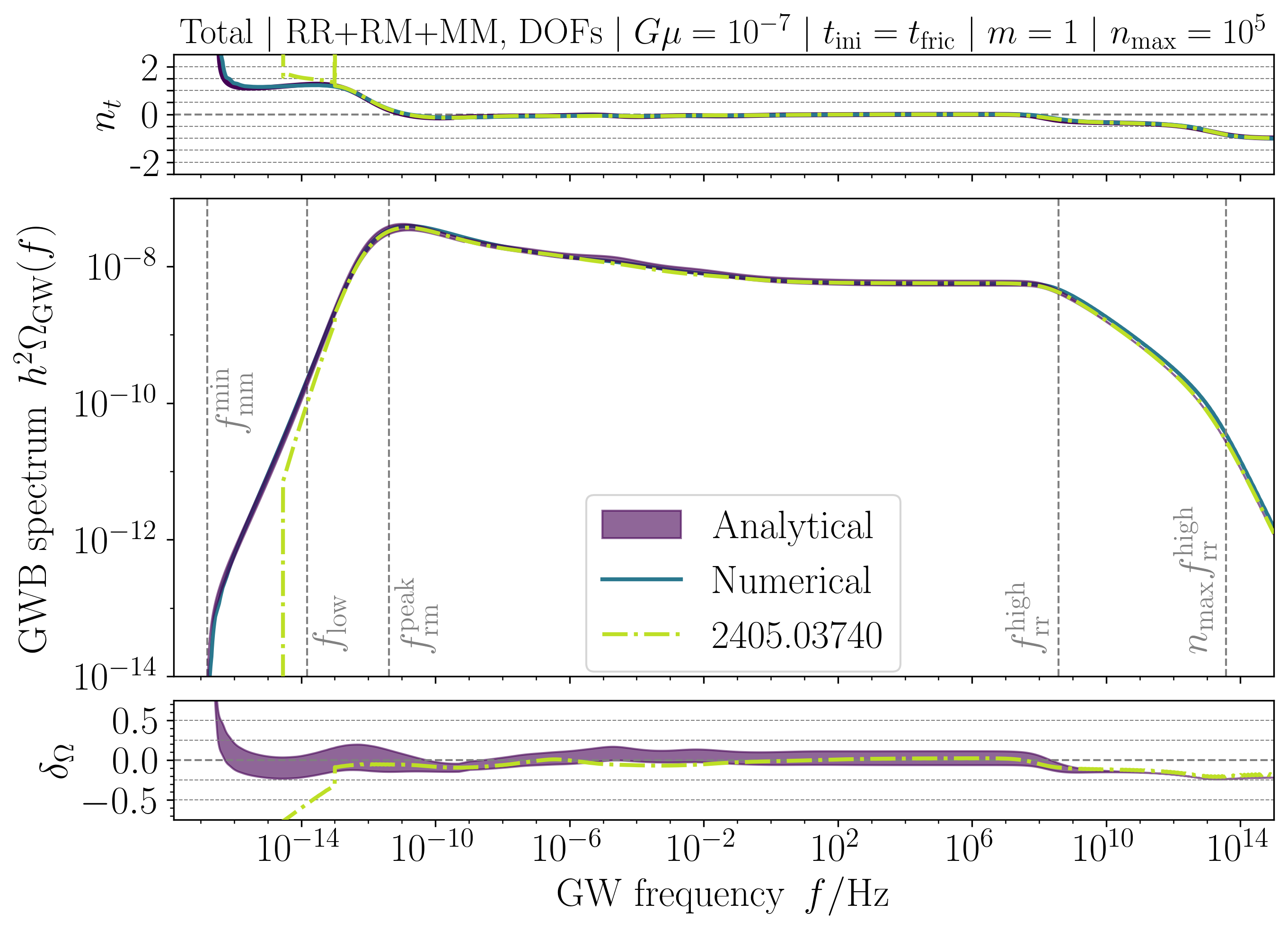}
\end{overpic}
\caption{ }
\label{fig:Full_Total_5}
    \end{subfigure}
    \begin{subfigure}{0.49\textwidth}
    \begin{overpic}[width = \textwidth]{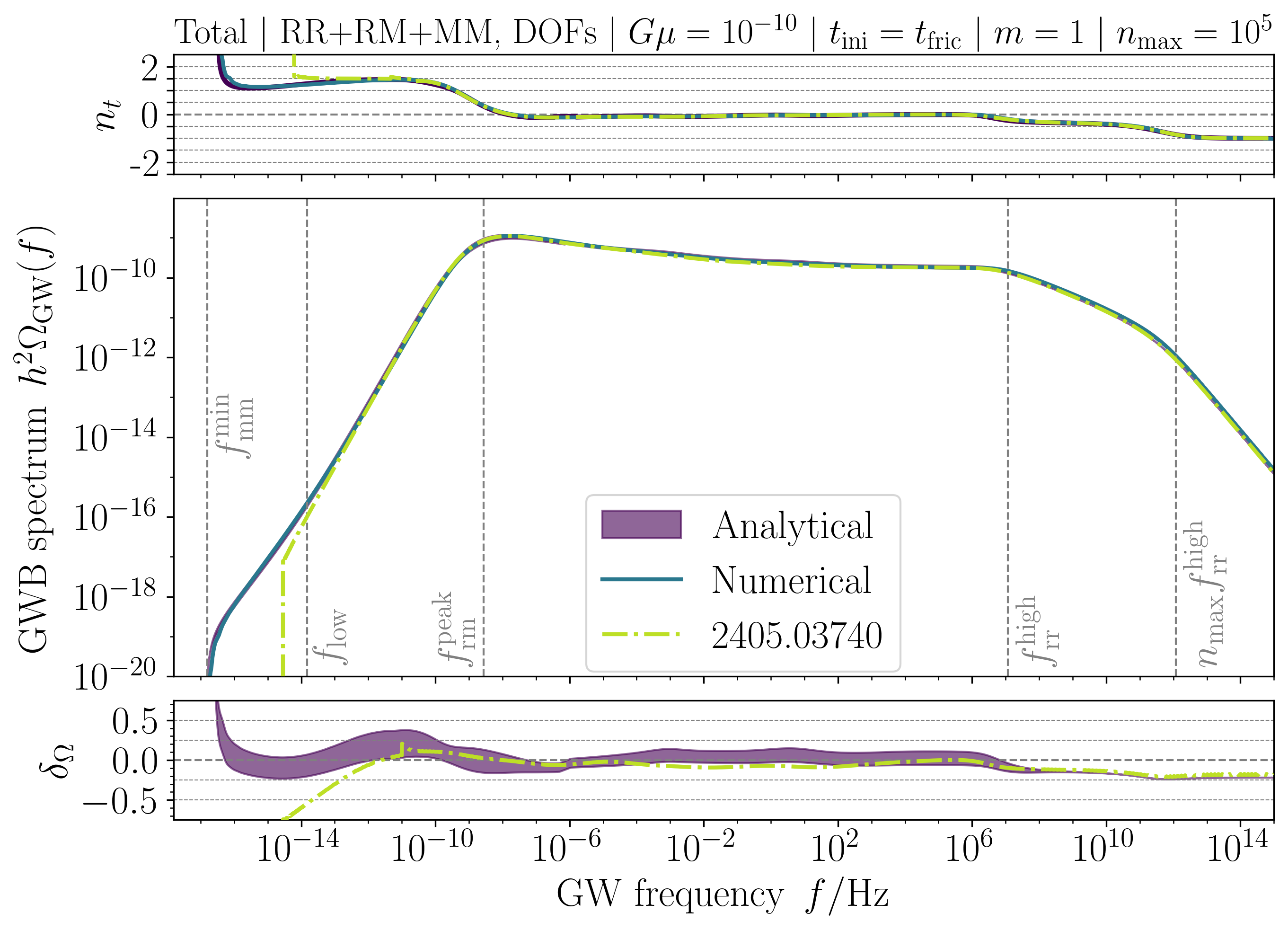}
\end{overpic}
\caption{  }
\label{fig:Full_Total_6}
    \end{subfigure}
    \caption{\footnotesize
     Same as in Fig.~\ref{fig:Full_Total} but variations in the effective number of relativistic DOFs are taken into account.}
     \label{fig:Full_Total_Varying}
\end{figure}

\subsection{Full spectrum}
We finally turn to a presentation of the complete spectra that can be obtained by combining the above results. A comprehensive summary of these can be found in Section \ref{sec:Summary_of_Formulae}. In Fig.~\ref{fig:Full_Total}, we present the four qualitatively different cases for a fixed effective number of relativistic DOFs, each covered by one benchmark scenario. In Fig.~\ref{fig:Full_Total_1}, none of the three contributions has a maximum frequency, in Fig.~\ref{fig:Full_Total_2}, the MM contribution has. Since the MM spectrum is subleading to the two other contributions, there is no visible qualitative difference though. In Fig.~\ref{fig:Full_Total_3}, both the MM and the RR spectrum have a maximum frequency, which implies that there is no plateau region and the overall spectrum is, apart from very low frequencies, the dominant one. Finally, in Fig.~\ref{fig:Full_Total_4}, all three contributions have a maximum frequency. {Therefore, this is a GWB spectrum from low-scale strings, as discussed in \cite{Schmitz_2024}, and we can clearly identify the characteristic oscillatory feature in the spectral index $n_t$ of the numerical spectrum. Our analytical description does not capture the oscillatory feature, but it does reproduce the general shape of the spectrum. This is visible as subtle oscillations in $\delta_\Omega$.} For Figs.~\ref{fig:Full_Total_1} and \ref{fig:Full_Total_2} in which the spectrum contains a plateau region, variations in the effective number of relativistic DOFs during radiation domination will be important. In Fig.~\ref{fig:Full_Total_Varying}, we show the full spectra taking this effect into account. We want to emphasize that our analytical expressions, even taking merely $m=1$, deviate by less than $\sim 25\%$ from the fully numerical spectrum (except upon approaching the minimum or maximum frequencies). Moreover, we are able to analytically describe for the first time the spectrum for the low-scale case, i.e., for low string tensions and late initial times as given, e.g., in Fig.~\ref{fig:Full_Total_4}.   

\begin{figure}
\tiny
    \centering
    \begin{subfigure}{0.49\textwidth}
    \begin{overpic}[width = \textwidth]{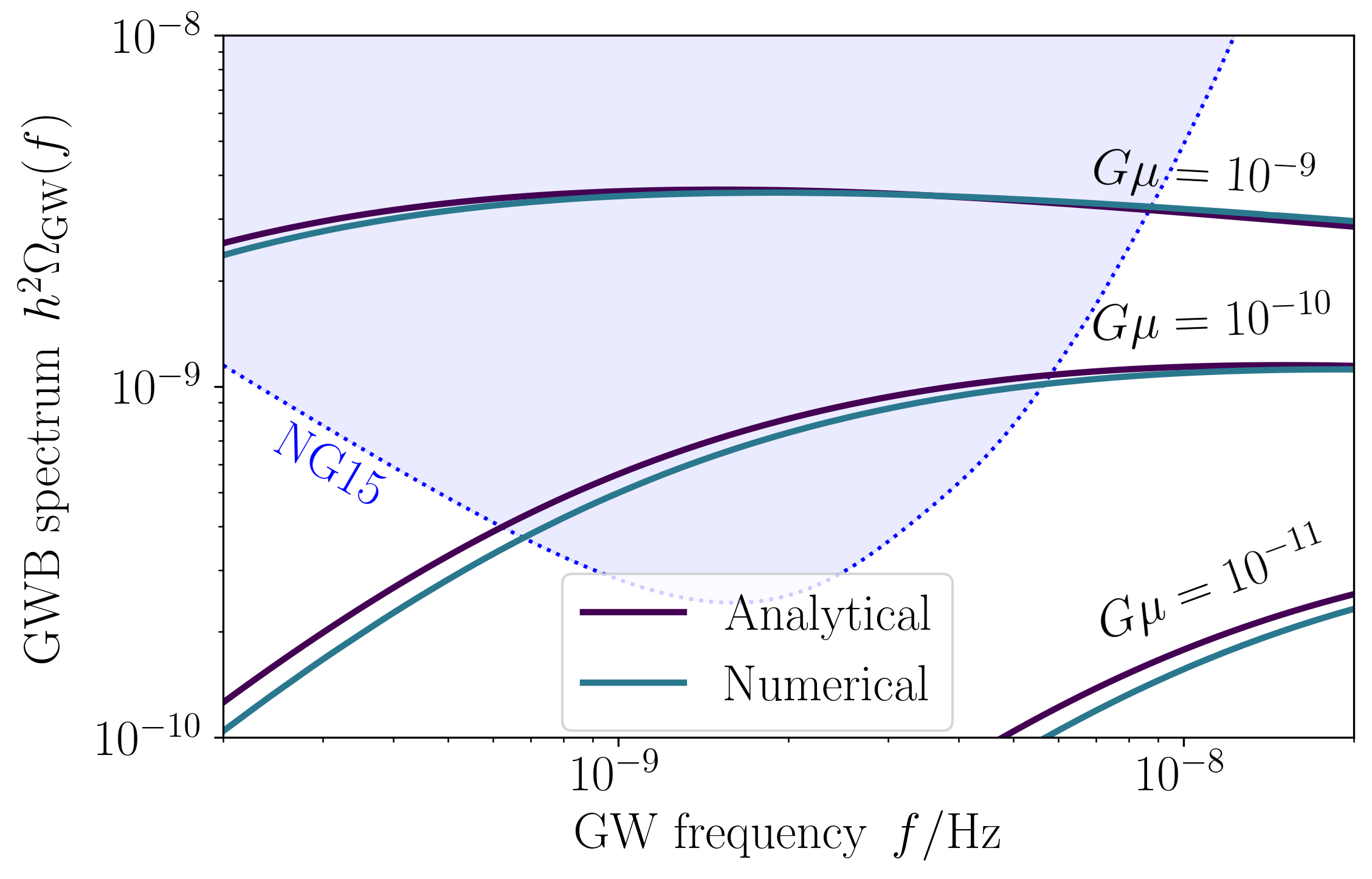}
\end{overpic}
\caption{ }
\label{fig:NG15}
    \end{subfigure}
    \begin{subfigure}{0.49\textwidth}
    \begin{overpic}[width = \textwidth]{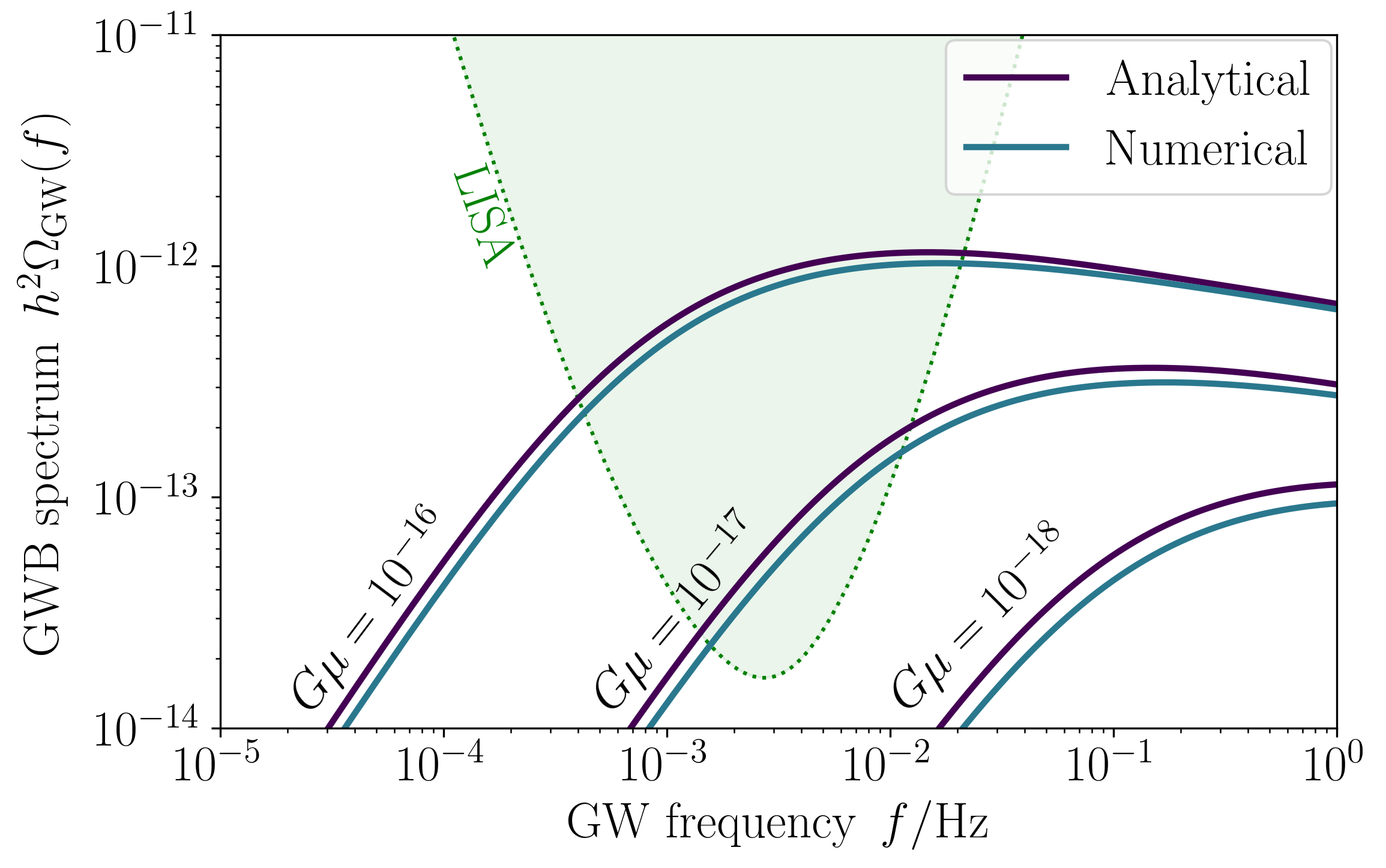}
\end{overpic}
\caption{ }
\label{fig:LISA}
    \end{subfigure}
    
\bigskip

    \begin{subfigure}{0.49\textwidth}
   \begin{overpic}[width = \textwidth]{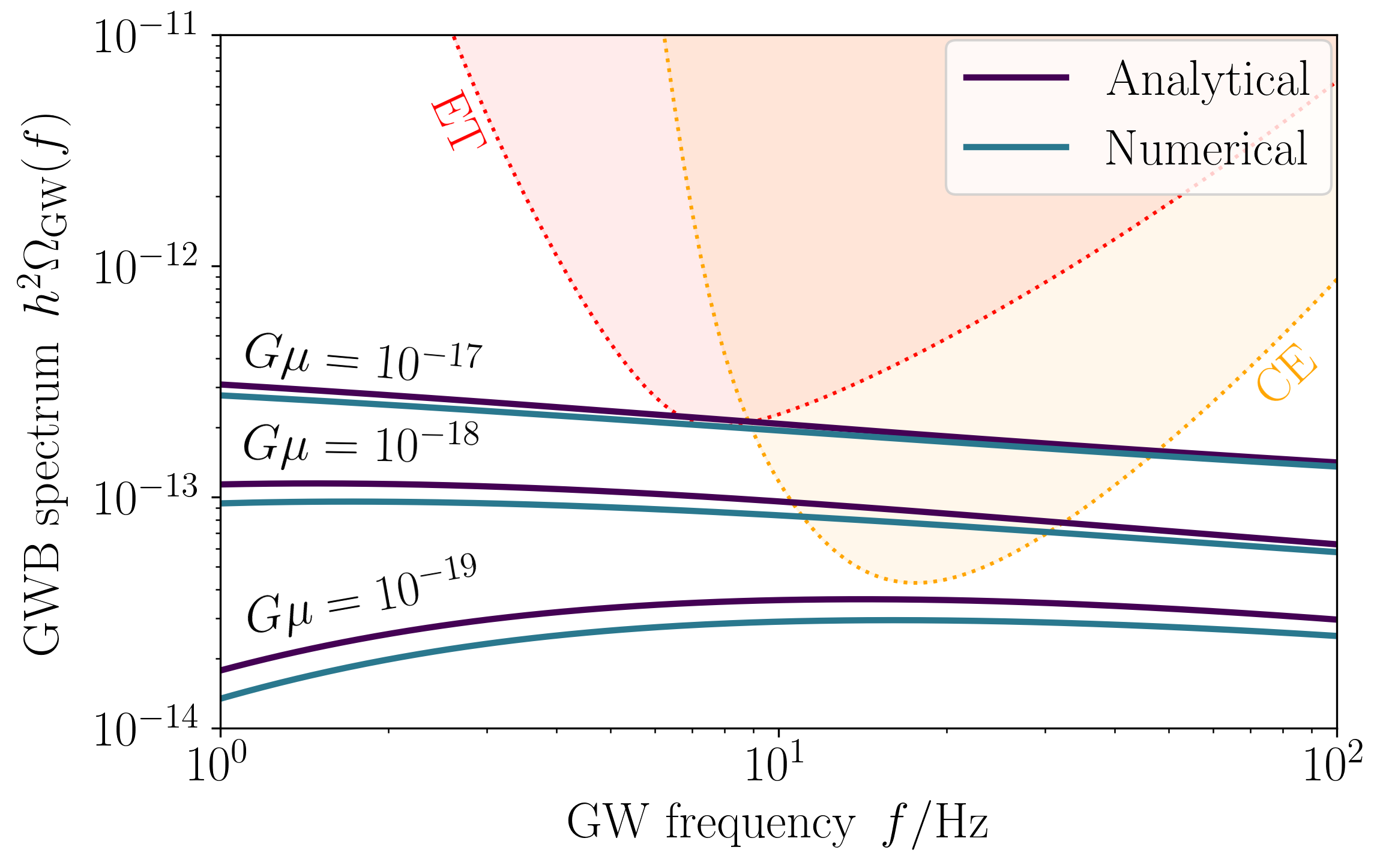}
\end{overpic}
\caption{  }
\label{fig:EinsteinTelescope}
    \end{subfigure}
       \begin{subfigure}{0.49\textwidth}
   \begin{overpic}[width = \textwidth]{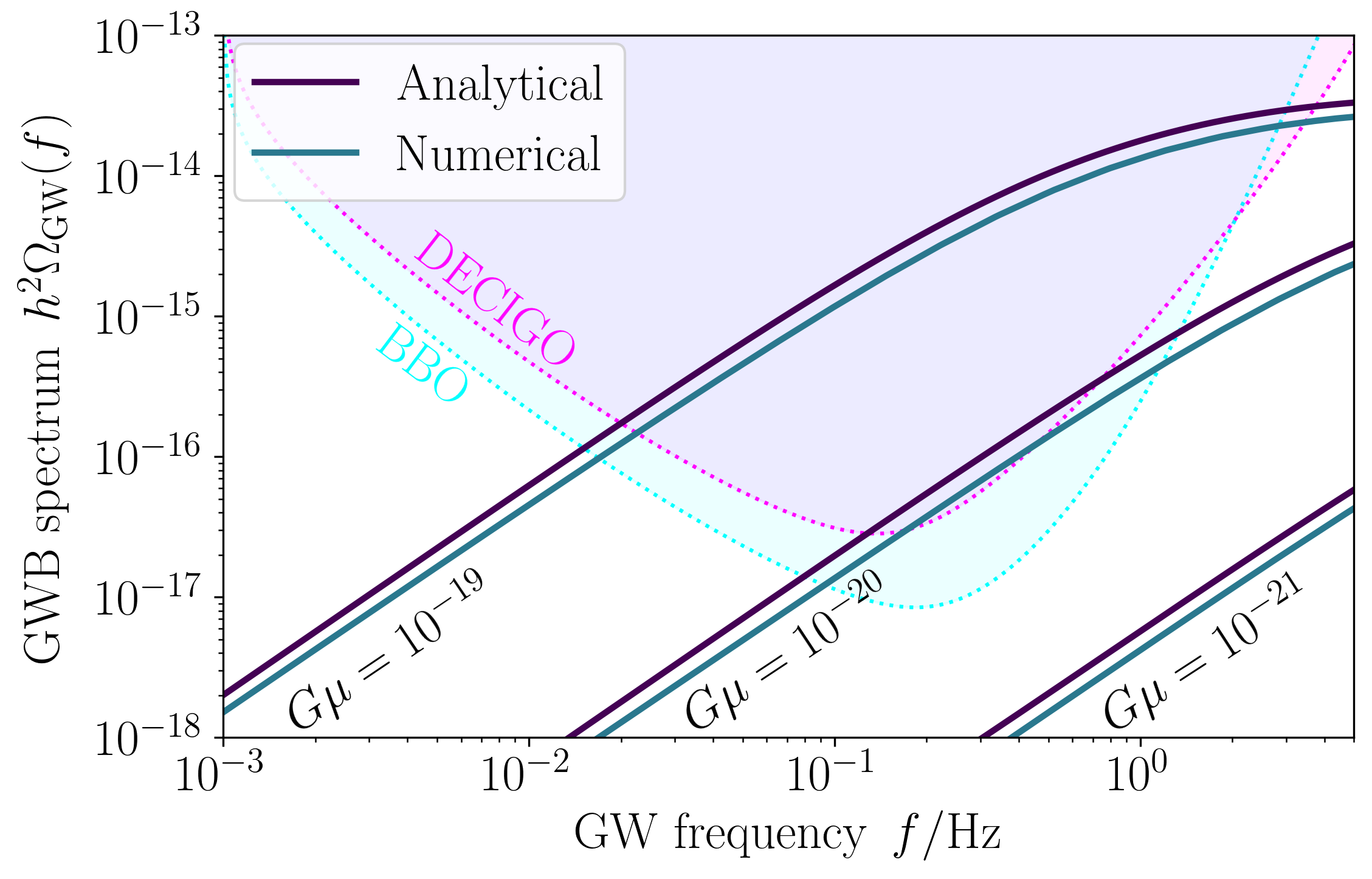}
\end{overpic}
\caption{  }
\label{fig:BBO}
    \end{subfigure}
    \caption{\footnotesize
     Plots showing the sensitivity curves of NANOGrav for their 15-year data (NG15) in panel \ref{fig:NG15}, LISA in panel \ref{fig:LISA}, Einstein Telescope (ET) and Cosmic Explorer (CE) in panel \ref{fig:EinsteinTelescope}, as well as Big Bang Observer (BBO) and DECIGO in panel \ref{fig:BBO}. In all plots, we show the SGWB spectra computed for three different string tensions $G\mu$ such that the lowest spectrum is just out of reach of the respective experiments. We fixed $t_{\rm ini}=t_{\rm fric}$. For each spectrum, we present our numerical spectrum (teal) and the spectrum calculated analytically in our formalism ({deep purple line obtained as the geometric mean of lower and upper bounds given in Eq.~\eqref{Upperandlowerbound}, evaluated for $m=1$ and $n_{\rm max}=10^5$.}) accounting for variations in the effective number of DOFs. Sensitivity curves were taken from Ref.~\cite{the_nanograv_collaboration_2023_8092346} (NG15), and Ref.~\cite{schmitz_2020_3689582} (all other).}
     \label{fig:Experiments}
\end{figure}

The provided expressions for the SGWB yield an excellent substitute for
the otherwise time-consuming computation necessary to numerically calculate the spectrum. This makes our results very useful for data analyses in current and future experiments. In Fig.~\ref{fig:Experiments}, we illustrate this by plotting different spectra (accounting for changes in DOFs) on top of the power-law integrated sensitivity curves for PTAs, ground-based, and space-borne interferometers. 

\section{Discussion and conclusions}
In this article, we presented a detailed derivation of analytical formulae for the stochastic gravitational wave background produced by a network of stable, local cosmic strings. Our treatment properly describes features of both the spectra from the fundamental mode and the total spectra from all modes. {In this way, we add to the existing literature (see~\cite{Sousa_2020, Blanco-Pillado:2024aca}), as our templates can account for some features that could not be described by expressions previously found in the literature.} In particular, this includes maximum and minimum frequencies above and below which the spectrum strictly vanishes. Furthermore, our formulae allow for the first time to analytically compute the gravitational wave background from strings with very low tensions for which qualitatively new spectral shapes arise. On top of that, we also provided expressions that analytically approximate the suppression of the spectrum at high frequencies due to a decrease in the effective number of DOFs with time in the early Universe. We also quantitatively discussed how the spectrum can be affected at ultrahigh frequencies that have wavelengths short enough to resolve the finite width of the strings. In difference to previous conclusions {(see \cite{Servant:2023tua})}, we found that the finite width affects all mode contributions at the same frequency and is, thus, mode-number-independent. 

Apart from providing a detailed understanding of different features of the gravitational wave spectra, our expressions for the total spectrum provide an easy way to compute the SGWB at all relevant frequencies without having to sum over a huge number of modes explicitly. This reduces computing times significantly. Our results will, therefore, be useful for gravitational wave searches with PTAs, ground-based interferometers, and space-based interferometers. 

{Using the spectra from our analytical template instead of our exact numerical spectra introduces a reconstruction error when parameters are inferred from data. As a simple estimate, we evaluated the numerical spectra shown in Fig.~\ref{fig:Experiments} at typical frequencies (where the shown experiments are most sensitive), 
and determined the string tension that our template requires to reproduce the numerically computed spectral amplitude. This procedure reveals a typical reconstruction error of $\lesssim 1\%$ in $\log_{10}(G\mu)$ and $\lesssim 30\%$ in $G\mu$.
}

While our studies are restricted to the case of cosmic strings produced in the spontaneous breaking of a local symmetry, {current modeling predicts that the GW signal from color-flux tubes can simply be obtained from the one of field strings by rescaling the amplitude of the GW spectrum with a constant prefactor~\cite{Yamada_2022, Yamada:2022aax}. The same is, at least in the range of nanohertz frequencies, true for cosmic superstrings, although more improved modeling shows this to be an oversimplification, in particular for a wider range of frequencies~\cite{Sousa_2016, Avgoustidis:2025svu}.} This makes our analytical results also relevant for those. In the future, the procedure laid out in this paper could be applied in a similar fashion to study the more complicated stochastic gravitational wave background from a network of metastable cosmic strings \cite{Preskill:1992ck, Buchmuller:2023aus}. 

As mentioned earlier, comparing the simple analytical results obtained in this paper to the very recent numerical spectra of Ref.~\cite{Wachter:2024zly} in Fig.~2 by eye, we can already tell that our analytical results will provide an accurate approximation up to corrections of maximally order $1$. {In future work \cite{Schmitz:2025abc}, we are going to compare our simple analytical results to the spectra of Ref.~\cite{Wachter:2024zly} more rigorously. In some preliminary work, a $\chi^2$ minimization between our spectra, rescaled with a fudge factor $\mathcal{C}_f$, and the spectra of Ref.~\cite{Wachter:2024zly}, weighted by the NANOGrav sensitivity curves from Ref.~\cite{the_nanograv_collaboration_2023_8092346}, was performed over the corresponding frequency band. For values of $G\mu \sim 10^{-11}\dots 10^{-9}$, this analysis yields fudge factors $\mathcal{C}_f \sim 0.7 \dots 1.0$, which indicates quantitatively good agreement between our analytical model and the most recent simulations within these parameter and frequency ranges.}

\acknowledgments
The authors would like to thank Adeela Afzal for her initial collaboration at the early stages of this project and Ken Olum as well as Lara Sousa and Olivia Bitcon for valuable comments on the draft. The work of K.~S.\ and T.~S.\ is supported by Deutsche Forschungsgemeinschaft (DFG) through the Research Training Group (Graduiertenkolleg) 2149: Strong and Weak Interactions\,--\,from Hadrons to Dark Matter.

\clearpage
\section{Summary of formulae}
\label{sec:Summary_of_Formulae}
\begin{small}
\begin{tcolorbox}[breakable]
\textbf{Cosmological parameters}
\begin{align*}
    h^2\Omega_m =0.143  && h^2\Omega_r = 4.18 \times 10^{-5} && H_r^0 = 2.10\times 10^{-20} {\rm Hz} && H_m^0 = 1.23 \times 10^{-18} {\rm Hz}
\end{align*}
We split radiation domination into $N=4$ intervals, in which we assume the effective number of degrees of freedom and thus $\mathcal{G}(t)$ to be constant with
$a_{(0)} = a_{\rm ini}$, $a_{(4)}=a_{\rm eq}$, and 
\begin{align*}    
    a_{(1)}/a_0 =1.6\times 10^{-15} && a_{(2)}/a_0=3.4\times 10^{-12} && a_{(3)}/a_0= 2.0\times 10^{-9} && a_{(4)}/a_0= 2.9\times 10^{-4}\\
    \mathcal{G}_0 = 0.39 && \mathcal{G}_1 = 0.43&& \mathcal{G}_2 = 0.83&& \mathcal{G}_3 = 1
\end{align*}
If $a_{(i)}<a_{\rm ini}$ for any $i$ in the list, then $a_{(i)}$ and $\mathcal{G}_{i-1}$ need to be removed from below $a_{\rm ini}$ until it is the smallest scale factor. After removing $r$ list entries, the above formulae can be applied by relabeling $i\to i-r$ for $i>r$. Correspondingly, radiation domination is then only split into $N=4-r$ intervals.\\\\
\textbf{Standard values} (Sections \ref{subsec:String_Network_Evolution}, \ref{subsec:Loop_Evolution}, \ref{subsec:RR_Loops_Fundamental}, \ref{subsec:RM_Loops_Fundamental})
\begin{align*}
\Gamma &= 50 & \xi_{r}&=0.27 & \xi_m &=0.63 & \alpha &=0.37 \\
v_r&=0.66 & v_m&=0.58 & \tilde{c}&=0.23 & \mathcal{F} &= 0.1
\end{align*}
\textbf{Prefactors, amplitudes, and other expressions} (Sections \ref{subsec:Loop_Evolution}, \ref{subsec:RR_Loops_Fundamental}, \ref{subsec:MM_Loops_Fundamental}, \ref{subsec:FullFundamentalSpectrum})
\begin{align*}
    C_\beta = \mathcal{F}\frac{\tilde{c}}{\sqrt{2}\alpha} \frac{v_\beta}{\xi_\beta^4} (\alpha \xi_\beta + \Gamma G\mu)^{3(1-\beta)}  && \chi_{\beta} = \frac{\alpha \xi_{\beta}}{\Gamma G\mu} && \text{with } \beta = \begin{cases} 1/2 & \text{during RD} \\ 2/3 & \text{during MD}\end{cases}
\end{align*}
\begin{align*}
    \mathcal{A}_{\rm rr} &= \frac{128\pi}{9}\,C_r\,h^2\Omega_r \left(\frac{G\mu}{\Gamma}\right)^{1/2} &
    \mathcal{A}_{\rm rm} &= \sqrt{\frac{128}{3}} \pi C_r h^2 \Omega_m \left(\frac{{4} \Omega_r}{\Omega_m}\right)^{3/4} \left(\frac{G\mu}{\Gamma}\right)^{1/2}\\
    \mathcal{A}_{\rm mm} &= 18\pi\,C_m\,h^2\Omega_{m} \,G\mu 
\end{align*}
\textbf{Frequencies} (Sections \ref{subsec:RR_Loops_Fundamental}, \ref{subsec:VaryingDOFs}, \ref{subsec:RM_Loops_Fundamental}, \ref{subsec:MM_Loops_Fundamental}, \ref{subsec:FullFundamentalSpectrum}, \ref{subsec:FullVaryingDOFs})
\begin{align*}
f_r^i = \frac{4H_r^0}{\Gamma G\mu} \frac{a_0}{a(t_i)} &&
    f_k^{(i)} = \frac{4 \mathcal{G}_k^{1/2} H_r^0}{\Gamma G\mu} \frac{a_0}{a_{(i)}} && f_m^i = \frac{3H_m^0}{\Gamma G\mu} \left(\frac{a_0}{a(t_i)}\right)^{1/2}  && \tilde{f}_m^i = \frac{3H_m^0}{\Gamma G\mu} \left(\frac{a_0}{\tilde{a}_m(t_i)}\right)^{1/2} 
\end{align*}
with $\tilde{a}_m(t)/a_0 = \left(3/2 H_m^0 t\right)^{2/3} $.
\begin{align*}
f_{\rm rr}^{\rm min} &= \frac{f_r^{\rm eq}}{\chi_r} \,\,\,\,\,\,\,\,\,\,\,\,\,\,\,\,\,\,\,\,\,\,\,\, f^{{\rm start},*}_{\rm rr} = \frac{f_{r}^{\rm ini}}{\chi_r} &
f_{\rm rr}^{\rm max} &= \frac{f_r^{\rm eq}}{(1+\chi_r)\left(\frac{a_{\rm ini}}{a_{\rm eq}}\right)^2-1}   \\
    f_{A,i}^{\rm min} &= \frac{f_i^{(i+1)}}{\chi_r} &
    \hat{f}_{A,i}^{{\rm start}, *} &=\frac{f_i^{(i)}}{\chi_r\sqrt{1+\chi_r}} \\   
  \hat{f}_{B,ij}^{\rm start} &= \min\left\{f_i^{(i)}, f_i^{(j)}/\sqrt{1+\chi_r}\right\}& \ \hat{f}_{B, ij}^{\rm end} &= \max\left\{f_i^{(i+1)}, f_i^{(j+1)}/\sqrt{1+\chi_r}\right\} \\
    f_{\rm rm}^{\rm max} &= \frac{f_m^0}{\left(\frac{\tilde{a}_m^{\rm ini}}{a_0}\right)^{3/2}(1+\chi_r)-1} & f^{{\rm start}, *}_{\rm rm}&= \frac{f_m^{\rm eq}}{\left(\frac{\tilde{a}^{\rm ini}_{\rm m}}{a_{\rm eq}}\right)^{3/2}(1+\chi_r) - 1} \\
    f_{\rm mm}^{\rm min} &= \frac{f_m^0}{\chi_m} \,\,\,\,\,\,\,\,\,\,\,\,\,\,\,\,\,\,\,\,\,\,\,\,
    f_{\rm mm}^{{\rm start}, *} = \frac{\tilde{f}_m^{\rm eq}}{\chi_m} 
    & 
    f_{\rm mm}^{\rm max} &= \frac{f_m^0}{(1+\chi_m)\left(\frac{\tilde{a}_m^{\rm eq}}{a_0}\right)^{3/2} - 1} 
\end{align*}
\textbf{Mode numbers} (Sections \ref{subsec:Full_RR_Loops}, \ref{subsec:FullVaryingDOFs}, \ref{subsec:Full_RM_MM_Loops})
\begin{align*}
k_{\rm rr}^{\rm min}= \max\left\{m, f/f_{\rm rr}^{\rm max}\right\} &&
k_{\rm rr}^{\rm max}= \max\left\{\min\left\{n_{\rm max}, f/f_{\rm rr}^{\rm min}\right\},k_{\rm rr}^{\rm min}\right\}\\ 
k_{\rm rr}^{\rm start} = \min\left\{\max\left\{k_{\rm rr}^{\rm min}, f/f_{\rm rr}^{{\rm start}, *}\right\}, k_{\rm rr}^{\rm max}\right\} &&
    k_{A,i}^{\rm max} = \max\left\{\min\left\{n_{\rm max}, f/f_{A,i}^{\rm min}\right\}, m\right\} \\
    k_{A,i}^{\rm start} = \min\left\{ \max\left\{m, f/\hat{f}_{A,i}^{{\rm start},*}\right\}, k_{A,i}^{\rm max}\right\} &&
    k_{B,ij}^{\rm max} = \max\left\{\min\left\{n_{\rm max}, f/f_{\rm rr}^{\rm min}\right\}, m\right\} \\
   k_{\rm rm}^{\rm min}= \max\left\{m, f/f_{\rm rm}^{\rm max}\right\} && k_{\rm rm}^{\rm max} =\max\left\{\min\left\{n_{\rm max}, f/f_{\rm mm}^{\rm min}\right\}, k_{\rm rm}^{\rm min}\right\}\\  
   k_{\rm rm}^{\rm start} = \min\left\{\max\left\{k_{\rm rm}^{\rm min}, f/f_{\rm rm}^{{\rm start}, *}\right\}, k_{\rm rm}^{\rm max}\right\} &&
   k_{\rm mm}^{\rm min}= \max\left\{m, f/f_{\rm mm}^{\rm max}\right\} \\ k_{\rm mm}^{\rm max} = \max\left\{\min\left\{n_{\rm max}, f/f_{\rm mm}^{\rm min}\right\},k_{\rm mm}^{\rm min}\right\}&&  k_{\rm mm}^{\rm start} = \min\left\{\max\left\{k_{\rm mm}^{\rm min}, f/f_{\rm mm}^{{\rm start},*}\right\}, k_{\rm mm}^{\rm max}\right\}
\end{align*}
\textbf{Frequency ratios and auxiliary functions} (Sections \ref{subsec:RR_Loops_Fundamental}, \ref{subsec:VaryingDOFs}, \ref{subsec:RM_Loops_Fundamental}, \ref{subsec:MM_Loops_Fundamental}, \ref{subsec:FullVaryingDOFs})
\begin{align*}
  x_r^i = f/f_r^i &&  x_k^{(i)} = f/f_k^{(i)} && x_m^i = f/f_m^i && \tilde{x}_m^i = f/\tilde{f}_m^i
\end{align*}
\begin{align*}
\phi\left(y\right) &= 27/2\,y -1 + 3^{3/2} \left(27/4\,y^2 - y\right)^{1/2} \\
 \varphi_2\left(x,\chi\right) &= \left[1/4 + \left(1+\chi\right)x^2\right]^{1/2} - 1/2\\
\varphi_3\left(x,\chi\right)&= 1/3\left[\phi^{1/3}\left((1+\chi)x^3\right) + \phi^{-1/3}\left((1+\chi)x^3\right) -1\right]
\end{align*}
\begin{align*}
    x_{A, i}^{\rm start} &= \max\left\{\chi_r^{-1}, \varphi_2(x_i^{(i)},   \chi_r)\right\} &  x_{A, i}^{\rm end} &= x_i^{(i+1)} \\
    x_{B,ij}^{\rm start} &= \max\left\{x_i^{(i)}, \varphi_2(x_i^{(j)}, \chi_r)\right\} & x_{B, ij}^{\rm end} &= \min\left\{x_i^{(i+1)}, \varphi_2(x_i^{(j+1)}, \chi_r)\right\}\\
 x_{\rm rm}^{\rm start} &= \max\left\{x_m^{\rm eq},\varphi_3\left(\tilde{x}_m^{\rm ini},\chi_r\right)\right\}    & x_{\rm mm}^{\rm start} &= \max\big\{\chi_m^{-1},\varphi_3\left(\tilde{x}_m^{\rm eq},\chi_m\right)\big\}
\end{align*}
\begin{align*}
     \hat{x}_{B,ij}^{\rm start (end)} = f/\hat{f}_{B,ij}^{\rm start (end)}
\end{align*}

\textbf{Shape functions} (Sections \ref{subsec:RR_Loops_Fundamental}, \ref{subsec:RM_Loops_Fundamental}, \ref{subsec:MM_Loops_Fundamental}, \ref{subsec:Full_RR_Loops}, \ref{subsec:Full_RM_MM_Loops})\\
We use $_2F_1$ to denote the ordinary hypergeometric function and $H_{n_{\rm max}}^{(q)}$ for the $n_{\rm max}$-th generalized harmonic number of order $q$.
\begin{align*}
S_{\rm rr}^{(1)}\left(x\right) &= \left(\frac{x}{1+x}\right)^{3/2} \;\;\;\;\;\;\;\;\;\; \;\;\;\;\;\;\;\;\;\;
S_{\rm rm}^{(1)}\left(x\right) = \frac{2+3\,x}{(x_m^0)^{1/2}\left(1+x\right)^{3/2}} \\
S_{\rm mm}^{(1)}\left(x\right) &=\frac{1}{(x_m^0)^2}\left( \frac{x^2+x-1}{1+x} - 2\,\ln\left(1+x\right)\right) \\
    S_{\rm rr,1}(k, x) &= \begin{cases} \frac{2  \Gamma (2-q)}{(q-1) (2 q+1) \Gamma (1-q)} \frac{x^{3/2}}{k^{1/2+q}} & \text{if } x< 10^{-3} \\
    -\frac{{_2F_1}\left[\frac{3}{2}, 1-q, 2-q, -\frac{k}{x}\right]}{(q-1)k^{q-1} }  & \text{else}\end{cases}   \\
    S_{\rm rr, 2}(k, x) &= \begin{cases} 
   \frac{x^3}{k^{q+2}} \left(-\frac{1}{q+2}+\frac{3}{ q+4}\frac{x^2}{k^2}-\frac{6}{q+6}\frac{x^4}{k^4}\right) & \text{if } \frac{x}{\sqrt{1+\chi_r}} <10^{-8} \lor f_{\rm rr}^{\rm max}>0  \\
    S_{\rm rr, 1}(k, x) & \text{else} 
    \end{cases} \\
    S_{\rm rr,3}(k) &= -\frac{S_{\rm rr}^{(1)}(\chi_r^{-1})}{(q-1)k^{q-1}}
\end{align*}
\begin{align*}
S_{\rm rm, 1}(k,x) &= \begin{cases}  \frac{3}{2(3+2q)k^{3/2+q}\sqrt{x_m^0}} x^3 & \text{if } x_m^0 <10^{-5} \lor f_{\rm rm}^{\rm max}>0 \\
S_{\rm rm, 2}(k, x) & \text{else} 
\end{cases}\\
    S_{\rm rm, 2}(k,x) &=\begin{cases} \frac{3k^{-1/2-q} x^2}{\left(2+4q\right) \sqrt{x_m^0}} & \text{if } x_m^0 <10^{-5} \lor f_{\rm rm}^{\rm max}>0\\ \frac{2k^{2-q}+2 (2 q-1) (k+x) \left(-x\right)^{1-q} \, _2F_1\left[\frac{1}{2},q-1;\frac{3}{2};\frac{k+x}{x}\right]}{\sqrt{x_m^0 (k+x)}} & \text{if } 10^{-5}\leq x_m^0< 10^{10} \land f_{\rm rm}^{\rm max}<0\\ \frac{3 k^{2-q}}{\left(2-q\right) \sqrt{x_m^0 x}} & \text{else}
    \end{cases}\\
\tilde{S}_{\rm mm, 1}(k, x) &= \left(\frac{x}{x_m^0}\right)^2 \frac{k^{1-q} \left(\, _2F_1\left(1,1-q;2-q;-\frac{k}{x}\right)-2 \left(\frac{k}{x}\right)^2 \ln \left(1+\frac{x}{k}\right)-\left(\frac{k}{x}-1\right)^2\right)}{3-q}  \\
    \tilde{S}_{\rm mm, 3}(k,x ) &=  \frac{k^{3-q}}{(q-3) \left(x_m^0\right)^2} \Biggl\{1-{_2}F_1\left(1,2-\frac{2 q}{3};3-\frac{2 q}{3};-\left(\frac{k}{x}\right)^{3/2}\right) +\\+ 2 \ln &\left(1+\left(\frac{x}{k}\right)^{3/2}\right)-\frac{2}{2q-3}\left(\frac{x}{k}\right)^{3/2} \left(3  {_2}F_1\left(1,\frac{2 q}{3}-1;\frac{2 q}{3};-\left(\frac{x}{k}\right)^{3/2}\right)+q-3\right)\Biggr\}
    \end{align*}
    \begin{align*}
    S_{\rm mm,1}(k,x) &= \begin{cases} \left(\frac{x}{x_m^0}\right)^2\left(-\frac{x}{3q k^q}+\frac{x^2}{2(1+q)k^{1+q}}\right) & \text{if } x_m^0 < 10^{-2}\\
    \tilde{S}_{\rm mm,1}(k,x), & \text{if } 10^{-2} \leq x_m^0 < 10^6\\
    \frac{k^{2-q}}{(x_m^0)^2}\left(\frac{x}{2-q}-\frac{2k}{(3-q)^2} - \frac{2k}{3-q}\ln\left(\frac{x}{k}\right) - \frac{3k^2}{(4-q)x}\right) & \text{else}
    \end{cases} \\
     S_{\rm mm, 2}(k) &= \begin{cases}
         \frac{k^{3-q}}{3-q}\left(\frac{1}{x_m^0}\right)^2\left(\frac{\chi_m^{-3}}{3}-\frac{\chi_m^{-4}}{2}\right) & \text{if } x_m^0 < 10^{-2} \\
         \frac{k^{3-q}}{3-q}\left(\frac{1}{x_m^0}\right)^2\left(\frac{\chi_m^{-3}}{3}-\frac{\chi_m^{-4}}{2}-1\right) & \text{else}
     \end{cases} \\
    S_{\rm mm, 3}(k,x) &= \begin{cases}
   - \frac{2x^{9/2}}{3(3+2q)k^{3/2+q}(x_m^0)^2} & \text{if } x_m^0 < 10^{-2} \\  
    \tilde{S}_{\rm mm, 3}(k, x) & \text{if } 10^{-2} \leq x_m^0 < 10^3\\
        S_{\rm mm, 1}(k,x) & \text{else} 
    \end{cases}
\end{align*}

\textbf{RR contribution} (Sections \ref{subsec:VaryingDOFs}, \ref{subsec:Full_RR_Loops}, \ref{subsec:FullVaryingDOFs})\\
We denote by $N$ the number of intervals during radiation domination for which we assume the effective number of degrees to be roughly constant (see \textit{Cosmological parameters})
\begin{align*} 
   h^2\Omega_{\rm rr}^{(1)} = \mathcal{A}_{\rm rr}\,\sum_{i=0}^{N-1}\Bigg(& \mathcal{G}_i\, S_{\rm rr}^{(1)}(x)\Big\vert_{x_{A,i}^{\rm start}}^{x_{A,i}^{\rm end}} \:\Theta(x_{A,i}^{\rm end} - x_{A,i}^{\rm start}) +\\ &+ \sum_{j=0}^{i-1}\mathcal{G}_j^{3/4}\mathcal{G}_i^{1/4} \, S_{\rm rr}^{(1)}(x)\Big\vert_{x_{B,ij}^{\rm start}}^{x_{B,ij}^{\rm end}}\: \Theta(x_{B,ij}^{\rm end} -x_{B, ij}^{\rm start}) \Bigg) 
\end{align*}
\\\\
If $f_{\rm rr}^{\rm max}\geq 0$:
\begin{align*}
    h^2 \Omega_{\rm rr}^{m, {\rm upper}} =  &\frac{1}{H_{n_{\rm max}}^{(q)}}\sum_{k=1}^m \frac{1}{k^q} h^2\Omega_{\rm rr}^{(1)}\left(\frac{f}{k}\right) +\\ &+ \frac{\mathcal{A}_{\rm rr}}{H_{n_{\rm max}}^{(q)}} \left\{S_{\rm rr,1}(k, x_r^{\rm eq})\big\vert_{k_{\rm rr}^{\rm min}}^{k_{\rm rr}^{\rm max}} - S_{\rm rr,2}(k, \sqrt{1+\chi_r}x_r^{\rm ini})\big\vert_{k_{\rm rr}^{\rm min}}^{k_{\rm rr}^{\rm start}} - S_{\rm rr,3}\left(k\right)\big\vert_{k_{\rm rr}^{\rm start}}^{k_{\rm rr}^{\rm max}} \right\} 
\end{align*}
\\
If $f_{\rm rr}^{\rm max} <0$:
\begin{align*}
    h^2 &\Omega_{\rm rr}^{m, \text{upper}} = \frac{1}{H^{(q)}_{n_{\rm max}}} \sum_{k=1}^m \frac{1}{k^q} h^2 \Omega_{\rm rr}^{(1)}\left(\frac{f}{k}\right) +  \\\nonumber &+ \frac{\mathcal{A_{\rm rr}}}{H^{(q)}_{n_{\rm max}}} \sum_{i=0}^{N-1}\Theta_i^A \mathcal{G}_i \left\{S_{\rm rr,1}\left(k, x_i^{(i+1)}\right)\Big\vert_{m}^{k_{A,i}^{\rm max}} - S_{\rm rr,1}\left(k, \sqrt{1+\chi_r}x_i^{(i)}\right)\Big\vert_{m}^{k_{A,i}^{\rm start}} - S_{\rm rr,3}\left(k\right)\Big\vert_{k_{A,i}^{\rm start}}^{k_{A,i}^{\rm max}} \right\}+\\\nonumber &+\frac{\mathcal{A}_{\rm rr}}{H_{n_{\rm max}}^{(q)}} \sum_{i=1}^{N-1}\sum_{j=0}^{i-1}\Theta_{ij}^B \mathcal{G}_i^{1/4}\mathcal{G}_j^{3/4} \biggl\{ S_{{\rm rr}, 1}\left(k,\hat{x}_{B,ij}^{\rm end}\right)\Big\vert_{m}^{k_{B,ij}^{\rm max}} - S_{{\rm rr},1}\left(k,\hat{x}_{B,ij}^{\rm start}\right)\Big\vert_{m}^{k_{B,ij}^{\rm max}} \biggr\} 
\end{align*}
with
\begin{align*}
    \Theta_i^A = \Theta\left(\frac{a_{(i+1)}}{a_{(i)}}-\sqrt{1+\chi_r}\right) && \text{and} && \Theta_{ij}^B = \Theta\left(\hat{f}^{\rm start}_{B, ij} - \hat{f}^{\rm end}_{B, ij}\right) 
\end{align*}

\textbf{RM contribution} (Section \ref{subsec:RM_Loops_Fundamental}, \ref{subsec:Full_RM_MM_Loops})
\begin{align*}
h^2&\Omega_{\rm rm}^{(1)} = \mathcal{A}_{\rm rm} \Theta\left(f-f_{\rm mm}^{\rm min}\right)\Theta\left(x_m^0 - x_{\rm rm}^{\rm start}\right) \left[S_{\rm rm}^{(1)}(x_{\rm rm}^{\rm start}) -S_{\rm rm}^{(1)}(x_{m}^0)\right] \\
   h^2 &\Omega_{\rm rm}^{m, \text{upper}}=\frac{1}{H_{n_{\rm max}}^{(q)}}\sum_{k=1}^m \frac{1}{k^q} h^2 \Omega_{\rm rm}^{(1)}\left(\frac{f}{k}\right) + \\ \nonumber &+\frac{\mathcal{A}_{\rm rm}}{H_{n_{\rm max}}^{(q)}}\left\{ S_{\rm rm, 1}\left(k, (1+\chi_r)^{1/3}\tilde{x}_m^{\rm ini}\right)\big\vert_{k_{\rm rm}^{\rm min}}^{k_{\rm rm}^{\rm start}} + S_{\rm rm, 2}\left(k, x_{m}^{\rm eq}\right)\big\vert_{k_{\rm rm}^{\rm start}}^{k_{\rm rm}^{\rm max}} - S_{\rm rm, 2}\left(k, x_m^0\right)\big\vert_{k_{\rm rm}^{\rm min}}^{k_{\rm rm}^{\rm max}}\right\}
\end{align*}
\textbf{MM contribution} (Section \ref{subsec:RM_Loops_Fundamental}, \ref{subsec:MM_Loops_Fundamental}, \ref{subsec:FullFundamentalSpectrum}, \ref{subsec:Full_RM_MM_Loops})
\begin{align*}
h^2&\Omega_{\rm mm}^{(1)} = \mathcal{A}_{\rm mm} \Theta\left(x_m^0 - x_{\rm mm}^{\rm start}\right) \left[S_{\rm mm}^{(1)}(x_m^0) - S_{\rm mm}^{(1)}(x_{\rm mm}^{\rm start})\right] \\
 h^2 &\Omega_{\rm mm}^{m , \text{upper}} =\frac{1}{H_{n_{\rm max}}^{(q)}}\sum_{k=1}^m \frac{1}{k^q} h^2 \Omega_{\rm mm}^{(1)}\left(\frac{f}{k}\right)+\\&\nonumber+\frac{\mathcal{A}_{\rm mm}}{H_{n_{\rm max}}^{(q)}}\left\{S_{\rm mm,1}\left(k, x_m^0\right)\big\vert_{k_{\rm mm}^{\rm min}}^{k_{\rm mm}^{\rm max}} - S_{\rm mm,3}\left(k, (1+\chi_m)^{1/3}\tilde{x}_m^{\rm eq}\right)\big\vert_{k_{\rm mm}^{\rm min}}^{k_{\rm mm}^{\rm start}} - S_{\rm mm, 2}(k)\big\vert_{k_{\rm mm}^{\rm start}}^{{k_{\rm mm}^{\rm max}}} \right\} 
\end{align*}
\textbf{Full spectrum}
\begin{align*}
    h^2\Omega_{\rm GW}^{m, {\rm upper}} = h^2 \Omega_{\rm rr}^{m, {\rm upper}} + h^2 \Omega_{\rm rm}^{m, {\rm upper}} + h^2 \Omega_{\rm mm}^{m, {\rm upper}}
\end{align*}
To calculate the full spectrum, one furthermore needs to choose values for $q$, $m$, $n_{\rm max}$, and $t_{\rm ini}$. For our plots, we used $q=4/3$, $m=1$, and $n_{\rm max}=10^5$. {The parameter $n_{\rm max}$ appears only in the maximum mode numbers and the limit $n_{\rm max}\to \infty$ can be obtained by removing it from these expressions. For example, for $k_{\rm rr}^{\rm max}$ this is achieved via the replacement $\min\left\{n_{\rm max}, f/f_{\rm rr}^{\rm min}\right\}\to f/f_{\rm rr}^{\rm min}$ and analogously for all other mode numbers.} A typical choice is $t_{\rm ini}=t_{\rm Pl}/(G\mu)^2$.
$h^2 \Omega_{\rm GW}^{m, {\rm lower}}$ is obtained by replacing $(m,n_{\rm max})\mapsto (m+1, n_{\rm max}+1)$ in the mode numbers. If one is interested in only one value of the spectrum instead of an upper and lower bound, one may refer to either of these bounds or ideally take the geometric mean $h^2\Omega_{\rm GW}^{m} = \sqrt{ h^2\Omega_{\rm GW}^{m, {\rm upper}}  h^2\Omega_{\rm GW}^{m, {\rm lower}}}$.  
\end{tcolorbox}
\end{small}

\newpage
\bibliography{references}
\bibliographystyle{JHEP}

\end{document}